\documentclass{aastex62}
\usepackage{graphicx}
\usepackage{subfigure}
\usepackage{epstopdf}
\usepackage{amsmath} 
\usepackage{amsmath}
\usepackage{bm}
\usepackage{graphicx}
\usepackage{multirow}
\usepackage{textcomp}
\usepackage{multirow}
\usepackage{natbib}
\usepackage{color}
\hypersetup{colorlinks,
linkcolor=blue,
citecolor=blue,
filecolor=blue,
urlcolor=blue}

\usepackage{soul}

\usepackage{enumerate}
\usepackage{textcomp}

\usepackage{multirow}
\usepackage{enumerate}

\shorttitle{Post-Newtonian corrections ..}
\shortauthors{Kazemi et al.}

\begin{document}

\title{Post-Newtonian corrections to Toomre's criterion}

\correspondingauthor{Mahmood Roshan}
\email{mroshan@um.ac.ir}

\author{Ali Kazemi}
\affil{Department of Physics, Ferdowsi University of Mashhad, P.O. Box 1436, Mashhad, Iran}

\author{Mahmood Roshan}
\affiliation{Department of Physics, Ferdowsi University of Mashhad, P.O. Box 1436, Mashhad, Iran}

\author{Elham Nazari}
\affiliation{Department of Physics, Ferdowsi University of Mashhad, P.O. Box 1436, Mashhad, Iran}

\begin{abstract}

The gravitational stability of a two-dimensional self-gravitating and differentially rotating gaseous disk in the context of post-Newtonian (hereafter PN) theory is studied. Using the perturbative method and applying the second iterated equations of PN approximation, the relativistic version of the dispersion relation for the propagation of small perturbations is found. We obtain the PN version of Toomre's local stability criterion by utilizing this PN dispersion relation. In other words, we find relativistic corrections to Toomre's criterion in the first PN approximation.
Two stability parameters $\eta$ and $\mu$ related to gravity and pressure are introduced. We illustrate how these parameters determine the stability of the Newtonian and PN systems. Moreover, we show that, in general, the differentially rotating fluid disk is more stable in the context of PN theory relative to the Newtonian one. Also, we  explicitly show that although the relativistic PN corrections destabilize non-rotating systems, they have the stabilizing role in the rotating thin disks. Finally, we apply the results to the relativistic disks around hypermassive neutron stars (HMNSs), and find that although Newtonian description predicts the occurrence of local fragmentations, PN theory  remains in agreement with the relevant simulations, and rules out the existence of local fragmentations.

\end{abstract}


\section{INTRODUCTION}

It has been an interesting challenge to achieve a criterion for gravitational instability of fluid and stellar disks since almost 60 years ago. The first fundamental studies can be found in \cite{safronov1960gravitational} and \cite{toomre1964gravitational}.
In these primary works, gravitational instability was studied for understanding the dynamics of disks and particularly the structure of spiral arms.
On the other hand, it is believed that star formation occurs in spiral arms. This process influences the main characteristics of the galactic disk and its interstellar gas (\citealp{kormendy2004secular,sellwood2014secular}).
Consequently, nowadays it is more interesting to achieve a relation between gravitational instability and star formation rate (\citealp{mckee2007theory,leroy2008star}). 
Lin and Shu in 1960 claimed that the spiral structure is a density wave. This picture has been not changed and consequently the linear analysis could be applied to study the gravitational stability of the disk.

On the other hand, because of the long-range nature of gravitational force, it is very difficult to find a criterion for gravitational stability. In fact, only for a special case this task can be done analytically, see \cite{kalnajs1972equilibria}.
In fact,  this feature of the gravitational force leads to some difficulties to compute the gravitational potential of a spiral arm.
Consequently, the global stability analysis of a galactic disk cannot be done straightforwardly (\citealp{kalnajs1977dynamics,jalali2007unstable}). 
However, for tightly wound density waves, one can ignore the long-range effects and find an analytical approach. This is a local approach, known as WKB approximation, in which the relevant analysis does not depend on the given model for the background disk.

Using this approximation one can find a criterion for gravitational stability of  a one component rotating razor-thin disk (\citealp{binney2008galactic}). Furthermore a more complex case is a zero thickness galactic disk consisting of two components, i.e., stars and gas. This case has been widely studied, for example, see \citealp{jog1984two,bertin1988global,wang1994gravitational,jog1996local,rafikov2001local}
 and \cite{shadmehri2012gravitational} for a two-component study where the gaseous component is turbulent.
In addition, there is a variety of analytical and numerical studies for disks with finite thickness, for example, see   \cite{vandervoort1970equilibria}; \cite{romeo1990thesis,romeo1992stability,romeo1994faithful}.

The effect of magnetic field on the gravitational stability of thin disks can be found in
\cite{elmegreen1987supercloud}, \cite{elmegreen1994supercloud}; \cite{gammie1996linear}; \cite{fan1997swing} and \cite{kim2001amplification}. Also, the effect of viscosity on the local stability has been studied in \cite{gammie1996linear}. 

In this paper, we are interested in relativistic corrections to Toomre's criterion. More specifically, in relativistic situations, the standard gravitational stability criterion needs to be changed. In fact, astrophysical instabilities in relativistic systems may substantially influence the dynamics of the system. For example during the merger of binary neutron stars the Kelvin-Helmholtz instability appears in the system (\citealp{anderson2008magnetized,obergaulinger2009semi,goodman1994parasitic}). 
Furthermore, the dynamical bar-mode instability can occur in differentially rotating neutron stars. In fact, the bar-mode instability in supernova collapse or binary neutron star mergers could lead to strong and observable gravitational waves \citep{shibata2000bar}. As another example see \cite{siegel2013magnetorotational}, where the magnetorotational instability (MRI) in the evolution of HMNS has been investigated.  

The relativistic corrections to the Jeans instability have been investigated in \cite{nazari2017post}. It is shown that applying the first relativistic corrections to the equations of hydrodynamics, a new Jeans mass is obtained. More specifically, the new Jeans mass is smaller than the standard case. Also, it is interesting that in the PN limit, the pressure and the internal energy can support the instability of the system (\citealp{nazari2017post}).

Since rotation plays an important role in relativistic disks, here we generalize \cite{nazari2017post} results to a rotating razor-thin disk. The result would be important in the differentially rotating and relativistic disks around HMNSs. Recently, \cite{ellis2017search} have shown that the existence of matter fragmentations in the above-mentioned disks, can effectively influence the power spectrum of the gravitational waves produced during the merger of two neutron stars. From this perspective, finding an analytical criterion for the growth of local fragmentations in relativistic disks seems interesting. However, the main difficulty is that general relativity's (GR) field equations cannot be handled analytically for arbitrary matter configurations. So it is helpful to use numerical approaches as well as approximative methods. The PN approximation is one of the powerful methods for this aim. Within this framework, the relativistic corrections are added to the Newtonian description in an iterative manner. Consequently, this approach is appropriate for systems where the velocities are small compared to the velocity of light. On the other hand, gravitational fields are strong enough which one cannot use Newtonian gravity, and  are weak enough to ignore higher order general relativistic effects. For an excellent introduction to PN theory, we refer the reader to \cite{poisson2014gravity}. 

This method introduced by  Einstein, Infeld, and Hoffmann in the 1930s and developed by Fock (\citealp{fock1959theory}) and Chandrasekhar (\citealp{chandrasekhar1965post,chandrasekhar1967post,chandrasekhar1969conservation}; \citealp{chandrasekhar1969Second}; \citealp{chandrasekhar1970212}).
 Clifford Will extended this approach in a subsequent series of papers with emphasis on the experimental foundations of GR (\citealp{thorne1971theoreticali,will1971theoreticalii,will1971theoreticaliii}). Furthermore, Blanchet developed the PN formalism in the direction suitable to investigate the gravitational emission from inspiralling compact binaries (\citealp{blanchet1989post,blanchet1989higher,blanchet1995gravitational}).

One can find some important applications of  PN theory in the literature. For example, it can be used to study the equation of motion of binary pulsars (\citealp{blandford1976arrival,epstein1977binary,hulse1975deep,damour1991orbital}), general relativistic tests of GR in the solar system (\citealp{thorne1987300, will1994proceedings}), the normal modes of relativistic systems \cite{rezania2000normal}, gravitational radiation reaction (\citealp{chandrasekhar1970212,burke1971gravitational,blanchet2006}), Jeans analysis (\citealp{nazari2017post}), and also accretion disks around black holes (\citealp{demianski1997dynamics}). Furthermore, it has been significantly useful for modeling the gravitational waves. Most recently, \cite{abbott2017gw170817} announced the first detection of gravitational waves from the inspiral of a binary neutron star system. On the other hand, for analyzing  the observations of the gravitational-wave detectors, the high accuracy templates predicted by GR theory is required. For relativistic systems, such as inspiral of a binary neutron star, the high-order PN framework can accurately model the gravitational wave emission (\citealp{blanchet2014gravitational}). This framework has been also used in \cite{abbott2017gw170817} to interpret the observed signal. For a comprehensive review of  PN theory and its applications, we refer the reader to \cite{will2014confrontation}.

In this paper, we use the PN approximation up to the first PN order (1\tiny PN \normalsize) to find a new version of Toomre's criterion including the first relativistic corrections to the standard form. This criterion can be written in terms of the standard Toomre's parameter. In this case, the differences between the Newtonian and the PN theories will be more clear.

The map of this paper is as follows. In Sec. \ref{pnhydro} we introduce the hydrodynamic equations of the gaseous disk in the context of PN theory. In Sec. \ref{dispersion_relation_section}, using a perturbative method, we linearize the PN equations of hydrodynamics, and then find the PN potentials of a tightly wound spiral pattern. Finally, we obtain the dispersion relation for WKB density waves.
The PN version of Toomre's criterion is found in Sec. \ref{Toomre}. In this section, introducing two dimensionless stability parameters $\eta$ and $\mu$, we derive the PN Toomre's criterion, as our main result in this paper.
In Sec. \ref{etamu} we introduce an exponential toy model and study its local stability.  Moreover, in this section, we study the growth rate of unstable modes and find the boundary between stable and unstable modes.
Sec. \ref{hmns} is devoted to the application of our results to disks around HMNS.
Finally, we summarize the results and conclusions in Sec. \ref{conclusion}.

\section{Post-Newtonian dynamics of a gaseous disk}\label{pnhydro}

In this section, we introduce the equations of hydrodynamics in the first PN approximation for a two-dimensional fluid. The system is supposed to be perfect fluid, and all the dissipative effects are ignored. The disk has no thickness and also background disk is axisymmetric and static. Furthermore, the plane of the disk situated in the $x-y$ plane and the cylindrical coordinates system can be used in this case.

Since we are interested in general relativistic effects, it is necessary to find the space-time metric, $g_{\mu\nu}$, in the PN approximation. In fact, metric appears in the relativistic equations of hydrodynamics. Fortunately, this metric is already known. Let us briefly review the PN metric. To find the PN metric, using the Landau-Lifshitz formulation of general relativity, one may rewrite the Einstein field equations as a flat space-time wave equation. Then using a suitable gauge condition, which is known as the harmonic gauge condition, it is possible to solve the wave equation by an iterative method. 

The solution will be a retarded integral over the past light cone of the field point. The domain of the integration can be conveniently partitioned into a near zone domain and a wave zone domain. In the near zone, the field point distance from the source $r=|\bm{x}|$ is small compared with a characteristic wavelength $\lambda_c=c\, t_c$ where $t_c$ is the characteristic time scale of the dynamics of the source, and $c$ is the speed of light. The boundary of the near and wave zones is situated roughly at an arbitrary radius $\mathcal{R}$. This radius is of the same order of magnitude as $\lambda_c$. The integrals can vanish at this boundary. 
In fact, by  partitioning the domain of integration into these two zones, all integrations will be finite, and one can show that the contributions from the near-zone spatial integrals containing $\mathcal{R}$ (where diverge at $\mathcal{R}\rightarrow\infty$) will be canceled by corresponding terms from the wave zone integrals (\citealp{will1996gravitational}). As a result, this calculation does not depend on the arbitrary boundary radius $\mathcal{R}$. Keeping in mind these facts, one can find the PN metric as below (\citealp{poisson2014gravity})
\begin{equation}\label{metric}
\begin{split}
&g_{00} = -1+\frac{2}{c^2}U+\frac{2}{c^4}(\Psi-U^2)+O(c^{-6})\\&
g_{0j} = -\frac{4}{c^3}U_j+O(c^{-5})\\&
g_{jk}=\delta_{jk} \left( 1+\frac{2}{c^2}U\right) +O(c^{-4})
\end{split}
\end{equation}
where $\Psi$ is defined as follows
\begin{equation}\label{big Psi}
\Psi = \psi+\frac{1}{2}\frac{\partial^2 X}{\partial t^2}
\end{equation}
 Potentials that appear in the metric components are given by
\begin{align}\label{potentials}
\nonumber & U(t,\bm{x}) = G\int \frac{\Sigma^{*} d^2y}{\left| \bm{x-y}\right| } \\
&\psi(t,\bm{x}) = G \int \frac{\Sigma^{*} (\frac{3}{2}v^{2}-U+\Pi+3p/\Sigma^{*})}{\left|\bm{ x-y}\right| }d^2y\\\nonumber
& X(t,\bm{x}) = G \int \Sigma^{*} \left| \bm{x-y}\right|  d^2y\\\nonumber
& U^j(t,\bm{x}) = G \int \frac{\Sigma^{*}v^{j}}{\left|\bm{ x-y}\right| } d^2y
\end{align}
where  quantities in integrands are evaluated at time $t$ and position $\bm{y}$. The matter variables in the PN approach are $  \left\lbrace  \Sigma^*,p,\Pi,\bm{v}  \right\rbrace$, where $\Sigma^*$ is the conserved surface mass density defined as
\begin{equation}
\Sigma^* = \sqrt{-g}\gamma \Sigma = \sqrt{-g} \Sigma \frac{u^0}{c}
\end{equation}
in which $\Sigma$ is the proper surface density and $g$ is the determinant of the metric tensor. Here $u^0$ is the zeroth component of the velocity four-vector $u^{\alpha}$. It should be noted that in the Newtonian limit there is no difference between $\Sigma^*$ and the proper surface density $\Sigma$.
Moreover, $p$ is the pressure, $\Pi=\epsilon/\Sigma$ is the internal energy per unit mass, and $v^j$ is the fluid's velocity field defined with respect to the time coordinate $t$. Note that $\epsilon$ is the proper surface density of internal energy and $U$ is the gravitational potential for surface density $\Sigma^*$. In the following subsection, we will introduce the Newtonian potential in terms of $\Sigma$. Using the PN metric and the conservation equation of the energy-momentum tensor, i.e., $\nabla_{\mu}T^{\mu\nu}=0$, one can find the PN equations of hydrodynamics (\citealp{poisson2014gravity}). In fact in the 1\tiny PN \normalsize approximation, the relativistic corrections appear as terms proportional to order $c^{-2}$ in the equations. The continuity equation in the PN limit takes the following form
\begin{equation}\label{continuity_eq}
\frac{\partial\Sigma^*}{\partial t}+\nabla\cdot(\Sigma^*\bm{v})=0
\end{equation}
In the cylindrical coordinate system $(R, \varphi,z)$, where the velocity is $\bm{v}=v_R\bm{e}_R+v_\varphi\bm{e}_\varphi$, Eq. \eqref{continuity_eq} is given by
\begin{equation}\label{Sigma}
\frac{\partial\Sigma^*}{\partial t}+\frac{1}{R}\frac{\partial}{\partial R}(R\Sigma^*v_R)+\frac{1}{R}\frac{\partial}{\partial\varphi}(\Sigma^* v_\varphi)=0
\end{equation} 
Moreover, the Euler equation in the PN approximation can be written as follows 
\begin{align}\label{Euleroriginal}
\frac{d\bm{v}}{dt}=&-\frac{\nabla p}{\Sigma^*}+\nabla U+\frac{1}{c^2}\bigg\lbrace\frac{1}{\Sigma^*}\bigg[ \Big( \frac{1}{2}v^2+U+\Pi+\frac{p}{\Sigma^*}\Big) \nabla p-\bm{v}\frac{\partial p}{\partial t}\bigg]+\Big[ (v^2-4U)\nabla U-\bm{v}(3\,\frac{\partial U}{\partial t}+4\bm{v}\cdot\nabla U)\nonumber\\
&+4\frac{\partial \bm{U}}{\partial t}-4\bm{v}\times(\nabla\times\bm{U})+\nabla\Psi\Big]\bigg\rbrace +O(c^{-4})
\end{align}
The terms inside the braces are PN corrections. In the 1\tiny PN \normalsize approximation, we ignore terms proportional to order $c^{-4}$ and higher. The left-hand side (L.H.S.) of Eq. \eqref{Euleroriginal} can be written as $\partial\bm{v}/\partial t+(\bm{v}\cdot\nabla)\bm{v}$. After some algebraic manipulations and using vector analysis in the cylindrical coordinate system, one can decompose Eq. \eqref{Euleroriginal} to components $R$ and $\varphi$. The $R$ component of the Euler equation reads
\begin{align}\label{RcompEuler}
\nonumber& \frac{\partial v_R}{\partial t}+ v_R\frac{\partial v_R}{\partial R}+\frac{v_\varphi}{R}\frac{\partial v_R}{\partial\varphi}-\frac{v_\varphi^2}{R}=-\frac{1}{\Sigma^*}\frac{\partial p}{\partial R}+\frac{\partial U}{\partial R} +\frac{1}{c^2}\bigg\lbrace\frac{1}{\Sigma^*}\bigg[ \bigg( \frac{1}{2}v^2+U+\Pi+\frac{p}{\Sigma^*}\bigg) \frac{\partial p}{\partial R}-v_R \frac{\partial p}{\partial t}\bigg]
+\bigg[ (v^2-4U)\frac{\partial U}{\partial R}\\
& -v_R\bigg( 3\,\frac{\partial U}{\partial t}+4 v_R\frac{\partial U}{\partial R}+\frac{4v_\varphi}{R}\frac{\partial U}{\partial\varphi} \bigg) +4\,\frac{\partial U_R}{\partial t}-\frac{4 v_\varphi}{R}\bigg( \frac{\partial(RU_\varphi)}{\partial R}-\frac{\partial U_R}{\partial\varphi}\bigg) +\frac{\partial\Psi}{\partial R}\bigg]\bigg\rbrace +O(c^{-4})
\end{align}
Also the $\varphi$ component of the Euler equation can be written as 
\begin{align}\label{Eulphi}
&\frac{\partial v_\varphi}{\partial t}+ v_R\frac{\partial v_\varphi}{\partial R}+\frac{v_\varphi}{R}\frac{\partial v_\varphi}{\partial\varphi}+\frac{v_\varphi v_R}{R}=-\frac{1}{\Sigma^*R}\frac{\partial p}{\partial\varphi}+\frac{1}{R}\frac{\partial U}{\partial\varphi}
+\frac{1}{c^2}\bigg\lbrace\frac{1}{\Sigma*}\bigg[ \bigg( \frac{1}{2}v^2+U+\Pi+\frac{p}{\Sigma^*}\bigg) \bigg( \frac{1}{R}\frac{\partial p}{\partial \varphi}\bigg) -v_\varphi \frac{\partial p}{\partial t}\bigg]\\\nonumber
& +\bigg[ (v^2-4U)\frac{1}{R}\frac{\partial U}{\partial\varphi}-v_\varphi\bigg( 3\, \frac{\partial U}{\partial t}+4 v_R\frac{\partial U}{\partial R}+\frac{4v_\varphi}{R}\frac{\partial U}{\partial\varphi} \bigg)+ 4\,\frac{\partial U_\varphi}{\partial t}+\frac{4 v_R}{R}\bigg( \frac{\partial(RU_\varphi)}{\partial R}-\frac{\partial U_R}{\partial\varphi}\bigg) +\frac{1}{R}\frac{\partial\Psi}{\partial \varphi}\bigg]\bigg\rbrace +O(c^{-4})
\end{align}
In order to have a complete set of differential equations, the above-mentioned equations should be joined with the first law of thermodynamics and an equation of state (hereafter EOS). The first law of thermodynamics for a perfect fluid can be written as
\begin{equation}\label{fistlaw}
\frac{d\Pi}{dt}=\frac{p}{\Sigma^{*2}}\frac{d\Sigma^*}{dt}+O(c^{-2})
\end{equation}
On the other hand, the pressure and density of the fluid can be related to each other via an EOS. A simple case suitable for astrophysical aims is the barotropic equation, given as
\begin{equation}\label{EOS}
p = p(\Sigma^*)
\end{equation}
where the conserved surface density is 
\begin{equation}\label{sigmastar}
\Sigma^*=\Sigma\left(1+\frac{v^2}{2\,c^2}+\frac{3\,U}{c^2}\right)
\end{equation}

Therefore Eqs. \eqref{Sigma}, \eqref{RcompEuler}, \eqref{Eulphi}, \eqref{fistlaw} and \eqref{EOS} are the main equations for a self-gravitating fluid system in the PN limit. As we know in the Newtonian case, there is one more equation, namely the Poisson's equation which gives the Newtonian gravitational potential $U$. In the PN limit, we have more potentials which are given by the following equations (\citealp{poisson2014gravity})
\begin{equation}\label{poi1}
\nabla^2U=-4\pi G\Sigma^*\delta(z)
\end{equation}
\begin{equation}\label{X}
\nabla^2X=2U
\end{equation}
\begin{equation}\label{psi}
\nabla^2\psi=-4\pi G\Sigma^*\delta(z)\left( \frac{3}{2}v^2-U+\Pi+\frac{3p}{\Sigma^*}\right) 
\end{equation}
\begin{equation}\label{Uj}
\nabla^2U^j=-4\pi G\Sigma^*v^j\delta(z)
\end{equation}
where $\delta$ is the Dirac delta function.
We have already written the solutions of these equations in \eqref{potentials} and included them in the Euler equation.

Now we are in a position to study the gravitational stability of the rotating disk  against local perturbations. For a non-rotating system, we have investigated the Jeans analysis in the PN limit in \cite{nazari2017post}. Therefore the current study can be considered as a complementary investigation to \cite{nazari2017post}. More specifically in the mentioned paper, we have studied the stability of an infinite and non-rotating fluid, and here our aim is to find a generalized version of Toomre's criterion for stability of a differentially rotating thin disk in the 1\tiny PN \normalsize approximation.

To do so, we use the WKB approximation, and by linearizing the governing equations, we find the PN potentials for tightly wound spiral density perturbations. In other words, we study the evolution and propagation of the matter density waves on the surface of the rotating disk by finding the relevant dispersion relation from the first-order perturbative analysis. As usual, the dispersion relation helps to find a stability criterion. In a Newtonian system, In an infinite homogeneous fluid system in Newtonian gravity, the gas pressure has stabilizing effects and plays against gravitational instability. However, a relativistic situation is more complicated. For example, high gas pressure can trigger the instability \cite{nazari2017post}. In the following sections, assuming adiabatic perturbations, we explore the effects of the PN correction terms on the gravitational stability of a rotating fluid. Furthermore, we assume that the fluid lies in the PN realm, and consequently the Newtonian description does not work for it.

\section{DISPERSION RELATION FOR A SELF-GRAVITATING FLUID DISK IN PN APPROXIMATION}\label{dispersion_relation_section}

In this section, we find the dispersion relation for the propagation of adiabatic perturbations on the surface of a self-gravitating and differentially rotating fluid disk in the context of PN theory by a three-step process. First, we linearize PN equations of hydrodynamics introduced in the previous section.
On the other hand, to investigate the local stability analysis we apply the tight-winding or WKB approximation. For a $m$-fold symmetric tightly wound density wave with wavenumber $k$, one can show that $\left|k R/m\right|\gg1$. Therefore the terms proportional to $1/R$ compared to the analogous terms proportional to $k$ can be neglected. This approximation provides us with a substantial simplicity in the calculations. Furthermore, we deal with a static and axisymmetric fluid disk which has no mean radial movement, i.e., for the background we have $v_{R}=0$.
In the second step, we find the PN potentials for the desired perturbations. To do so, we use the  above-mentioned WKB approximation,  and simplify the differential equations. Finally, we solve the linearized PN equations to find the dispersion relation of the perturbations. Using this relation, we investigate the stability of the system against local gravitational perturbations.

\subsection{The linearized post-Newtonian equations of hydrodynamics}\label{THE LINEARIZED}

Our aim is to solve ten scalar equations introduced in Sec. \ref{pnhydro}, Eqs. \eqref{Sigma}, \eqref{RcompEuler}, \eqref{Eulphi}, \eqref{fistlaw}, \eqref{EOS}, and \eqref{poi1}-\eqref{Uj}, for first-order perturbations, and find the dispersion relation in the PN limit. Hereafter by prime sign we mean derivative with respect to $R$, and show the perturbed quantities as $Q_1=Q_{a} \exp\left[i(\bm{k}\cdot\bm{x}-\omega t+m\varphi)\right]$, where $\omega$ is the oscillation frequency of the disturbances, $k$ is the wavenumber, and as usual $m$ is a positive integer which specifies the rotational symmetry of the disturbance. Note that in the rest of this paper, perturbed quantities are labeled by subscripts "$1$", and background quantities do not have any subscript. In other words, each quantity can be collectively written as $\tilde{Q}=Q+Q_1$ where $Q_1/Q\ll 1$.

It is instructive to find the PN version of the rotation speed, circular frequency and also two Oort constants for the background system. In other words, let us find the PN corrections to these background quantities. We need the PN background quantities in the subsequent sections. To do so, we use the radial component of the Euler equation (\ref{RcompEuler}) for the background disk. By assuming that the unperturbed disk is static,  axisymmetric, and $v_{R}=0$, Eq. (\ref{RcompEuler}) can be written as
\begin{align}
\frac{v_{\varphi}^2}{R}=\frac{p'}{\Sigma^* }-U'-\frac{1}{c^2}\bigg\lbrace(v_{\varphi}^2-4U) U'-\frac{4 v_{\varphi}}{R}(R\, U_{\varphi})'+\psi'+\frac{p'}{\Sigma^* }\Big(\frac{p}{\Sigma^* }+\Pi +U+\frac{v_{\varphi }^2}{2}\Big)\bigg\rbrace
\end{align}
We solve this algebraic equation and derive $v_{\varphi}$. Then we expand the solution in powers of $c^{-1}$ and drop  all the terms of order $c^{-4}$ and higher. Finally, we obtain
\begin{align}\label{vphi0}
v_{\varphi}&\simeq  \sqrt{F}+\frac{1}{c^2}\bigg\lbrace 2(R\,U_{\varphi })'+\frac{R}{\sqrt{F}}\Big(\frac{1}{2} R \left(U'\right)^2+2\, U U'-\frac{p'}{4\, \Sigma^* } \Big[\frac{R\, p'}{\Sigma^* }+\frac{2\, p}{\Sigma^* }+2\, \Pi +R\,U'+2\,U\Big]-\frac{\psi '}{2}\Big)\bigg\rbrace+O(c^{-4})
\end{align}
where $F=\frac{R\,p'}{ \Sigma^* }-R\,U'$.
Now let us decompose the Newtonian and PN terms. Here, we have not ignored the pressure effects on the circular velocity in the first approximation. The first term on the right-hand side (R.H.S.) includes both Newtonian and PN contributions. In fact, $U$ is the gravitational potential for $\Sigma^*$ which includes both Newtonian and PN contributions, see Eq. \eqref{sigmastar}. To separate these contributions, we use Eq. \eqref{sigmastar} to  define the scalar potentials $U_{\text{N}}$, $U_{\text{v}}$, and $U_{\text{u}}$ as follows
\begin{equation}\label{U0}
U=U_{\text{N}}+\frac{1}{c^2}\left(U_{\text{v}}+U_{\text{u}}\right)
\end{equation}
in which
\begin{align}\label{defU0}
U_{\text{N}} = G\int \frac{\Sigma}{\left| \bm{x-y}\right| } d^2y,~~~~
U_{\text{v}} = G \int \frac{\Sigma R^2 \Omega^2}{2\,\left|\bm{ x-y}\right|} d^2y, ~~~~~
U_{\text{u}} = G \int \frac{3\,\Sigma U_\text{N}}{\left| \bm{x-y}\right|} d^2y
\end{align}
In this set of equations, $U_{\text{N}}$ is the Newtonian potential constructed from proper density $\Sigma$.  Moreover, $\Omega$ is the circular frequency in Newtonian gravity. In order to complete this decomposition, we should also separate the Newtonian and PN contributions in the pressure. To do so, let us expand the pressure as follows
\begin{equation}\label{p_PN}
p=p_{\text{N}}+\frac{1}{c^2}p_{\text{c}}
\end{equation}
where $p_{\text{N}}$ is the Newtonian pressure which is a function of $\Sigma$. After specifying EOS, we will introduce $p_{\text{c}}$ as the PN correction to the pressure.
Now using Eqs. (\ref{sigmastar}), (\ref{U0}) and (\ref{p_PN}), we rewrite Eq. (\ref{vphi0}) as follows
\begin{align}\label{vphi0PN}
v_{\varphi}  \simeq & \sqrt{F_{\text{N}}}+\frac{1}{c^2}\bigg\lbrace 2(R\,U_{\varphi })'+\frac{R}{\sqrt{F_{\text{N}}}}\Big(\frac{1}{2} R \left(U_{\text{N}}'\right)^2+2 U_{\text{N}} U_{\text{N}}'
-\frac{p_{\text{N}}'}{4\, \Sigma } \Big[\frac{R\, p_{\text{N}}'}{\Sigma }+\frac{2\, p_{\text{N}}}{\Sigma}+2\, \Pi +R\,U_{\text{N}}'+8\, U_{\text{N}}+R^2\Omega^2\Big]\\\nonumber
& +\frac{p_{\text{c}}'}{2\,\Sigma}-\frac{U_{\text{v}}'}{2}-\frac{U_{\text{u}}'}{2}-\frac{\psi '}{2}\Big)\bigg\rbrace+O(c^{-4})
\end{align}
 in which $F_{\text{N}}=\frac{R\,p_{\text{N}}'}{ \Sigma}-R\,U_{\text{N}}'$.
This is the circular velocity in the PN approximation. It is worth mentioning that this equation represents a balance between the centrifugal force, the pressure gradient and the gravitational force including PN corrections. Naturally, by removing the 1\tiny PN \normalsize correction terms, the Newtonian rotation speed is recovered as follows
\begin{equation}\label{OmegaN}
v_{\varphi}\simeq\sqrt{\frac{R\,p_{\text{N}}'}{ \Sigma}-R\,U_{\text{N}}'}=R\,\Omega
\end{equation}
It is also instructive to define a PN circular frequency $\Omega_{\text{p}}$. To do so, let us rewrite the PN circular speed $v_{\varphi}$ as  $v_{\varphi}\simeq R\,\Omega_{\text{p}}$. Using this definition and substituting Eq. \eqref{OmegaN} into Eq. \eqref{vphi0PN}, we eventually arrive at
\begin{align}\label{OmegaPN}
\Omega_{\text{p}}&\simeq \Omega-\frac{1}{c^2}\bigg\lbrace\,\frac{1}{2\,R\,\Omega}\left(U_{\text{v}}'+U_{\text{u}}'\right)+\frac{1}{2}R^2\Omega^3-\frac{2\,U_{\varphi }}{R}
-2 U_{\varphi}'+2\,\Omega\,U_{\text{N}}
+\frac{\psi'}{2\,R\,\Omega}-\frac{H}{R}\,\bigg\rbrace + O(c^{-4})
\end{align}
where the function $H$ is defined as
\begin{align}
H=\frac{p_{\text{c}}'}{2\, \Sigma\,  \Omega }-\frac{p_{\text{N}}}{2\, \Sigma\,\Omega}\Big(R \,\Omega^2 +U_{\text{N}}'\Big)-R^2\, \Omega\, U_{\text{N}}'-\frac{\Pi}{2\, \Omega }\Big( R\, \Omega^2 +U_{\text{N}}'\Big)
\end{align}
Before moving on to discuss the first order perturbations, let us introduce two PN Oort constants that will be useful to simplify equations in the subsequent calculations. By substituting Eq. (\ref{OmegaPN}) into the definition of two Oort constants (see \citealp{binney2008galactic}), we find the following constants in the PN limit 
 \begin{align}
  A_{\text{p}}= & -\frac{R\,\Omega '}{2}+\frac{1}{c^2}\bigg\lbrace\,\frac{U_{\varphi}}{R}+R\,\Omega\,U_{\text{N}}'+\frac{1}{2} R^2\,\Omega ^3-\frac{{\psi}'}{4\,R\,\Omega}
 +R\,U_{\text{N}}\,\Omega '+\frac{3}{4} R^3\,\Omega ^2\,\Omega '-\frac{U_{\text{u}}'}{4\,R\,\Omega}-\frac{\Omega '\,U_{\text{u}}'}{4\,\Omega ^2}-\frac{{\psi}'\,\Omega '}{4\,\Omega ^2}
 \\\nonumber
 &-\frac{U_{\text{v}}'}{4\,R\,\Omega}+\frac{U_{\text{u}}''}{4\,\Omega }-\frac{\Omega '\,U_{\text{v}}'}{4\,\Omega ^2} - U_{\varphi}'+\frac{{\psi}''}{4\,\Omega }-R\,U_{\varphi}''+\frac{U_{\text{v}}''}{4\,\Omega }+\frac{H}{2\,R}-\frac{H'}{2} \, \bigg\rbrace+ O(c^{-4})
 \end{align}
 and
 \begin{align}\label{PN Oort}
 \nonumber
 B_{\text{p}} = &-\Omega-\frac{R\,\Omega '}{2}+\frac{1}{c^2}\bigg\lbrace 2\,U_{\text{N}}\,\Omega + R^2\,\Omega ^3-\frac{U_{\varphi}}{R} 
+\frac{3}{4} R^3\,\Omega ^2\,\Omega '+R\,\Omega \,U_{\text{N}}'+R\,U_{\text{N}}\,\Omega '+\frac{{\psi}'}{4\,R\,\Omega}-\frac{{\psi}'\,\Omega '}{4\,\Omega ^2}+\frac{U_{\text{u}}'}{4\,R\,\Omega }\\
&+\frac{U_{\text{v}}'}{4\,R\,\Omega }-\frac{\Omega '\,U_{\text{u}}'}{4\,\Omega ^2}-\frac{\Omega '\,U_{\text{v}}'}{4\,\Omega ^2}-3\,U_{\text{$\varphi
   $}}'
 +\frac{U_{\text{u}}''}{4\,\Omega }+\frac{{\psi}''}{4\,\Omega }+\frac{U_{\text{v}}''}{4\,\Omega }-R\,U_{\varphi}''-\frac{H}{2\,R}-\frac{H'}{2}\bigg\rbrace +O(c^{-4})
 \end{align}
 If we neglect the 1\tiny PN \normalsize terms, these constants will coincide with two Oort constants $A$ and $B$ in the Newtonian gravity, respectively.

We now return to our discussion on the linearized PN hydrodynamics. By using first order expansion for all quantities as $\tilde{\Sigma}^*=\Sigma^*+\Sigma_1^*$, $\tilde{v}_{R}=v_{R}+v_{R1}=v_{R1}$, $\tilde{v}_{\varphi}=v_{\varphi}+v_{\varphi1}$, $\tilde{p}=p+p_1$, $\tilde{\Pi}=\Pi+\Pi_1$, $\tilde{U}=U+U_1$, $\tilde{\Psi}=\psi+\Psi_1$, $\tilde{U}_{R}=U_{R}+U_{R1}$, and $\tilde{U}_{\varphi}=U_{\varphi}+U_{\varphi1}$ and keeping only the first order terms, we linearize Eqs. (\ref{RcompEuler}) and (\ref{Eulphi}). The linearized form of Eq. (\ref{RcompEuler}) is 
 \begin{align}\label{per R com}
\frac{\partial\,v_{R1}}{\partial\,t}  +\,\Omega_{\text{p}}\frac{\partial\,v_{R1}}{\partial\,\varphi}-2\,\Omega_{\text{p}}\,v_{\varphi 1}=&-\frac{p_1'}{\Sigma^*}+\frac{\Sigma_1^*}{\Sigma^{*2}}p'+U_1'
+\frac{1}{c^2}\bigg\lbrace\frac{\sigma p_1'}{\Sigma^*}+
\Big(R\,\Omega_{\text{p}}v_{\varphi 1}+U_1+\Pi_1+\frac{p_1}{\Sigma^*}-\frac{\Sigma^*_1}{\Sigma^{* 2}}p\Big)\frac{p'}{\Sigma^*}\\\nonumber
&-\frac{\Sigma_1^*\,\sigma}{\Sigma^{*2}}p'+\left(R^2\,\Omega_{\text{p}}^2-4\,U\right)U_1'+\left(2\,R\,\Omega_{\text{p}}\,v_{\varphi 1}-4\,U_1\right)U'+4\,\frac{\partial\,U_{R 1}}{\partial\,t}-4\,\frac{v_{\varphi 1}}{R}\,(R\,U_{\varphi})'\\\nonumber
&-4\,\Omega_{\text{p}}\,(R\,U_{\varphi 1})'+4\,\Omega_{\text{p}}\,\frac{\partial\,U_{R 1}}{\partial\,\varphi}+\Psi_1'\bigg\rbrace+O(c^{-4})
 \end{align}
and the perturbed form of $\varphi$ components of the Euler equation (\ref{Eulphi}) becomes
 \begin{align}\label{per phi com}
  \frac{\partial\,v_{\varphi 1}}{\partial\,t} & -2\,B_{\text{p}}\,v_{R 1}+\Omega_{\text{p}}\,\frac{\partial\,v_{\varphi 1}}{\partial\,\varphi}=-\frac{1}{\Sigma^*\,R}\frac{\partial\,p_1}{\partial\,\varphi}+\frac{1}{R}\frac{\partial\,U_1}{\partial\,\varphi} +\frac{1}{c^2}\bigg\lbrace\frac{\sigma}{R \Sigma^*}\frac{\partial\,p_1}{\partial\,\varphi}-\frac{R\,\Omega_{\text{p}}}{\Sigma^*}\,\frac{\partial\,p_1}{\partial\,t}+(R^2\Omega_{\text{p}}^2-4\,U)\frac{1}{R}\frac{\partial\,U_1}{\partial\,\varphi}  \\\nonumber
 &-R\,\Omega_{\text{p}}\Big( 3\,\frac{\partial\,U_1}{\partial\,t}+4\,v_{R 1} U'+4\,\Omega_{\text{p}}\,\frac{\partial\,U_1}{\partial\,\varphi}\Big)+4\frac{\partial\,U_{\varphi 1}}{\partial\,t}  +\frac{4v_{R 1}}{R}\,\left(RU_{\varphi}\right)'+\!\frac{1}{R}\,\frac{\partial\,\Psi_1}{\partial\,\varphi}\!\bigg\rbrace+O(c^{-4})
 \end{align}
where $\sigma$ is defined as
\begin{eqnarray}
\sigma= \frac{1}{2}\,R^2\,\Omega_{\text{p}}^2+U+\Pi+\frac{p}{\Sigma^*}
\end{eqnarray}
Before linearizing the continuity equation let us apply the WKB or tight-winding approximation to Eqs. \eqref{per R com} and \eqref{per phi com}. In this approximation, the surface density perturbation  $\Sigma_1^*$ can be considered as a radially propagating spiral wave. More specifically, it can be written as $\Sigma^*_1=\Sigma^*_a\exp(ik R+i m\varphi-i\omega t)$ for which $|k R/m|\gg 1$ (\citealp{binney2008galactic}). For such a perturbation the spiral arms are tightly wound and the pitch angle is small. With this approximation, one can find the local stability criterion independent of the physical properties of the background disk. Now, we use this Fourier form for all first-order perturbations as $Q_1=Q_a \exp(ik R+i m\varphi-i\omega t)$. However, we do not apply the condition $|kR/m|\gg 1$ until the next section. In this case Eq. (\ref{per R com}) takes the following form
\begin{align}\label{Eul R}
  v_{Ra} \left(i m \Omega _{\text{p}}  -i \omega \right)=&2\,\Omega_\text{p}v_{\varphi a}-i\, k \left(\frac{p_{a}}{\Sigma^*}-U_{a}\right)+\frac{\Sigma_a^*}{\Sigma^{* 2}}p'
 + \frac{1}{c^2}\bigg\lbrace\Big(R\,\Omega_{\text{p}}v_{\varphi a}+U_a+\Pi_a+\frac{p_a}{\Sigma^*}-\frac{\Sigma_a^*}{\Sigma^{* 2}}p-\frac{\Sigma_a^* \sigma}{\Sigma^{*}}\Big)\frac{p'}{\Sigma^*}\\\nonumber 
 & -\frac{v_{\varphi a}}{R} \Big(4\,\left(R\,U_{\varphi}\right)'-2\,R^2\,\Omega _{\text{p}}U'\Big)+i\,k\,\Big(\frac{p_{a}\,\sigma}{\Sigma^*}-4\,U\,U_{a}+R^2\,\Omega _{\text{p}}^2\,U_{a}-4\,R\,\Omega _{\text{p}}\,U_{\varphi a}+\Psi _{a} \Big)\\\nonumber
&-4\,U_{a}\,U'-4\,i\,\omega\,U_{Ra}-4\,\Omega _{\text{p}} \,U_{\varphi a}+4\,i\,m\,\Omega _{\text{p}}\,U_{Ra}\bigg\rbrace+O(c^{-4})
\end{align}
and Eq. (\ref{per phi com}) is displayed as follows
\begin{align}\label{Eul phi}
 v_{\varphi a} \left(im\Omega _{\text{p}}-i\omega \right)=\nonumber &2 B_{\text{p}} v_{Ra} +\frac{i m}{R}\left( U_{a}-\frac{p_{a}}{\Sigma^*}\right) +\frac{1}{c^2}\bigg\lbrace
\frac{4 v_{Ra}}{R}\Big(U_{\varphi}+R\left( U_{\varphi}'-R\, \Omega _{\text{p}} U'\right)  \Big)-i\omega\Big(4 U_{\varphi a}
-R\,\Omega_{\text{p}}\big(\frac{p_{a}}{\Sigma^*}+3U_a\big)\Big)\\
& +\frac{i m}{R}\Big(\frac{p_{a}\sigma}{\Sigma^*} -4\,UU_a-3\,R^2\Omega_\text{p}^2 U_a  +\Psi_a \Big)\bigg\rbrace +O(c^{-4})
\end{align}
One may easily solve Eqs. \eqref{Eul R} and \eqref{Eul phi} in order to find $v_{\varphi a}$ and $v_{Ra}$. To do so, we use Eqs. (\ref{sigmastar}), (\ref{U0}), (\ref{p_PN}), (\ref{OmegaPN}), and (\ref{PN Oort}) and then expand the solutions up to order $c^{-2}$. 
Although we do not need the exact form of the solutions, we mention that both perturbations are proportional to $\Delta^{-1}$, where $\Delta$ is given by
\begin{equation}
\Delta = \kappa^2- \left(m\,\Omega-\omega\right)^2
\end{equation}
 in which $\kappa=\sqrt{4\,\Omega^2+ 2\, R \Omega\Omega'}$ is the Newtonian epicycle frequency. It is important to note that there is a singularity in $v_{\varphi\text{a}}$ and $v_{Ra}$ when $\Delta=0$. This is the case also in the Newtonian system. In other words, there are some  special radii at which $\omega=m\,\Omega\pm\kappa$. This is recognized as the Lindblad resonance condition in Newtonian gravity.
In fact, the response of the differentially rotating disk to a weak perturbation is very strong when the condition for the Newtonian Lindblad resonance is established and also our analysis will not work in the vicinity of these radii. One may expect that in the PN limit the PN epicycle frequency and angular frequency must appear in $\omega=m\,\Omega\pm\kappa$ instead of the Newtonian quantities. However, interestingly the PN corrections do not directly appear in the definition of the singularity.  Nevertheless, the singularity happens at different radii in a PN system compared with a Newtonian system. In fact, as we will show, $\omega$ is a function of radius and is different in  the Newtonian and PN systems. In other words, the PN corrections are collected in $\omega$. 

We now linearize the continuity equation (\ref{Sigma}) as follows
 \begin{eqnarray}
 \frac{\partial\Sigma_1^*}{\partial t}+\frac{\left(R\Sigma^*v_{R1}\right)'}{R}+\frac{1}{R}\frac{\partial}{\partial\varphi}\left(\Sigma^*v_{\varphi 1}+R\,\Omega_{\text{p}}\Sigma_1^*\right)=0
 \end{eqnarray}
 Applying the Fourier expansion and using the WKB approximation, this equation takes the following form 
\begin{equation}\label{vRa}
 \Sigma_{a}^*\,\left(m\,\Omega_{\text{p}}-\omega\right)+k\,\Sigma^*\,v_{Ra}=0
\end{equation}
On the other hand, the linearized versions of the first law of thermodynamics and the EOS are given by
\begin{equation}\label{Pia}
\Pi_{a}=\frac{p}{\Sigma^{*2}}\,\Sigma_{a}^*
\end{equation}
and
\begin{equation}\label{pa}
p_{a}=c_{\text{s}}^{*2}\,\Sigma_{a}^*, ~~~~c_{\text{s}}^{*2}=\left(\frac{d\,p}{d\,\Sigma^*}\right)_{\Sigma^*}
\end{equation}
respectively. Where $c_{\text{s}}^*$ is the sound speed. Finally let us linearize the modified Poisson equations \eqref{poi1}-\eqref{Uj}. One may simply verify that
\begin{equation}\label{U1}
\nabla^2U_1=-4\,\pi\,G\,\Sigma_1^*\,\delta(z)
\end{equation}
\begin{equation}\label{U2}
\nabla^2X_1=2\,U_1
\end{equation}
\begin{align}\label{U3}
\nabla^2\psi_1  =-4\pi G\delta(z)\bigg\lbrace\Sigma_1^*\left( \frac{3}{2}\,R^2\,\Omega_{\text{p}}^2-U+\Pi+\frac{3p}{\Sigma^*}\right)+\Sigma^*\Big(3\,R\,\Omega_{\text{p}}v_{\varphi 1}-U_1+\Pi_1+\frac{3\,p_1}{\Sigma^*}-\frac{3\,p}{\Sigma^{*2}}\Sigma_1^*\Big)\bigg\rbrace
\end{align}
\begin{align}\label{U4}
\nabla^2U_{R1}-\frac{U_{R1}}{R^2}-\frac{2}{R^2}\frac{\partial\,U_{\varphi 1}}{\partial\,\varphi}=
-4\pi G\Sigma^* v_{R1}\delta(z)
\end{align}
\begin{align}\label{U5}
\nabla^2U_{\varphi 1}-\frac{U_{\varphi 1}}{R^2}+\frac{2}{R^2}\frac{\partial U_{R1}}{\partial\varphi}=-4\pi G\delta(z)\left(\Sigma^* v_{\varphi 1}+\Sigma_1^* v_{\varphi}\right)
\end{align}

Now we have linearized all the governing equations. In fact, Eqs. \eqref{Eul R}, \eqref{Eul phi}, and \eqref{vRa}-\eqref{U5} are the main equations that we need to find the dispersion relation. In the next subsection, we find all the Fourier coefficients $Q_a$ and the relevant dispersion relation. More specifically, we calculate all the PN potentials with respect to $\Sigma^*_a$ by solving differential Eqs. \eqref{U1}-\eqref{U5}.

\subsection{Post-Newtonian potentials of a tightly wound spiral pattern}\label{PN POTENTIALS}
In order to complete the set of equations that locally describe a self-gravitating and differentially rotating disk, let us find the PN gravitational potentials in this subsection. As we mentioned before, to have an analytic expression for the dispersion relation, we apply the WKB approximation. In this approximation, the density of the spiral pattern is expressed as a radially propagating plane wave, i.e., $\Sigma^*_1=\Sigma^*_a\exp(ik R+i m\varphi-i\omega t)$. In the following, we find the corresponding gravitational potentials.

To solve Eq. \eqref{U1}, with no loss of generality, we assume that $\bm{k}=k\,\bm{e}_x$. Where $\bm{e}_x$ is the unit vector pointing along the $x$-axis in the Cartesian coordinates. The solution $U_1$ should satisfy $\nabla^2 U_1=0$ for $z\neq0$, and $U_1=U_a \exp\left[i(kx-\omega t) \right] $ for $z=0$. So we can write
\begin{equation}
U_1=U_{a}\, e^{i\,\left(k\,x-\omega\,t\right)-\left| k\,z\right|}
\end{equation}
 Using the Gauss's theorem, one may easily verify that $U_{a}$ and $\Sigma_{a}^*$ are related as (for more details see \citealp{binney2008galactic})
\begin{equation}\label{Ua}
U_{a}=\frac{2\,\pi\,G\,\Sigma_{a}^*}{|k|}
\end{equation}

To obtain $X_1$ from Eq. \eqref{U2}, we differentiate this differential equation as
\begin{equation}
\nabla^2\left(\nabla^2\,X_1 \right)=2\,\nabla^2\,U_1
\label{mnew}
\end{equation}
and after using Eq. (\ref{U1}), Eq. \eqref{mnew} can be rewritten as follows
\begin{equation}\label{dif4X}
\nabla^2\left(\nabla^2 X_1 \right) = -8 \pi G \Sigma_1^*\delta(z)
\end{equation}
One may guess the solution to this equation as $X_1=X_{a}\exp\left[i\left(k x-\omega t\right)\right]\exp\left[-\left| k'z\right|\right]$. In the following, by finding $k'$ and $X_{a}$, we show that this guess leads to a unique solution. Outside of the razor-thin disk $\left(z\neq0\right)$, we have $\nabla^2\left(\nabla^2 X_1 \right) = 0$. Substituting $X_1$ into this equation we find that $k'=|k|$. Now, we find the superpotential $X_1$ on the surface of the disk $\left(z=0\right)$. To do so, Let us integrate Eq. (\ref{dif4X}) with respect to $z$. The result is
\begin{equation}
\lim\limits_{\zeta\rightarrow 0}\int\limits_{-\zeta}^{+\zeta}\nabla^2\left(\nabla^2 X_1 \right)\,dz =-8\pi G \Sigma_1^*
\end{equation}
where $\zeta$ is a positive constant, and to restrict this calculation to the plane $z=0$, we let $\zeta\rightarrow 0$. This equation can be written as
\begin{align}\label{lim}
\lim\limits_{\zeta\rightarrow 0}\int\limits_{-\zeta}^{+\zeta}  \bigg(\frac{\partial^4 X_1}{\partial x^4}+\frac{\partial^4 X_1}{\partial y^4}+\frac{\partial^4 X_1}{\partial z^4}+2\frac{\partial^4 X_1}{\partial x^2 \partial y^2} +2\frac{\partial^4 X_1}{\partial x^2\partial z^2}+2\frac{\partial^4 X_1}{\partial y^2\partial z^2}\bigg) dz=-8\pi G\Sigma_1^*
\end{align}  
Using the suggested solution for $X_1$, one can easily simplify the above equation. For example, the first term on the R.H.S. can be written as
\begin{align}
\lim\limits_{\zeta\rightarrow 0}  \int\limits_{-\zeta}^{+\zeta}\left(\frac{\partial^4 X_1}{\partial x^4} \right) dz=k^4 X_a e^{i(k x-\omega t)} \times \lim\limits_{\zeta\rightarrow 0}\Big(  \int\limits_{-\zeta}^{0}e^{k z}dz+ \int\limits_{0}^{+\zeta}e^{-kz}dz \Big)=0
\end{align}
In fact, $\nabla^2 X_1$ does not satisfy the Laplace equation at $z=0$ plane. Moreover it is easy to verify that there is a discontinuity at $z=0$. After some algebra,  Eq. \eqref{lim} reads
\begin{eqnarray}\label{limit}
\lim\limits_{\zeta\rightarrow0}\left\lbrace 2\,\frac{\partial^3X_1}{\partial x^2\partial z}
+\frac{\partial^3 X_1}{\partial\,z^3}\right\rbrace _{-\zeta}^{+\zeta}
=-8\pi G\Sigma_1^*
\end{eqnarray}
Finally, from this equation, we obtain
\begin{equation}\label{Xa}
X_{a}=-\frac{4 \pi G\Sigma_{a}^*}{|k|^3}
\end{equation}
Therefore, the final form of the perturbed superpotential $X_1$ in the PN approximation is
\begin{eqnarray}
X_1=-\frac{4 \pi G\Sigma_{a}^*}{|k|^3}\,e^{i \left(k x-\omega t\right)-\left| k z\right|}
\end{eqnarray}
This function is unique and satisfies all the relevant constraints. Now using the same procedure, we calculate $\psi_1$ using \eqref{U3}. So we assume $\psi_1=\psi_{a}\exp\left[i\left(k x-\omega t\right)\right]\exp\left[-\left| k'z\right|\right]$, and integrate \eqref{U3} along the $z$-axis in order to find $\psi_{a}$ with respect to $\Sigma_{a}^*$ and $v_{\varphi\text{a}}$. The result is
\begin{align}\label{psia}
\psi_{a}=\frac{2\pi G}{|k|}\bigg\lbrace\Sigma_{a}^*\left( \frac{3}{2} R^2\,\Omega_{\text{p}}^2-U+\Pi+\frac{3p}{\Sigma^*}\right) + \Sigma^*\left(3\, R\, \Omega_{\text{p}} v_{\varphi a}-U_a+\Pi_a+\frac{3 p_a}{\Sigma^*}-\frac{3 p}{\Sigma^{*2}}\Sigma_a^*\right)\bigg\rbrace
\end{align}
It should be noted that $p_{a}$, $\Pi_{a}$, and $U_{a}$ are also proportional to $\Sigma_{a}^*$. On the other hand, by linearizing Eq. (\ref{big Psi}), we immediately obtain $\Psi_{a}$ with respect to $\psi_{a}$ and $X_{a}$ as $\Psi_{a}=\psi_{a}-\frac{\omega^2}{2} X_{a}$.

Similarly one may solve Eqs. \eqref{U4} and \eqref{U5} for $U_{Ra}$ and $U_{\varphi a}$ and find the following solutions respectively
\begin{equation}\label{URa}
U_{Ra}=\frac{2 \pi G\Sigma^*}{|k|}\,v_{Ra},~~~~U_{\varphi a}=\frac{2 \pi G}{|k|}\left(\Sigma^*v_{\varphi a}+R \,\Omega_{\text{p}}\Sigma_{a}^*\right)
\end{equation}
 We have used the WKB approximation to ignore the contribution of the terms proportional to $R^{-2}$  on the L.H.S. of \eqref{U4} and \eqref{U5}. It is clear that if we find $v_{\varphi a}$ and $v_{Ra}$ with respect to $\Sigma_{a}^*$, then all the coefficients $Q_a$ will be proportional to $\Sigma_{a}^*$. We do this in the next subsection.

\subsection{Dispersion relation for WKB density waves in the post-Newtonian approximation}\label{dis PN}

In this subsection, we introduce the PN dispersion relation for tightly wound spiral density waves that propagate on the surface of self-gravitating and differentially rotating fluid disks. As mentioned earlier, to investigate the local stability of a differentially rotating disk analytically, we have to use the WKB approximation. To find the dispersion relation, we substitute $\Pi_{a}$, $p_{a}$, $U_{a}$, $U_{Ra}$,
 $U_{\varphi a}$, and $\Psi_{a}$, obtained in subsections \ref{THE LINEARIZED} and \ref{PN POTENTIALS}, into Eqs. \eqref{Eul R} and \eqref{Eul phi}. In this case, one can straightforwardly find $v_{Ra}$ and $v_{\varphi a}$ with respect to $\Sigma_{a}^*$. Since the related calculations are simple and the final expressions are too long, we do not write the exact form of $v_{Ra}$ and $v_{\varphi a}$. Finally, substituting $v_{Ra}$ into Eq. \eqref{vRa}, we find the dispersion relation as follows

\begin{align}\label{abb form omega}
\omega ^4 \left(A_4+\frac{B_4}{c^2}\right)+\omega ^3 \left(A_3+\frac{B_3}{c^2}\right)+\omega ^2 \left(A_2+\frac{B_2}{c^2}\right)
+\omega\left(A_1+\frac{B_1}{c^2}\right)+A_0+\frac{B_0}{c^2}=0
\end{align}
in which $A_i$ and $B_i$ are functions of background variables and wavenumber $k$. In Appendix \ref{exact quadratic function} we present the exact form of these functions. In fact, this quartic equation is the dispersion relation for density waves propagating on the surface of a PN fluid disk. We recall that  the corresponding relation is a quadratic equation in the Newtonian limit.

On the other hand, to find $\omega^2\left(k\right)$, our first task is to solve Eq. \eqref{abb form omega} and obtain $\omega^2$. Naturally, there are four solutions for $\omega$ and only two of them recover the Newtonian dispersion relation when we turn off the PN corrections. Unfortunately, these solutions are too long to be written here. However, the Newtonian part of the solutions is listed in Appendix \ref{exact quadratic function}.

It should be mentioned that even in the Newtonian case, the local stability of the system against non-axisymmetric perturbations is complicated, for example, see \cite{julian1966non} where it has been claimed that the system is stable against $m\neq 0$ perturbations. However, feedback between trailing and leading WKB density waves may cause unstable non-axisymmetric perturbations. In fact, the so-called Toomre's criterion is obtained for axisymmetric perturbations and there is no analytic criterion for non-axisymmetric perturbations in Newtonian gravity. Naturally, in the PN limit, the situation is much more complicated.
Therefore in the following, we restrict ourselves to axisymmetric perturbations and set $m=0$ in Eq. (\ref{abb form omega}).
After solving this equation, in which $m=0$, and again applying the WKB approximation, and also ignoring the terms of order $c^{-4}$ and higher, we find the frequency of the density waves with respect to $k$ and background variables.
In this case, we obtain 
\begin{align}\label{PN dis}
 \omega^2= \nonumber& k^2 c_{\text{s}}^{*2}-2\pi G \Sigma^* |k|+\kappa^2-\frac{1}{c^2}\bigg\lbrace k^2 c_{\text{s}}^{*2}\left(\frac{p}{\Sigma^*}+\Pi+\frac{R^2\Omega^2}{2}+U_{\text{N}}\right)+2\, R^2 \Omega^2 \kappa^2+2\pi G \Sigma^* |k|\Big(\frac{p}{\Sigma^* }+\Pi-\frac{3}{2} R^2\Omega^2\\\nonumber
&-5 U_\text{N}\Big)-2\, R^2 \Omega^4+\frac{3 \psi'}{R}+\frac{4\Omega\,U_{\varphi}}{R}+4 \kappa^2 U_\text{N}-4\Omega U_{\varphi}'+\frac{3}{R}\left(U_{\text{u}}'+U_{\text{v}}'\right)-R\,U_{\text{N}}'\left(\kappa^2+4\Omega^2\right)-\frac{H}{R\Omega}\big(\kappa^2+2\Omega^2\big)\\
&-2\Omega H'+U_{\text{u}}''+U_{\text{v}}''+\psi''-4 R \Omega U_{\varphi}''\bigg\rbrace
\end{align}
This is the PN dispersion relation for axisymmetric disturbances in the WKB approximation. As expected it depends on the properties of the background disk (e.g., PN potentials, $\Sigma^*$, $c_{\text{s}}^{*}$, etc.). In fact, this is the key equation that allows us to study the behavior of excited disturbances in the fluid disks in the context of PN theory. In order to abbreviate this equation, we follow \cite{nazari2017post}, and define the following parameters
\begin{eqnarray}\label{eff1}
 c_{\text{sp}}^2=\left(1-\alpha\right)c_{\text{s}}^{*2},~~~~~~
 G_{\text{p}}=\left(1+\beta\right)G
\end{eqnarray}
in which
\begin{eqnarray}
\alpha=\frac{1}{c^2}\left(\frac{p}{\Sigma^* }+\Pi+\frac{1}{2}R^2\,\Omega^2+U_{\text{N}}\right),~~~~~~
\beta=\frac{1}{c^2}\left(\frac{p}{\Sigma^* }+\Pi-\frac{3}{2}R^2\,\Omega^2-5 U_\text{N}\right)
\end{eqnarray}
In fact, $c_{\text{sp}}$ is an effective sound speed and $G_{\text{p}}$ is an effective gravitational constant. If we neglect terms of order $c^{-2}$ in the above definitions, we immediately retrieve the Newtonian versions of them (i.e., $c_{\text{sp}}^2\simeq c_{\text{s}}^2$ and $G_{\text{p}}\simeq G$). Furthermore, we define $\kappa_{\text{p}}$ as $\kappa_{\text{p}}^2=\kappa^2+\gamma_c$, where $\kappa$ is the Newtonian epicycle frequency and $\gamma_c$ is given by
\begin{align}\label{eff2}
\gamma_c= &\frac{1}{c^2}\bigg\lbrace 2R^2\Omega^4-2R^2 \Omega^2 \kappa^2-\frac{3}{R}\left(\psi'+U_{\text{u}}'+U_{\text{v}}'\right)-4\kappa^2 U_\text{N}+R\,U_{\text{N}}'\left(\kappa^2+4\Omega^2\right)+\frac{H}{R\Omega}\big(\kappa^2+2\Omega^2\big)+2\Omega H'\\\nonumber
&+4 \Omega\Big(R U_{\varphi}''+\,U_{\varphi}'-\frac{U_{\varphi}}{R}\Big)-U_{\text{u}}''-U_{\text{v}}''-\psi''\bigg\rbrace
\end{align}
Finally, with these new parameters, we rewrite Eq. (\ref{PN dis}) in a simple and compact form 
\begin{equation}\label{abb PN dis}
\omega^2=k^2c_{\text{sp}}^2-2\pi G_{\text{p}}\Sigma^*|k|+\kappa_{\text{p}}^2
\end{equation}
which although is reminiscent of the corresponding dispersion relation in Newtonian gravity, includes the first PN relativistic corrections. We can say that, this equation is the most important expression in this paper. As expected, by ignoring the PN corrections, one can easily find the Newtonian dispersion relation as
\begin{equation}
\omega^2=k^2\,c_{\text{s}}^2-2\pi G\Sigma\,|k|+\kappa^2
\end{equation}

In the next section, we find the stability criterion using Eq. \eqref{abb PN dis}, and investigate the effects of the PN contributions. In the absence of rotation, and for an infinite medium, one can simply decide about the stabilizing or destabilizing effects of PN corrections, for more details see \cite{nazari2017post}. However, here the situation is complicated, and as we shall see without determining the background properties, it is not possible to specify the contribution of PN corrections on the stability of the system. In other words, these effects depend on the background system. We clarify this point in the next section.

\section{ Toomre's Criterion in the PN limit}\label{Toomre}
In this section, we will find a criterion for stability of a differentially rotating, gaseous disk in PN gravity. Note that all the corrections and quantities on the R.H.S. of Eq. \eqref{abb PN dis} are real. Consequently $\omega^2$ is a real quantity. Hence the disk is unstable against perturbation with wavenumber $k$ if $\omega^2<0$ and stable if $\omega^2>0$.
In general, since the R.H.S. of Eq. \eqref{abb PN dis} is a quadratic equation for $k$, and the coefficient of $k^2$ is positive. In this case the minimum of $\omega^2$ occurs at $k_{\text{s}}$. Therefore if $\omega^2$ is positive for this wavenumber, it will be positive for all other modes. In this case, the gaseous disk will be stable against all local axisymmetric disturbances. In other words, $k_s$ is the most probable mode that may develop an instability in the system.

By differentiating the R.H.S. of Eq. \eqref{abb PN dis}, we find $k_{\text{s}}$ as follows
\begin{align}\label{kmin}
k_{\text{s}}=\frac{\pi\,G_{\text{p}}\Sigma^*\,\text{sign}(k)}{c_{\text{sp}}^2}
\end{align}
If the R.H.S. of Eq. \eqref{abb PN dis} for this wavenumber is positive, then the system will be stable against axisymmetric perturbations. Therefore, one may simply write the PN Toomre's criterion as
\begin{align}\label{Q PN}
Q_{\text{p}}=\frac{\kappa_{\text{p}}\,c_{\text{sp}}}{\pi\,G_{\text{p}}\,\Sigma^*}>1
\end{align}

As it is clear, we have defined the PN parameters in such a way that the PN Toomre's criterion is mathematically similar to the standard case. We recall that the standard Toomre stability criterion is written as $Q>1$, where $Q=\frac{\kappa\,c_{\text{s}}}{\pi\,G\,\Sigma}$. In order to make a comparison between Newtonian and PN limits, it is instructive to write Eq. \eqref{Q PN} with respect to $Q$ instead of $Q_{\text{p}}$. To do so, let us fix the EOS. Indeed, we should find the relation between $c_{\text{s}}^*$ and the Newtonian sound speed $c_{\text{s}}$. Here we assume that the EOS of the disk is polytropic as $p=K \Sigma^{*\gamma}$, where $K$ and $\gamma$ are constant real parameters. Hence, using the definition of sound speed and Eq. \eqref{sigmastar}, we immediately obtain
\begin{align}\label{cs star}
c_{\text{s}}^{*2}=c_{\text{s}}^2\left[ 1+\frac{\gamma-1}{c^2}\left(\frac{R^2\Omega^2}{2}+3U_{\text{N}}\right)\right]
\end{align}
also one can easily show that the PN correction to the pressure defined in Eq. (\ref{p_PN}) is written as
\begin{eqnarray}
p_{\text{c}}=K \gamma\Sigma^{\gamma}\left(\frac{R^2\Omega^2}{2}+3U_{\text{N}}\right)
\end{eqnarray}
Now by substituting the effective quantities $\kappa_{\text{p}}$, $c_{\text{sp}}$, $G_{\text{p}}$, and $\Sigma^*$ into Eq. \eqref{Q PN}, and expanding the result up to order $c^{-2}$, we can write Eq. \eqref{Q PN} in terms of $Q$ as 
\begin{align}\label{correctionto1}
 Q   >&1+\frac{(4 \pi  G \kappa  R \Sigma^2)^{-1}}{c^2}\Big(
p \left(8 R  c_\text{s} (\kappa ^2-\Omega ^2)-4 R^2 \Omega ~  c_\text{s} \Omega'\right) +\Pi (8 \pi  G
   \kappa  R \Sigma^2-4 R^2 \Sigma \Omega  c_\text{s} \Omega'-8 R \Sigma\Omega ^2 c_\text{s})\Sigma\\\nonumber
   & - H \left(4 R^2 \Sigma c_\text{s} \Omega'+12 R \Sigma \Omega  c_\text{s}\right) + U_\text{N} \left(8 \pi  G \kappa  R \Sigma^2-6 \pi  \gamma  G \kappa  R \Sigma^2\right)
          + 2R^2\Sigma U_\text{N}' \left(\pi  G \kappa  \Sigma-4 R \Omega  c_\text{s} \Omega'-12 \Omega^2 c_\text{s}
            \right)\\\nonumber         
            & +
           8 \Sigma \Omega  c_\text{s} U_{\varphi}-8 R \Sigma \Omega  c_\text{s} U_{\varphi}'-8 R^2 \Sigma \Omega  c_\text{s} U_{\varphi}''
           +6 \Sigma c_\text{s} \psi' +2 R \Sigma c_\text{s} \psi''+6 \Sigma c_\text{s} U_\text{u}'+2 R \Sigma c_\text{s} U_\text{u}''+6 \Sigma c_\text{s} U_\text{v}' \\\nonumber
            &+2 R \Sigma c_\text{s} U_\text{v}'' -4 R^3 \Sigma \Omega ^4 c_\text{s}-\pi  (\gamma -2) G \kappa R^3 \Sigma^2 \Omega ^2-4 R^2 \Omega  c_\text{s} H'\Big)
\end{align}
The terms proportional to order $c^{-2}$ on the R.H.S. of Eq. \eqref{correctionto1} are the PN corrections to the standard Toomre's criterion. Consequently, the stability analysis can be influenced by these PN corrections. We emphasize that Eqs. \eqref{abb PN dis} and \eqref{correctionto1} are the main result of this paper. Notice that the PN Toomre's criterion is fundamentally \textit{nonlocal} in the sense that it depends on integrals across the entire disk via functions like $U$, $\psi$ and so on. Albeit, this happens also in the standard case where one needs to calculate the epicycle frequency $\kappa$. More specifically to calculate $\kappa$, it is necessary to find the potential $U$. However, this nonlocality is much more evident in the PN limit.

It is interesting to determine the role of all physical quantities like the pressure, internal energy and angular momentum on the stability of the system. In our recent work (\citealp{nazari2017post}), we indicated that, in an infinite and non-rotating medium, the pressure and internal energy can make the system more unstable in the PN approximation. In other words, although in the Newtonian system pressure supports the stability, in the PN limit has destabilizing effects.
One may note that the terms which contribute a negative value to the R.H.S. of Eq. \eqref{correctionto1} will stabilize the system. However in the rotating systems studied here, unfortunately, it is not straightforward to specify the sign of all the terms. In practice, to investigate the role of each term in the local stability of the system, we first need to determine the surface mass density profile, $\Sigma$, of the disk. In other words, it is not simple to make a general analysis to determine the stabilizing/destabilizing behavior of all the terms. For example, the potential $\psi$ contains the pressure as well as the internal energy. Therefore, without taking into account the contributions of $p_N$ and $\Pi$ in $\psi$, it is not possible to determine their role.

\section{Exponential fluid disk in the PN approximation}\label{etamu}

In the following, let us choose a toy model for the disk in order to  clarify the effect of the physical quantities on the local stability of the system. As we know, the exponential profile of the surface mass density is widely used to describe galactic disks in astrophysics (\citealp{lin1987formation,silk1981dissipational,leroy2008star}). By considering this specific surface-density profile, in principle, one can calculate each term in the PN Toomre's criterion. Hereafter we choose an exponential disk as follows
\begin{align}\label{density profile}
\Sigma(y) = \Sigma_0\,e^{-2y}
\end{align}
where the dimensionless parameter $y$ is defined as $y=\frac{R}{2 R_{\text{d}}}$,  $R$ is the radial coordinate in the cylindrical coordinates system, $R_{\text{d}}$ is the disk's length scale, and $\Sigma_0$ is the central surface density.

Now, let us calculate all the  terms on the R.H.S. of Eq. \eqref{correctionto1} by using the above-mentioned surface density. One can easily derive the Newtonian potential of an axisymmetric exponential disk at the equatorial plane $z=0$, for more detail see \cite{binney2008galactic}. This potential is
\begin{align}\label{U0N}
U_{\text{N}}(y,0) =2 \pi G \Sigma_0 R_{\text{d}}\,y \left[I_0(y) K_1(y)-I_1(y) K_0(y) \right] 
\end{align}
where $I_n$ and $K_n$ for $n=0,1$ are modified Bessel functions of the first and second kinds respectively. Consequently, after inserting this equation into Eq. \eqref{OmegaN}, we have
\begin{align}\label{omegaN}
\Omega^2(y) =-\frac{K\gamma\Sigma_0^{-1+\gamma}e^{-2y(\gamma-1)}}{2R_{\text{d}}^2 y}+\frac{\pi G \Sigma_0}{R_{\text{d}}}\left[I_0(y) K_0(y)-I_1(y) K_1(y) \right] 
\end{align}
One should note that the first term on the R.H.S of Eq. \eqref{omegaN} comes from the pressure gradient. As we will discuss in this section, this term can be ignored in ordinary non-relativistic systems like spiral galaxies. To find the PN contributions of the gravitational potential $U$, i.e., $U_{\text{u}}$ and $U_{\text{v}}$, we insert Eqs. \eqref{density profile}, \eqref{U0N}, and \eqref{omegaN} into Eq. \eqref{defU0}. Then we solve these two integrals numerically. Note that some integrations in Eq. \eqref{potentials} should be done numerically. For example, the terms containing $U_{N}$ and $\Omega$ in $\psi$ lead to numerical integrations. Moreover, the last equation for $U_{\varphi}$ is solved numerically too.

After calculating all the PN corrections on the R.H.S. of Eq. \eqref{correctionto1}, we find a complicated equation in terms of $y$. This equation includes some numeric and analytic parts. Therefore we do not write this long equation here. However, in order to discuss the result more clearly, we define two helpful dimensionless parameters $\eta$ and $\mu$, using combinations of $c$, $G$, $R_{\text{d}}$, $\Sigma_0$, and $K$. These parameters are
\begin{align}\label{eta}
 \eta=\frac{G R_{\text{d}} \Sigma_0}{c^2}
\end{align} 
and 
\begin{align}\label{mu}
\mu=\frac{K\Sigma_0^{\gamma-1}}{c^2}
\end{align}
 Let us explicitly clarify how these parameters help to describe the stability of the system. In fact, the standard Toomre's criterion, i.e., $Q=\frac{c_{\text{s}} \kappa}{\pi G \Sigma}>1$, is conveniently interpreted as a competition between stabilizing effects of pressure $(c_{\text{s}})$ and angular momentum $(\kappa)$ against destabilizing effects of gravity $(G \Sigma )$. 
Using our toy model, one can easily show
\begin{align}
\frac{c_{\text{s}}^2}{c^2}=\mu\gamma e^{-2y(\gamma-1)}
\end{align}
As we see, $\mu$ is a representative of the fluid pressure. Note that, for some cases of interest in the Newtonian context, like galaxy-scale gas disks, the sound speed $c_{\text{s}}$ is much smaller than the rotational speed \cite{binney2008galactic}. One should note that this is not true in all Newtonian disks. For example, Newtonian accretion disks, like Advection Dominated Accretion Flows (ADAFs) (\citealp{narayan1994advection}), can have $c_{\text{s}}\simeq v_{\varphi}$. Therefore for systems with $c_{\text{s}}\ll v_{\varphi}$, the angular velocity for this case reads
\begin{align}\label{nnn1}
v_\varphi\simeq\sqrt{-R\frac{dU}{dR}}
\end{align}
and consequently, the first term vanishes in (\ref{omegaN}). However, in this paper, we deal with weakly relativistic systems. Therefore the pressure contributions cannot be ignored in the calculations.

In order to find an interpretation for $\eta$, we use Eq. (\ref{OmegaN}) to show that $\kappa \propto \sqrt{ (F_g+F_p)/R}$ where $F_g$ and $F_p$ denote respectively the gravitational force per unit mass, and the pressure force per unit mass.
Therefore one may consider $\kappa$ as a function related to gravity and pressure. From this perspective, one may write Toomre's parameter as $Q=c_{\text{s}}/c_g$ where $c_g=\pi G \Sigma/\kappa$. Now for our toy model in which the pressure gradient has been ignored, i.e., $|F_p/F_g|\ll 1$, one can show
\begin{align}
\frac{c_g ^2}{c^2}=   \frac{\pi\eta}{2}
\, e^{-4y}
\Big[ \big( 2 I_0(y)+y I_1(y)\big)  K_0(y)-\big( y I_0(y)+I_1(y)\big)  K_1(y)\Big] ^{-1}
\end{align}
In this special case, where we have ignored $F_p$ compared with $F_g$, we can describe Toomre's criterion, i.e., $Q\propto \sqrt{\mu/\eta}$, only as a competition between pressure and gravity. We emphasize that the pressure gradient's contribution to $\Omega$, i.e., the first term on the R.H.S of Eq. \eqref{omegaN}, is ignored. Notice that this does not mean that the fluid pressure is totally ignored. In fact its effect is still present in $c_\text{s}$. In this specific case, $\kappa$ is only related to gravity. So, $\eta$ and $\mu$ can be simply regarded as parameters related to the strength of gravity and pressure respectively. However, it should be stressed that when the pressure gradient is important, as we will see in the subsequent sections, we we cannot consider $\kappa$ as a function only of gravitational strength. In this case, it is more accurate to consider it as a function of $F_p$ and $F_g$. Consequently, we  lose the clear interpretation of the parameter $\eta$ as a representative of gravity. Therefore one needs to be careful when interpreting the stability criterion using these parameters.  Nonetheless, it is still helpful to use them, to simplify the analysis. 

Now, for our toy model by taking into account the pressure gradient, the Newtonian Toomre's parameter reads 
\begin{align}\label{QN}
Q = & \frac{\sqrt{\gamma}e^{-y(\gamma-3)}}{\sqrt{2}\pi} \Big[\frac{e^{-2y(\gamma-1)}(-3+2y(\gamma-1))\gamma}{y}\frac{\mu^2}{\eta^2}
+4\pi\frac{\mu}{\eta}\big((2 I_0(y)+y I_1(y)) K_0(y)-(y I_0(y)
+I_1(y)) K_1(y)\big)\Big]^{1/2}
\end{align}
From this expression we infer that in the Newtonian limit and for $y>3/2(\gamma-1)$, increasing $\eta$ helps the instability and increasing $\mu$ supports the stability. Albeit one should note that only the ratio  $\mu/\eta$ is important. In other words, increasing $\mu/\eta$ will stabilize the disk. However,  the situation is more complicated in the PN limit, and as we will show in what follows, it is not straightforward to guess the effects of these parameters on the stability of the system. In fact, in this limit, the R.H.S. of stability criterion (\ref{correctionto1}) is a function of $\eta$ and $\mu$ as well. Moreover, $\eta$ and $\mu$ do not appear as  the ratio $\mu/\eta$  on the R.H.S. This is the origin of complexity in interpreting the role of the stability parameters.
\begin{figure}[]
 \begin{center}
  \includegraphics[scale=0.86]{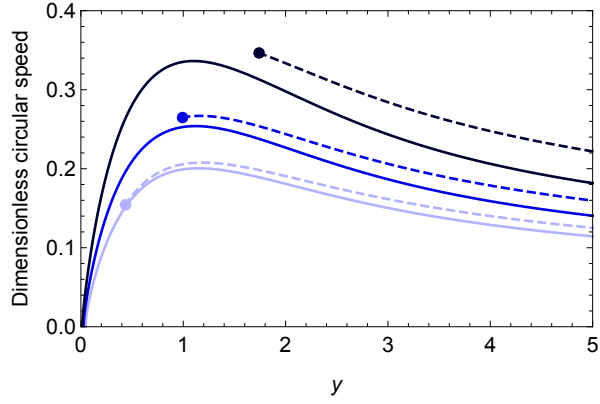}
  \caption{The dimensionless circular speed with respect to dimensionless radius ($y=R/2R_{\text{d}}$) for an adiabatic exponential fluid disk with $\gamma=5/3$.
 The solid (dashed) curves belong to the Newtonian (PN) case. Here $\mu$ is fixed, and given by $\mu=0.01$. The curves from top to bottom belong to $\eta=0.05$, $\eta=0.03$, and $\eta=0.02$ respectively. The points at the beginning of the PN curves indicate radii after which the PN approximation is valid.}\label{rotarion curve}
\end{center}
 \end{figure}

On the other hand, one may impose some restrictions on the magnitude of the parameters $\eta$ and $\mu$. As we know, PN theory is limited to the relatively slow motions as well as weak gravitational fields. More specifically, in the PN limit, we have 
\begin{eqnarray}
\frac{U_{\text{N}}}{c^2},~ \frac{v_{\varphi}^2}{c^2},~ \frac{c_{\text{s}}^2}{c^2},~ \frac{\Pi}{c^2},~ \text{and}~\frac{p_\text{N}}{\Sigma c^2}\ll 1
\end{eqnarray}
If we assume that these fractions cannot be larger than $0.1$, then using Eq. \eqref{U0N}, we can rewrite the condition $U_{\text{N}}/c^2 < 0.1$ as follows
\begin{align}\label{cond_eta}
\eta < \frac{0.1}{ 2\pi y} \left[I_0(y)K_1(y)-I_1(y)K_0(y)\right]^{-1}
\end{align}
In fact, by using this condition, we find a maximum value for $\eta$ at each radius to ensure that PN theory is valid. Similarly, to obtain constraints on $\mu$, we also rewrite the conditions $c_{\text{s}}^2/c^2<0.1$, $\Pi/c^2<0.1$, and $p/\Sigma c^2 <0.1$ as follows
\begin{align}\label{cond_mu}
&\mu< 0.1\text{min}(\gamma^{-1}, \left(\gamma-1\right), 1) \,e^{2y\left(\gamma-1\right)}
\end{align}
It turns out that for the cases studied in this paper, the constraint on $\mu$  which comes from $c_{\text{s}}^2/c^2<0.1$ is the most restrictive one. Moreover, considering $v_{\varphi}^2/c^2<0.1$, one can find a relation between $\eta$ and $\mu$ as follows
\begin{align}\label{cond_em1}
2  y  \Big(2 \pi  \eta y \big(I_0(y) K_0(y) -I_1(y) K_1(y)\big)-\gamma  \mu ~ e^{2y(1-\gamma)}\Big)<0.1
\end{align}
On the other hand, the existence of the pressure gradient which appears with a negative sign in the calculations may, in principle, yield to imaginary quantities. Therefore we still need extra constraints on $\eta$ and $\mu$ to control this possibility. More specifically, one can find two more constraints, assuming the fact that $v_\varphi$ and $\kappa$  must be real quantities. Regarding Eq. \eqref{OmegaN} and the definition of epicyclic frequency, and also using Eq. \eqref{omegaN}, we find two more independent constraints
\begin{align}\label{cond_em2}
 2 \pi \eta  y  \big(I_0(y) K_0(y)-I_1(y) K_1(y)\big)-\gamma ~ \mu ~ e^{2y(1-\gamma)}>0
\end{align}
and
\begin{align}\label{cond_em3}
\gamma ~  \mu ~ e^{-2 (\gamma -1) y} (2 (\gamma -1) y-3)+4 \pi  \eta y \Big(\big(2 I_0(y)+y I_1(y)\big) K_0(y)
  -\big(y I_0(y)+I_1(y)\big) K_1(y)\Big)>0
\end{align}
To summarize, we have found five conditions \eqref{cond_eta}-\eqref{cond_em3} which are necessary to ensure that results are reliable and the PN regime characteristics have been taken into account. Naturally, for given values of $\eta$ and $\mu$, there is a specific radius $y^*$  where for $y<y^*$ the PN conditions are not satisfied. In other words, we need higher PN corrections for this region. It is instructive to illustrate this fact by plotting the circular velocity of the exponential disk in PN gravity. As seen from Eq. \eqref{vphi0PN}, the contributions of PN potentials change the Newtonian circular speed. Now we compare the Newtonian circular velocity with the PN circular velocity. In Fig. \ref{rotarion curve}, we plot dimensionless circular velocity defined as $v_{\varphi}/c$ for $\mu=0.01$ and different values of $\eta$.
In this figure, for each value of $\eta$, the curve is drawn from the threshold radius after which the PN approximation is valid, and accordingly the forbidden region is omitted. One can see that, as expected, the threshold radius gets larger with increasing $\eta$. Moreover, this figure shows that for a fixed radius, increasing $\eta$ yields larger dimensionless circular velocity. Furthermore, the difference between Newtonian and PN curves increases by $\eta$. Regarding this case as an example, one can see how these dimensionless parameters can identify the PN reliable area.

\begin{figure}[!]
\begin{center}
 \includegraphics[scale=0.8]{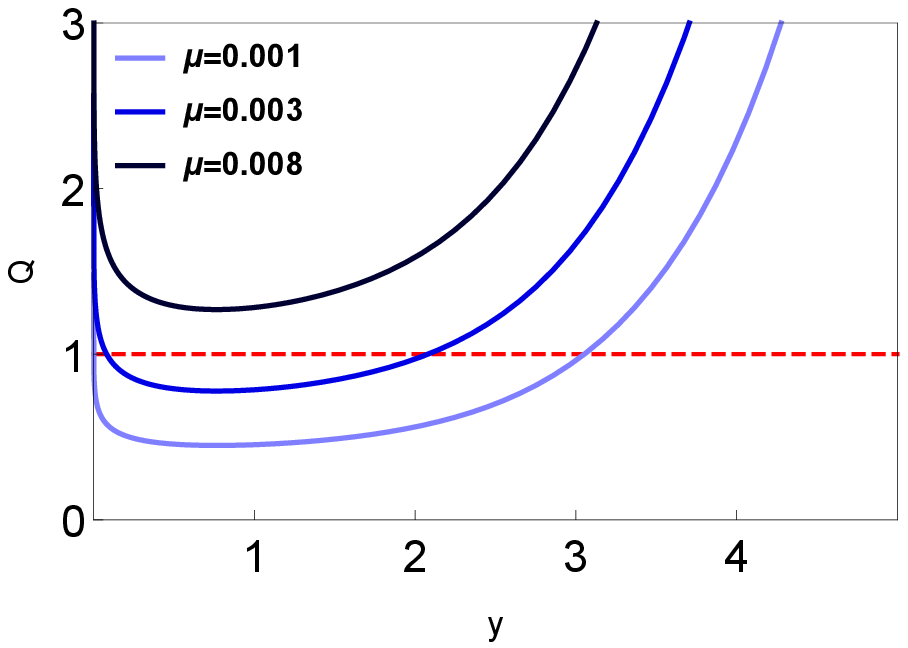}
 \hspace{2mm}
 \includegraphics[scale=0.8]{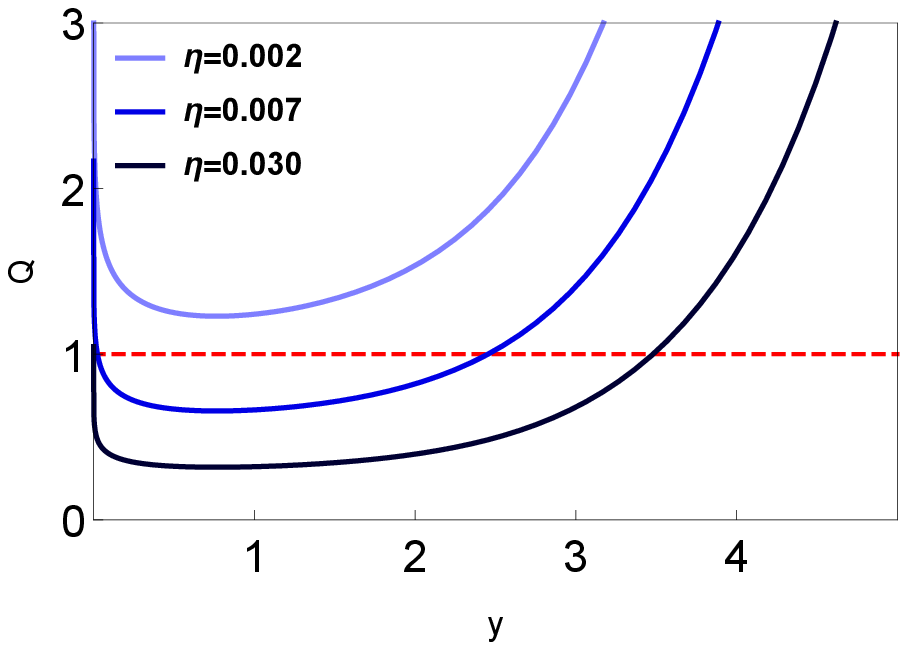}
\caption{This figure shows the quantity $Q$, when the pressure gradient is ignored, in terms of $y=R/2R_{\text{d}}$ for various values of the parameters. The left panel is plotted for a fixed $\eta$ ($=0.015$) and from up to down $\mu=0.008$, $\mu=0.003$, and $\mu=0.001$ respectively. The right panel belongs to $\mu=0.001$, and from up to down $\eta=0.002$, $\eta=0.007$, and $\eta=0.030$ respectively. \label{NTC}}
\end{center}
 \end{figure}
 %
\begin{figure*}[!]
\centering
\includegraphics[scale=0.45]{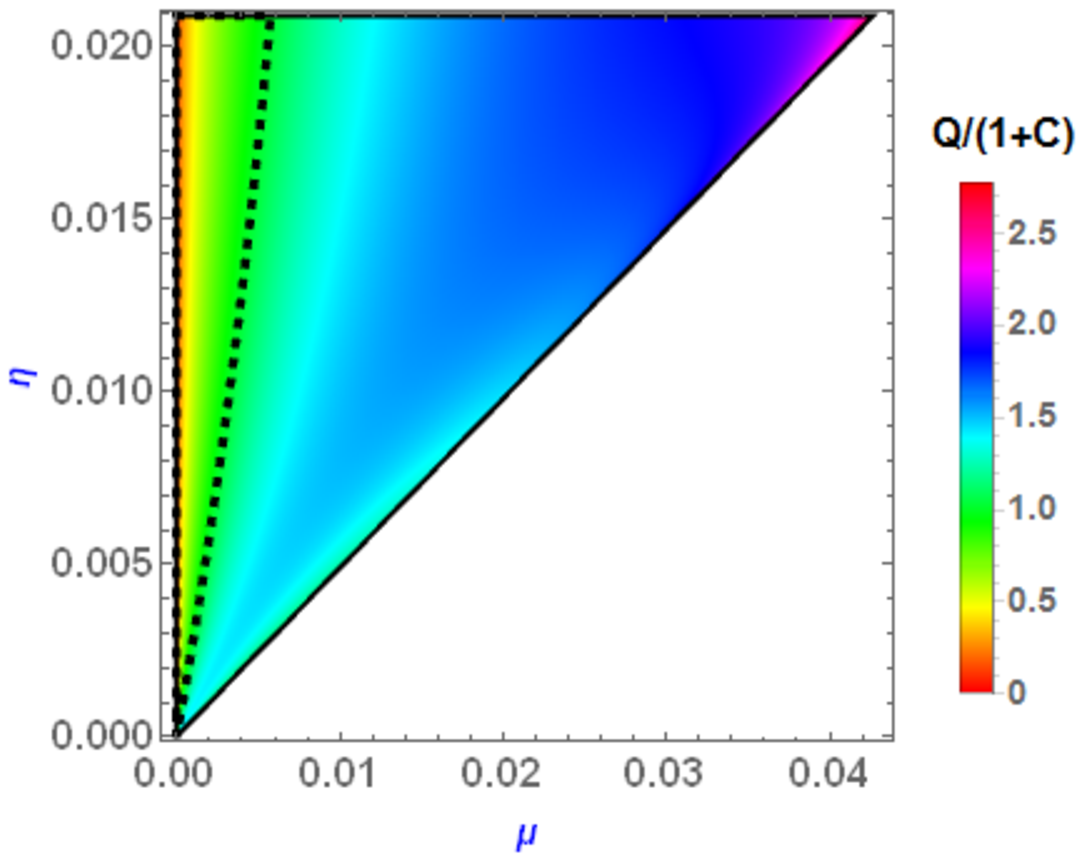}
\hspace{4mm}
\includegraphics[scale=0.45]{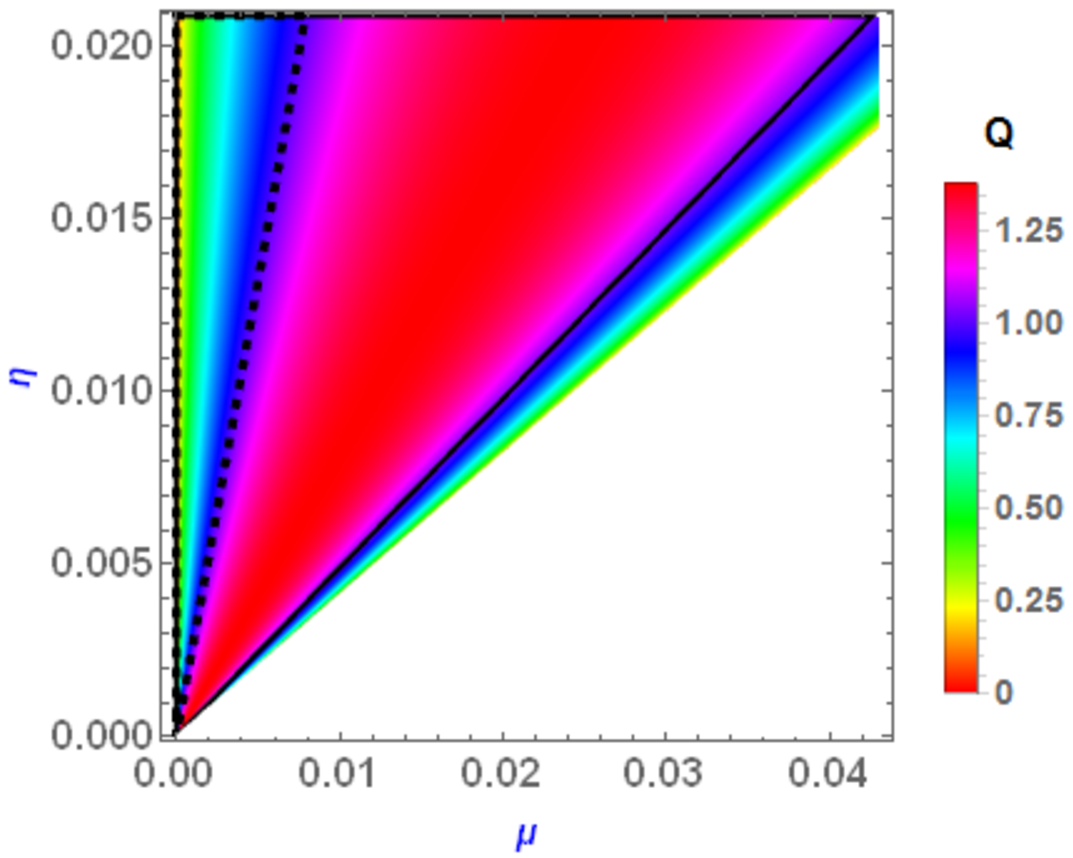}
\hspace{4mm}
\includegraphics[scale=0.45]{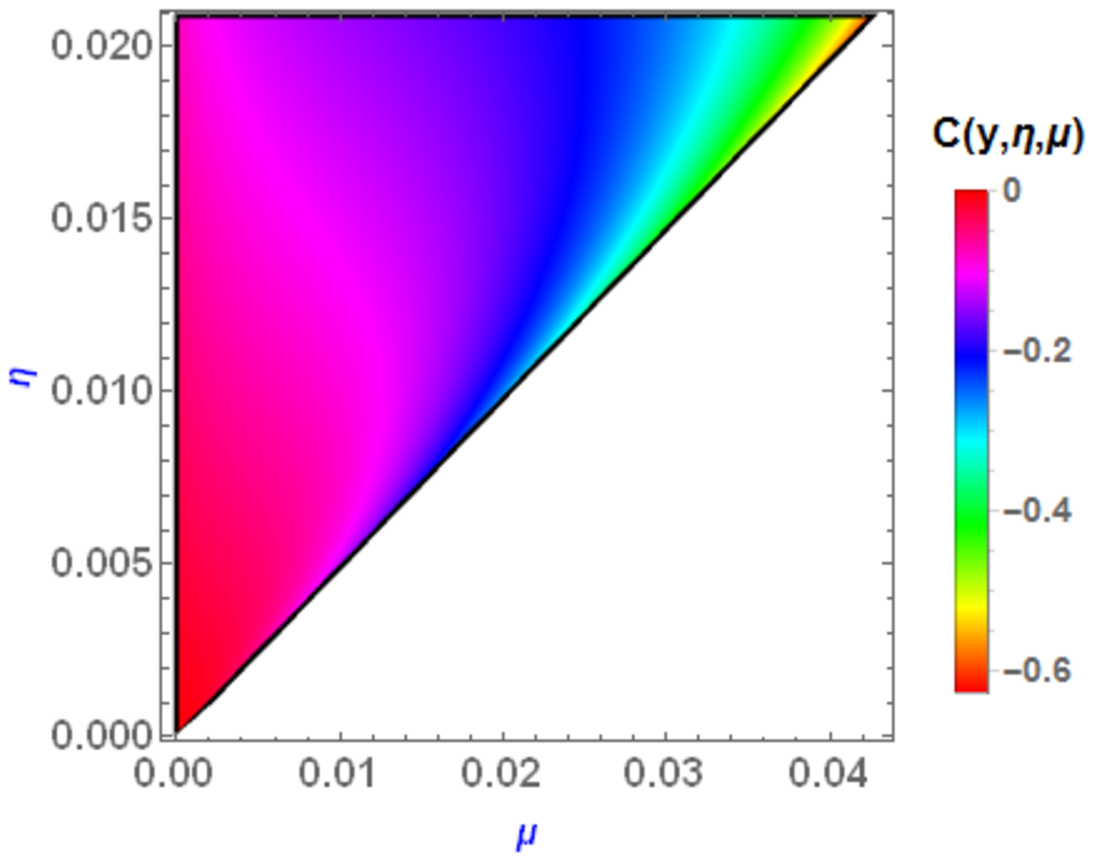}\vspace{3mm}
\includegraphics[scale=0.45]{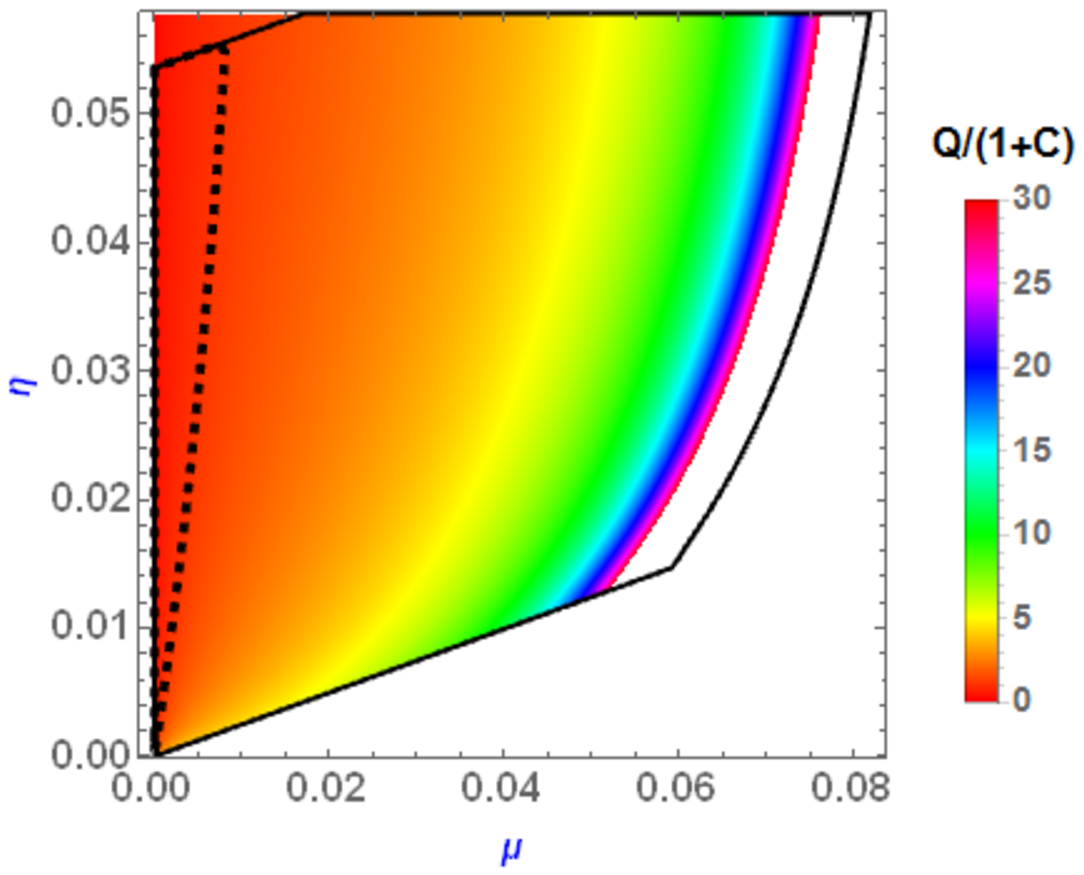}
\hspace{4mm}
\includegraphics[scale=0.45]{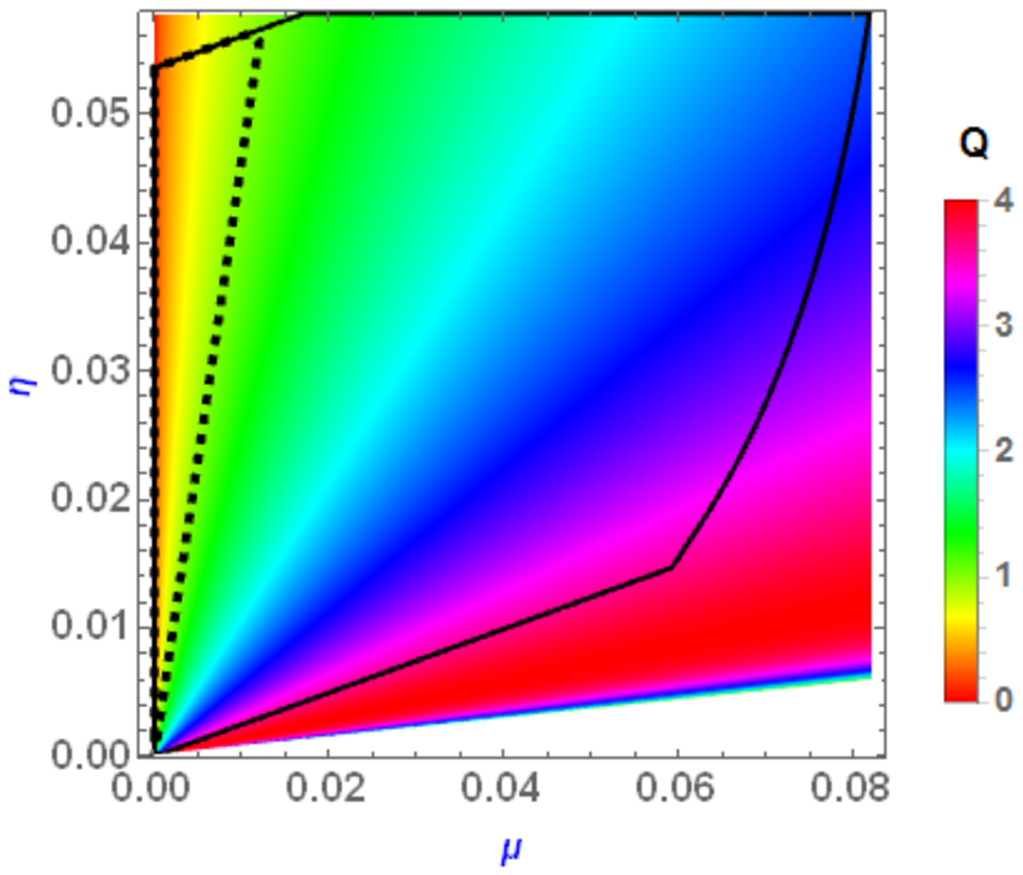}
\hspace{4mm}
\includegraphics[scale=0.45]{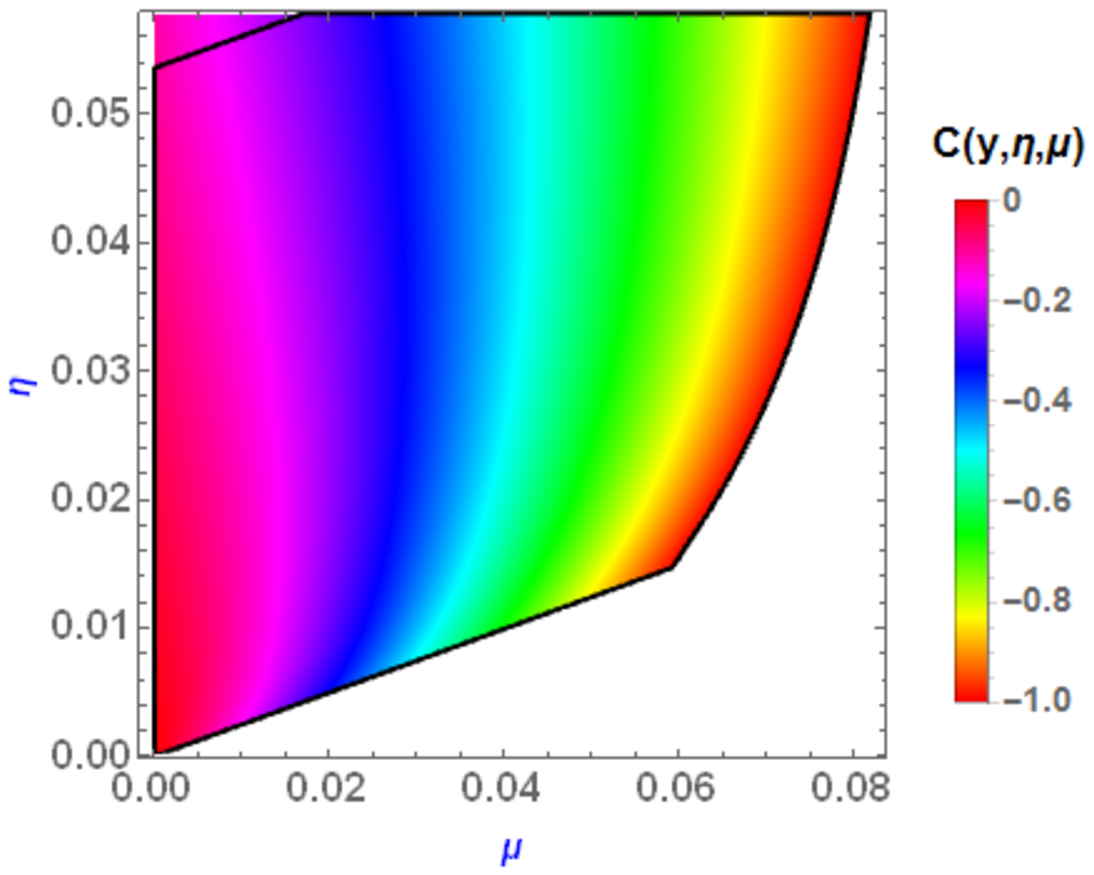}\vspace{3mm}
\includegraphics[scale=0.45]{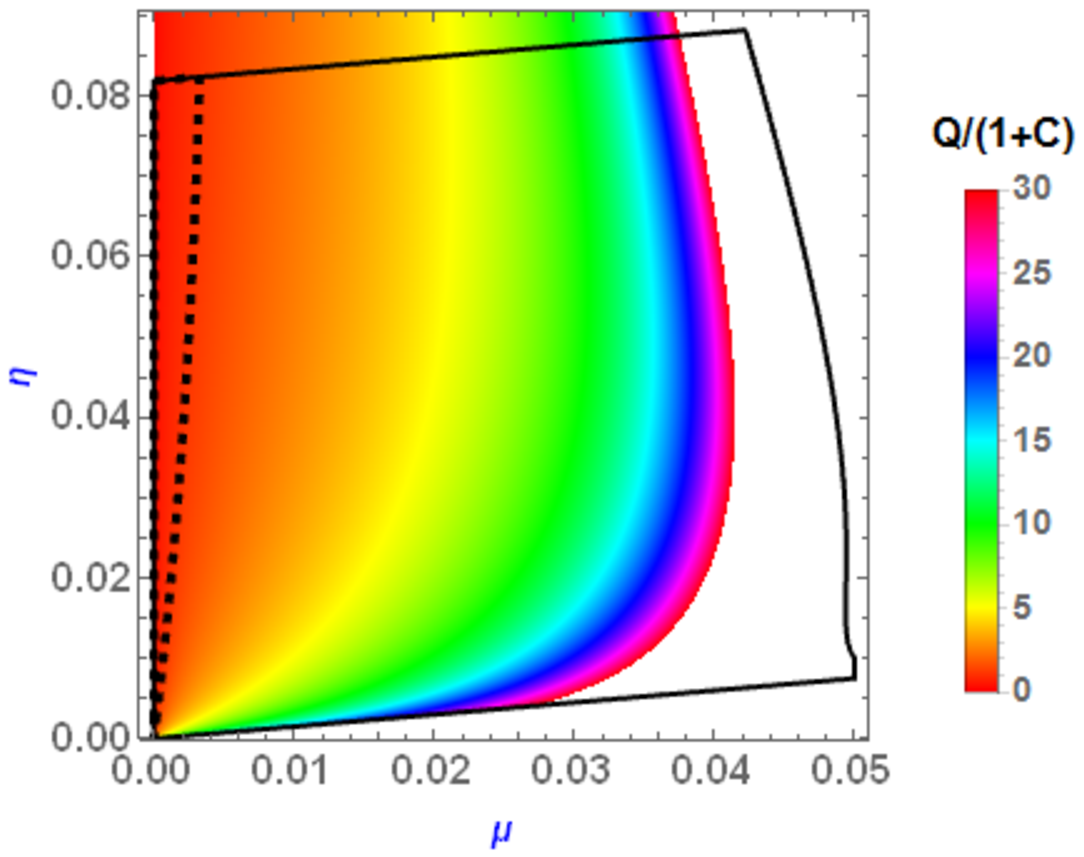}
\hspace{4mm}
\includegraphics[scale=0.45]{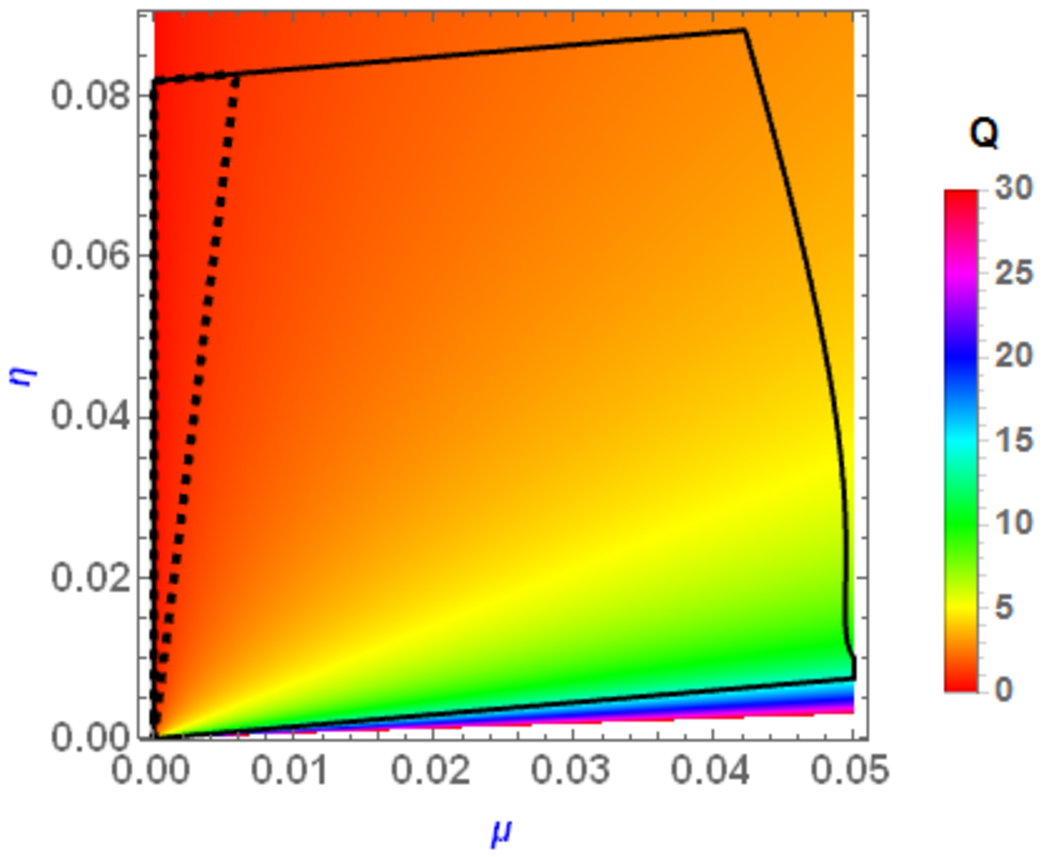}
\hspace{4mm}
\includegraphics[scale=0.45]{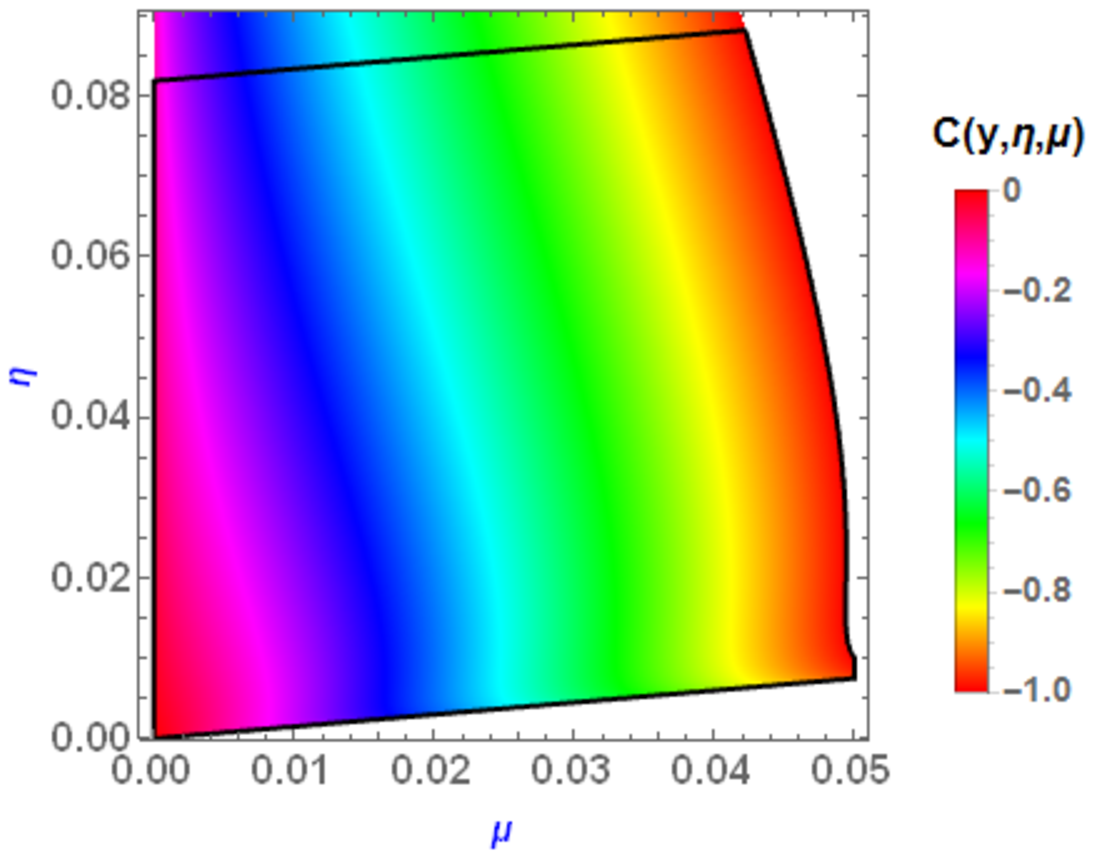}\vspace{3mm}
\includegraphics[scale=0.45]{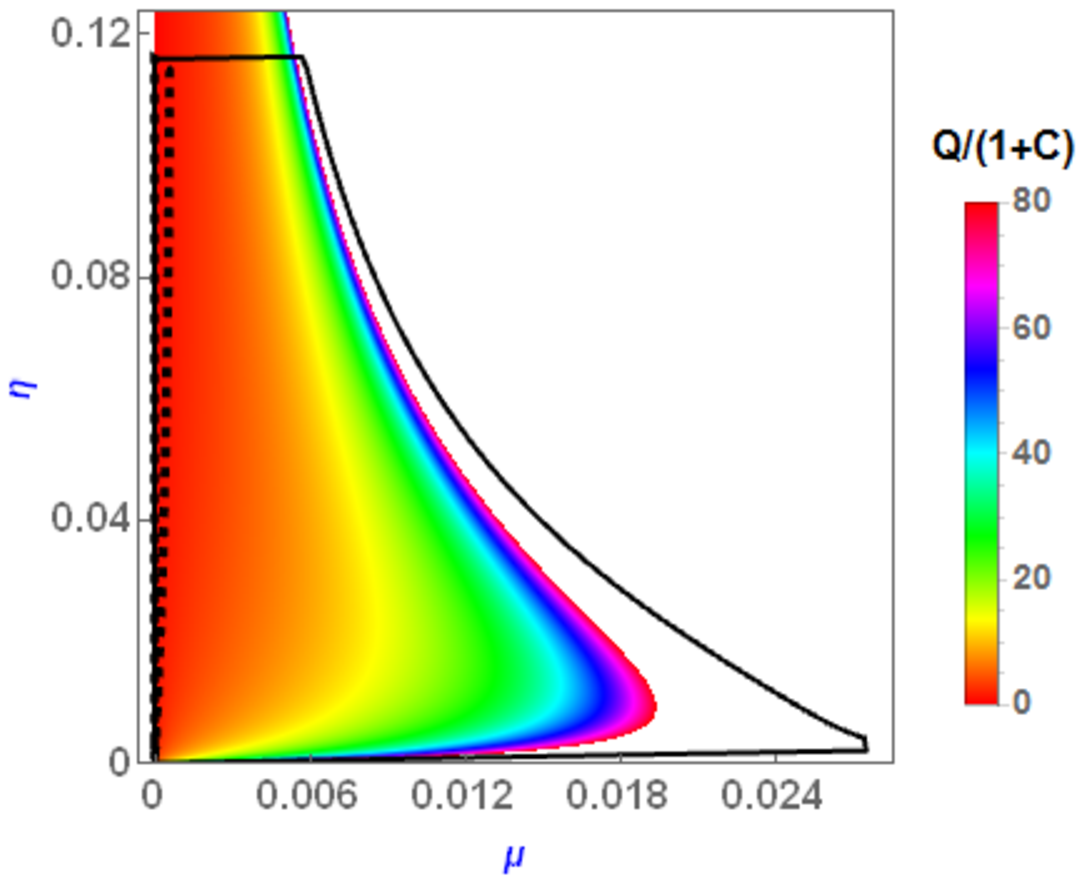}
\hspace{4mm}
\includegraphics[scale=0.45]{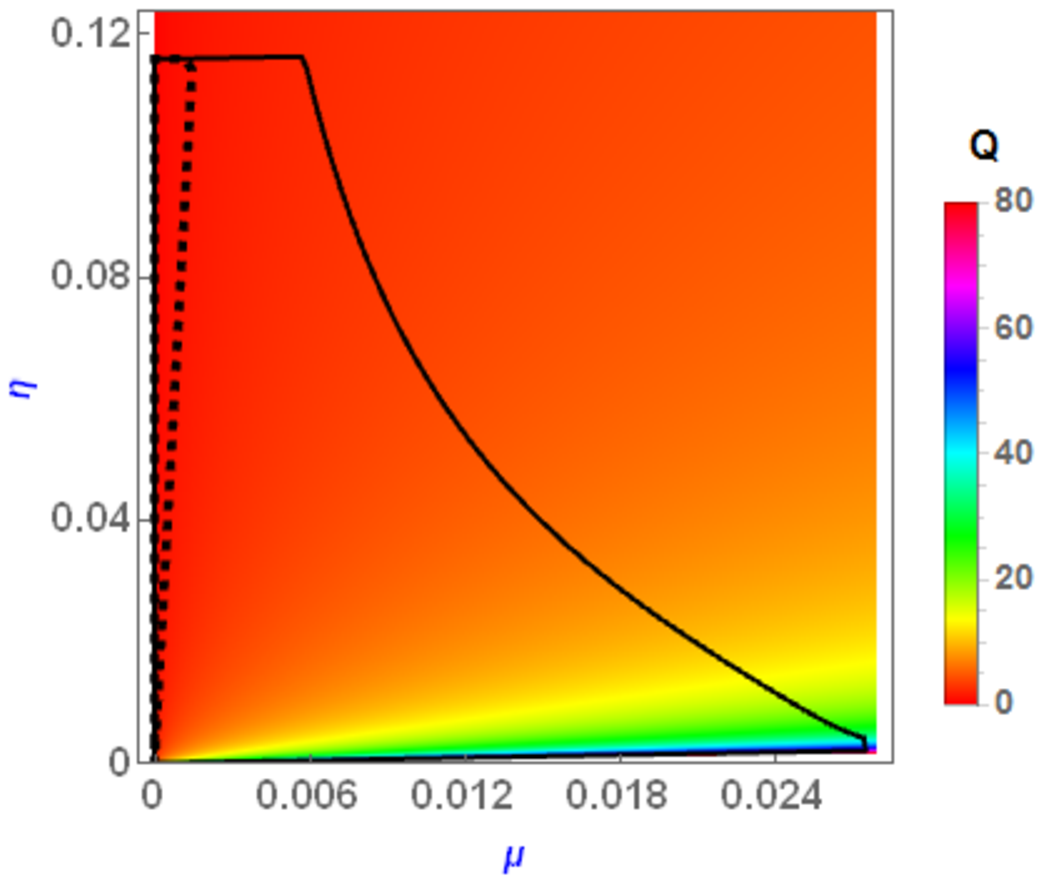}
\hspace{4mm}
\includegraphics[scale=0.45]{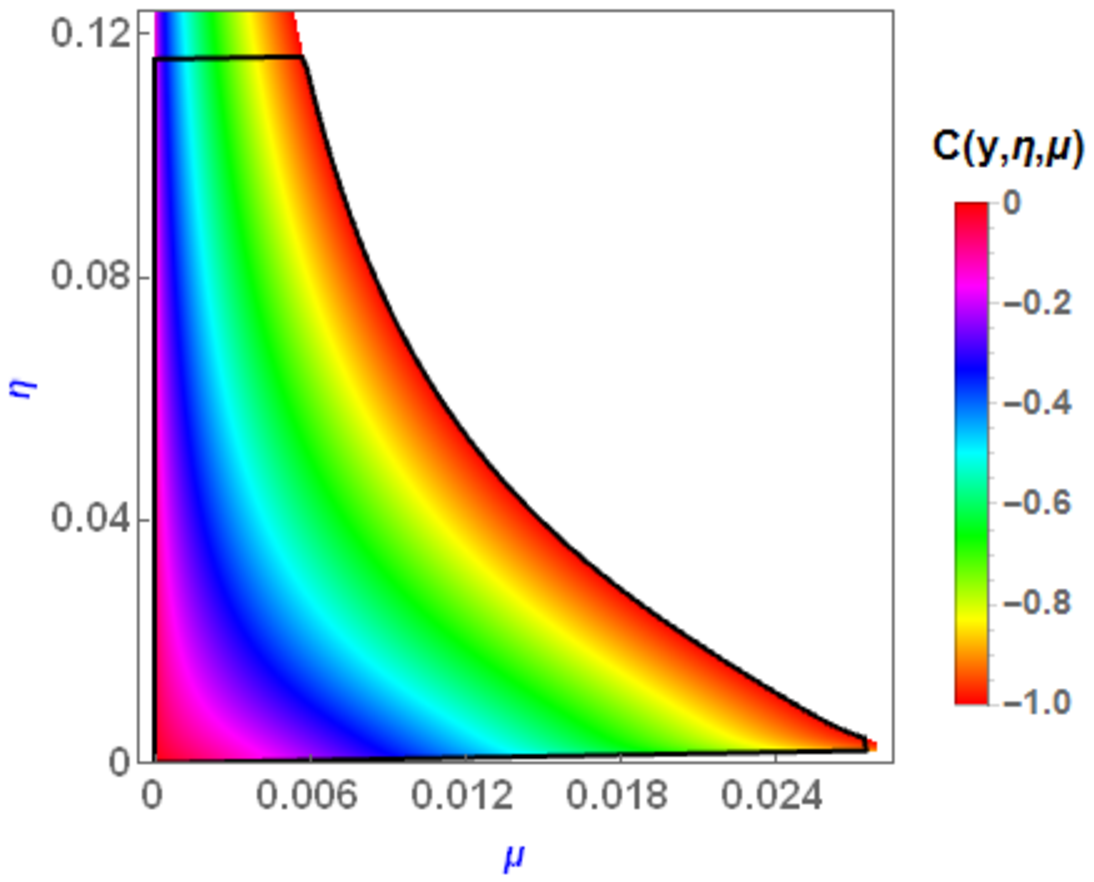}\vspace{3mm}
\includegraphics[scale=0.45]{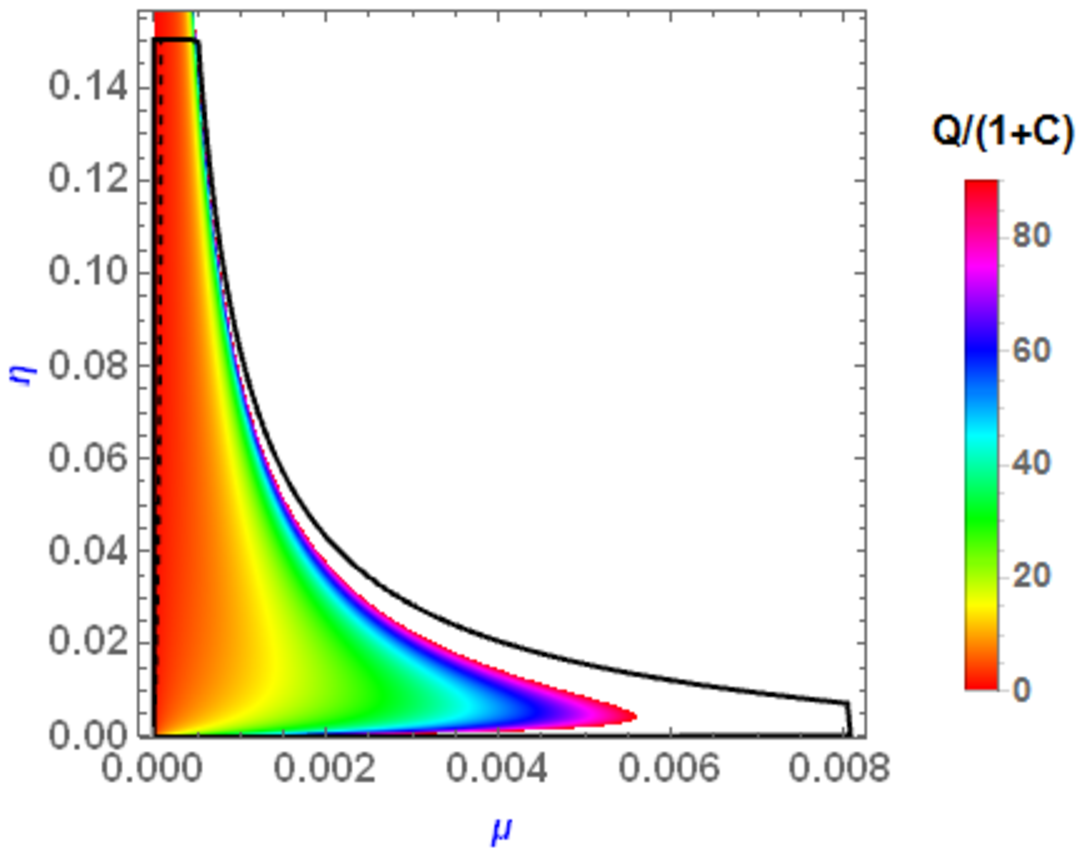}
\hspace{4mm}
\includegraphics[scale=0.45]{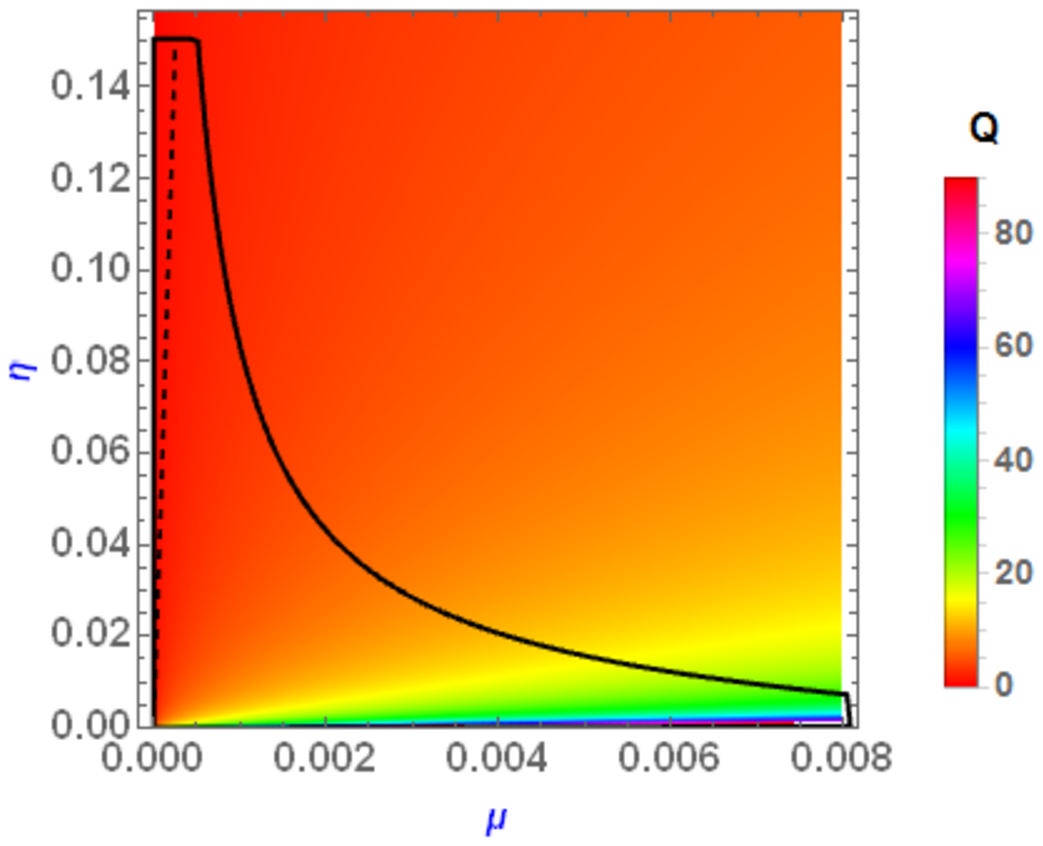}
\hspace{4mm}
\includegraphics[scale=0.45]{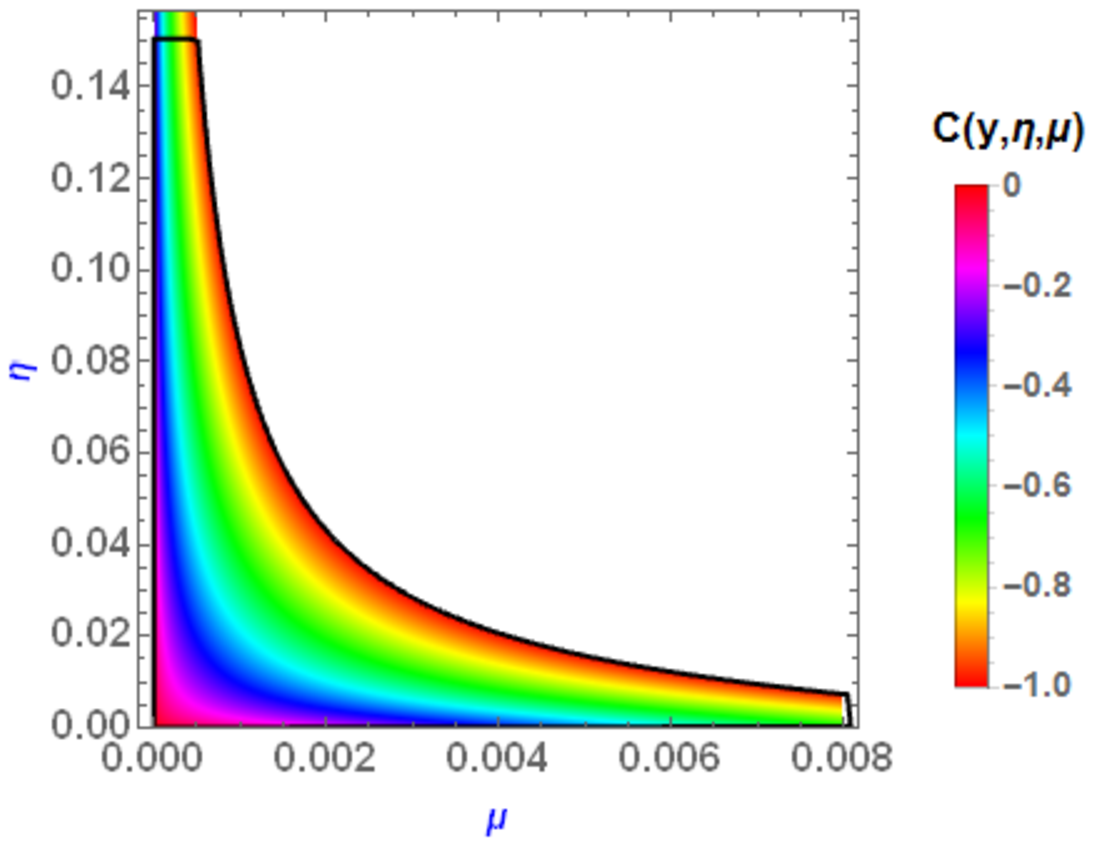}
\caption{In this figure, at each row (where devoted to a fixed $y$), from left to right the panels show $Q/\left( 1+C(y,\eta,\mu)\right) $, $Q$, and $C(y,\eta,\mu)$ respectively (in terms of $\eta$ and $\mu$). Furthermore, in each column, form up to down, panels belong to $y=0.5$,  $y=2$, $y=3$, $y=4$, and $y=5$ respectively. The black (solid) curves show the region where all the mentioned constraints are established. Also, the dashed curves in the first (second) column show the PN (Newtonian) unstable area ( i.e., $Q/\left( 1+C(y,\eta,\mu)\right)<1 $ and $Q<1$). \label{Dens_plots}}
\end{figure*}


\subsection{Local stability of the relativistic exponential disk in the PN limit}

Now let us return to our main aim in this paper. After calculating the PN corrections on the R.H.S. of Eq. \eqref{correctionto1} and using the dimensionless quantities $y$, $\eta$ and $\mu$, one can rewrite Eq. \eqref{correctionto1} as follows
\begin{align}\label{Q1C}
Q(y,\mu,\eta)>1+C(y,\eta,\mu)
\end{align}
where $C(y,\eta,\mu)$ is the combination of all PN corrections to the R.H.S. of Toomre's criterion.
As mentioned before, the final relation is long and complicated. Therefore we present some figures to reveal the main features in Eq. \ref{Q1C}.

Now let us explain the role of each parameter on the stability of the PN system. To distinguish the effects of $\eta$ ($\mu$), we we vary it while keeping the other parameter, i.e., $\mu$ ($\eta$), constant. For the sake of completeness,  we first briefly review the Newtonian Toomre's criterion for our exponential disk.  Fig. \ref{NTC} shows Toomre's parameter $Q$ in terms of $y$, for various values of $\eta$ and $\mu$. In this case, one may assume that the pressure gradient is negligible.
The left (right) panel is plotted for a constant $\eta$ ($\mu$) and different values of $\mu$ ($\eta$).
The left panel shows that by increasing $\mu$, as expected the system becomes more stable. Similarly, the right panel confirms that $\eta$ has destabilizing effects. So, it seems that these parameters are helpful to study the local stability of the system.

 Here we  return to the PN Toomre's criterion (\ref{correctionto1}). Using  Fig. \ref{Dens_plots}, we can practically study the PN system and compare it with the Newtonian case. Note that, in this figure, in addition to the  mentioned conditions \eqref{cond_eta}-\eqref{cond_em3} a new constraint is included. 
In fact, the magnitude of $C(y,\eta,\mu)$, the correction terms appeared on the R.H.S of Eq. \eqref{Q1C}, should be smaller than  unity.
The region in the parameter space ($\eta$, $\mu$) in which all these constraints are established, is denoted as a black solid curve. One can see the PN Toomre's criterion in terms of the stability parameters $\eta$ and $\mu$, in the left panel of each row in Fig. \ref{Dens_plots}.  In practice, we rewrite Eq. \eqref{Q1C} as $Q/\left(1+C(y,\eta,\mu) \right) >1$ to study the roles of the parameters more clearly. In this case the PN system will be stable if $Q/\left(1+C(y,\eta,\mu) \right)$ be larger than unity. For comparison, the middle panels show the standard Toomre's parameter for different values of $\eta$ and $\mu$.  Moreover, the right panels show the correction term $C(y,\eta,\mu)$ introduced in PN Toomre's criterion.  The sign of this term and its magnitude directly specify the deviations between Newtonian and PN descriptions. In other words, if $C<0$ ($C>0$), then the system is more stable (unstable) in the PN description compared to the Newtonian analysis. 

 From  the middle panel of the first row, one can see that increasing $\mu$  unexpectedly destabilizes the Newtonian system at small radii. In fact, using Eq. (\ref{QN}) one can simply show that the first term is negative for small radii $y<3/2(\gamma-1)$. This directly means that the pressure  becomes important at small radii, and can reduce the effect of the angular momentum in the system. Consequently, we have smaller rotational velocity and $\kappa$, i.e., larger $c_g$,  compared with a similar disk in which the pressure gradient is ignored. Therefore, in the Newtonian description, increasing the pressure parameter $\mu$  destabilizes the inner radii, in contrary to what might be expected from the usual sense about the stabilizing effects of the pressure. It turns out that this behavior also exists in the PN analysis. This point is not  very clear in Fig. \ref{Dens_plots}, however, we will discuss it in Fig. \ref{Boundary}. It should be noted that considering the necessary restrictions on the magnitude of the parameters, the unexpected behavior is seen in a small region.

It is important to mention that the unexpected behavior of the pressure  has also been seen in the local  gravitational instability  in PN non-rotating systems  (\citealp{nazari2017post}). In this case, the gravitational effects of the relativistic pressure can strengthen the gravitational force, and in principle supports the instability. For a detailed description of this point, we refer the reader to \cite{nazari2017post}.

\begin{figure}[!]
\begin{center}
 \includegraphics[scale=0.8]{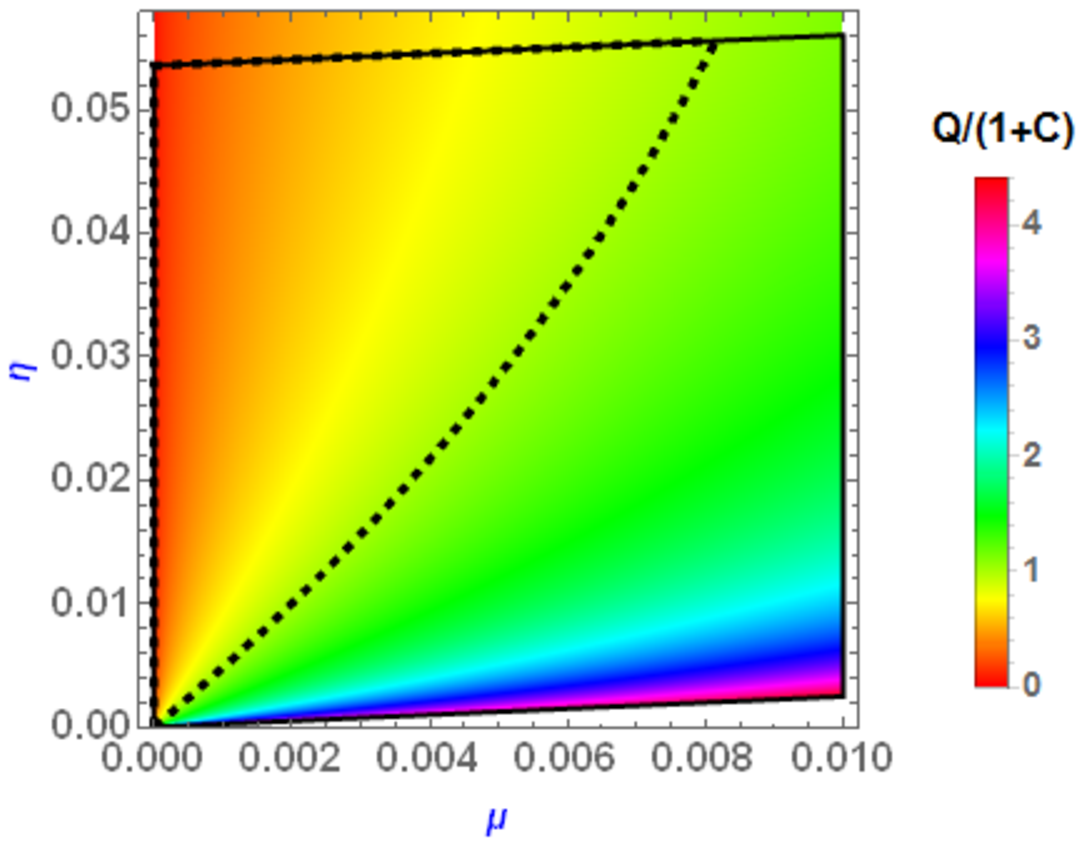}
 \caption{The PN Toomre's criterion  at $y=2$. It is clear that increasing the parameter $\eta$ ($\mu$)  makes the system more unstable (stable). \label{zoom}}
\end{center}
 \end{figure}

For larger radii, considering the two left panels at each row, one can see that increasing $\mu$ leads to stability in both Newtonian and PN limits. 
The response of the system to changes in $\eta$ is more straightforward in both descriptions. More specifically, increasing this parameter destabilizes the system in all radii. However, since we have to illustrate the whole allowed parameter space in Fig. \ref{Dens_plots}, the destabilizing effect of $\eta$ is not  very clear (except in the left panel  of the first row). We  illustrate the behavior of $\eta$ in a zoomed plot in Fig. \ref{zoom}.
One may conclude from Fig. \ref{Dens_plots} that, in the allowed range of the parameters, $Q$ and $Q/(1+C(y,\eta,\mu))$ are increasing functions with respect to $R$. In fact, as expected, moving to the edge of the disk, the system will become more stable.
Furthermore, regarding the right panels of Fig. \ref{Dens_plots}, one can compare the stability of the disk in the context of Newtonian and PN regimes. It is clear that the correction term $C$ is negative for all cases. This means that the system is more stable in the PN viewpoint  compared to the Newtonian one. This is one of our main results in this paper. Also, the difference between two theories gets insignificant for the small values of both parameters. Furthermore one can see that the disk can be unstable at all radii, for small values of $\mu$ in both Newtonian and PN limits.

It is interesting that the PN effects stabilize our rotating exponential thin disk, while  these relativistic effects decrease the Jeans mass in a non-rotating infinite medium \citep{nazari2017post} and consequently, destabilize it. Albeit it must be noted that one cannot  conclude that PN effects stabilize all rotating systems. As we mentioned before, deciding about this behavior is not straightforward and one has to fix the background system. Therefore, in principle, the outcome of the  stability analysis would not be the same in different systems.

\subsection{Growth rates in PN approximation}
Let us compare the growth rate of the perturbation in both regimes using the parameters $\eta$ and $\mu$. For unstable modes, $\omega^2<0$ and it is convenient to define a dimensionless growth rate $s=i\,\omega/\kappa$. Also, a dimensionless wavenumber can be defined as $q=k c_{\text{s}}/\kappa$. Using these parameters and inserting Eqs. \eqref{sigmastar} and \eqref{cs star} into Eq. \eqref{PN dis} we find
\begin{align}\label{PN growing rate}
s^2= &  \frac{2 q}{Q}-q^2-1+\frac{1}{c^2}\Big\{
 \frac{q}{Q} \Big(2 \Pi-4 U_\text{N}+\frac{2 p}{\Sigma } -2 R^2 \Omega ^2\Big)
 + q^2 \Big((4-3 \gamma ) U_\text{N}+\frac{p}{\Sigma }+\Pi -\frac{1}{2} (\gamma -2) R^2
    \Omega ^2\Big)\\\nonumber
    & -\frac{2 R \Omega  H'}{\kappa ^2}-H \Big(\frac{2 \Omega }{\kappa
       ^2}+\frac{1}{\Omega }\Big)-\Big(\frac{4 R \Omega ^2}{\kappa
       ^2}+R\Big) U_\text{N}'+4 U_\text{N}+2 R^2 \Omega ^2+\frac{1}{\kappa^2}\Big(
       -2 R^2 \Omega ^4 +U_\text{u}''+U_\text{v}''+\frac{3 U_\text{u}'}{R}\\\nonumber
       &+\frac{3 U_\text{v}'}{R}-4 R \Omega  U_{\varphi}''-4 \Omega  U_{\varphi}'+\frac{4 \Omega  U_{\varphi}}{R}+ \psi''+\frac{3 \psi'}{R}
       \Big)
\Big \}
\end{align}
As we have already mentioned for an exponential disk given by Eq. \eqref{density profile}, we can calculate all the terms in the above equation. It is natural that $\eta$ and $\mu$ appear in the result. Therefore using Eq. \eqref{PN growing rate} for given values of $y$, $\eta$, and $\mu$, we can plot $s^2$ with respect to $q$, and investigate the response of the system to unstable modes.

It is important to note that Eq. \eqref{PN growing rate}, in general, is not applicable for all wavenumbers. In fact, some wavenumbers may violate the necessary conditions on the PN corrections. Therefore, one can find the relevant constraint on $q$ by taking into account that the magnitude of the PN correction terms, in principle, should be smaller than the Newtonian terms. This limitation has also been  exploited in \cite{nazari2017post}.
\begin{figure}
\begin{center}
\includegraphics[scale=0.85]{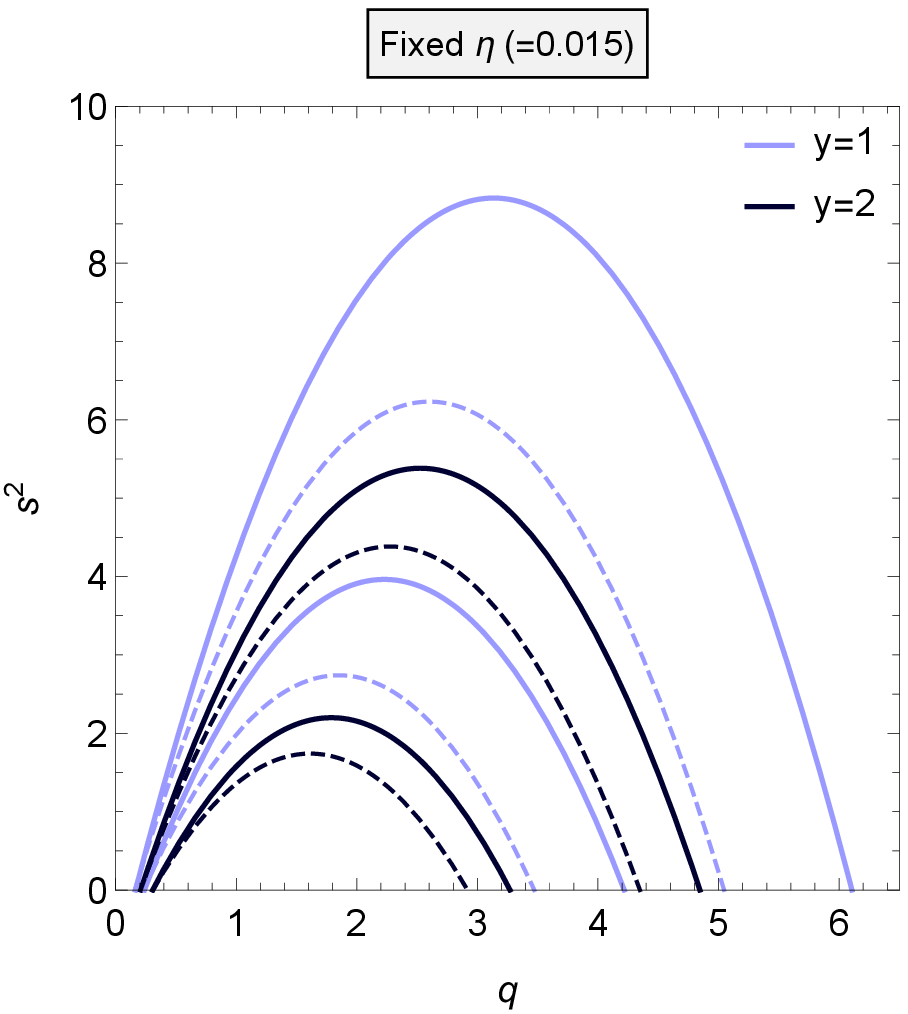}\hspace{10mm}\vspace{0.5cm}
\includegraphics[scale=0.85]{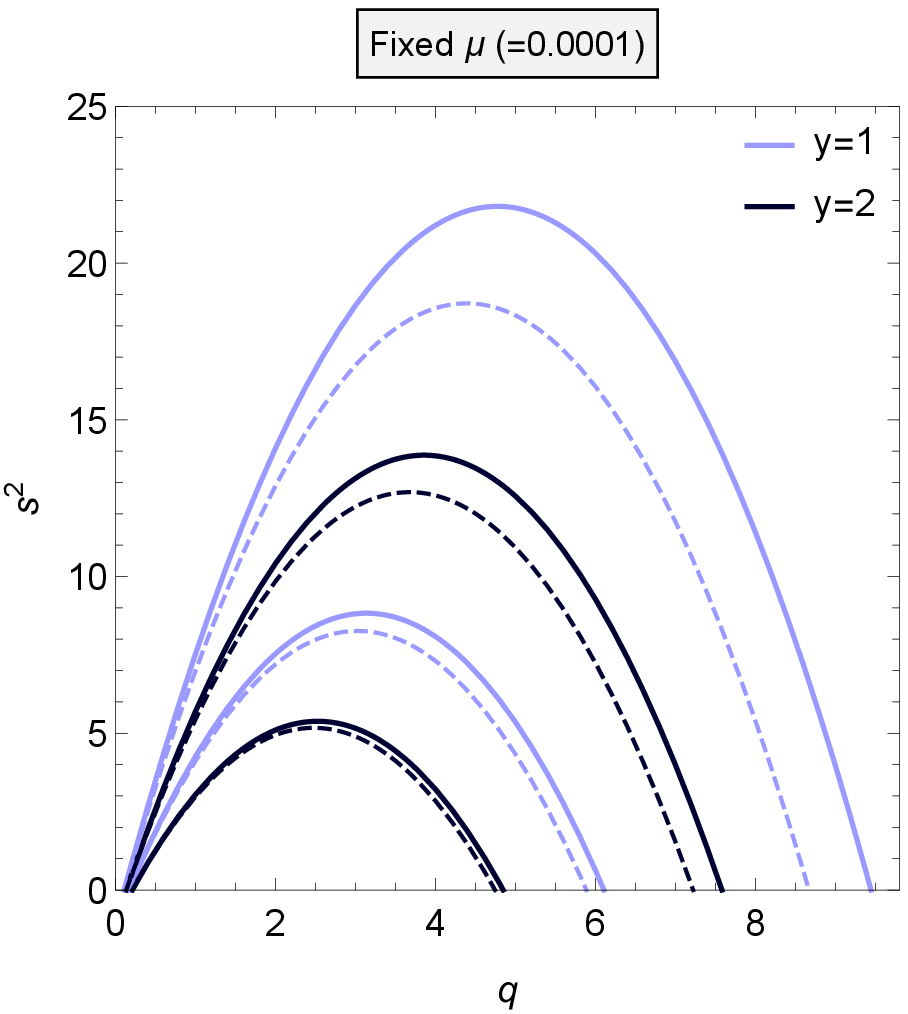}
\caption{The squared dimensionless  growth rate with respect to the dimensionless wavenumber $q=k c_{\text{s}}/\kappa$ for an adiabatic fluid with $\gamma=5/3$. The dashed (solid) curves belong to the PN (Newtonian) case. The left panel shows the growth rate for $\eta=0.015$. From up to down, for each radius, dashed/solid curves correspond to $\mu=5\times10^{-4}$ and  $\mu=10^{-3}$. The right panel shows  the growth rate for $\mu=10^{-4}$. From up to down, for each radius, dashed/solid curves correspond to $\eta=7\times10^{-3}$ and $\eta=3\times10^{-3}$. }\label{growingRate}
\end{center}
\end{figure}
%
 Fig. \ref{growingRate} shows the  growth rate of unstable modes for allowed values of $\eta$ and $\mu$ at different radii in Newtonian and PN limits.
To find the overall effect of $\mu$ on the local stability of the system, we keep $\eta$ fixed in the left panel of Fig. \ref{growingRate} and change $\mu$.  On the other hand, similarly, in the right panel, we keep $\mu$ constant and vary $\eta$. Furthermore, for the sake of completeness, we do this analysis at two different radii $y=1$ and $y=2$.
In both panels of Fig. \ref{growingRate}, the dashed (solid) curves belong to PN (Newtonian) description. The darker (lighter) curves belong to $y=2$ ( $y=1$). 
The left panel in Fig. \ref{growingRate} shows that the growth rate decreases by increasing $\mu$, at the fixed radius and $\eta$ ($=0.015$), in both Newtonian and PN gravities. Moreover, the instability interval shortens. This result is not surprising in the sense that $\mu$ is the representative of the pressure in the system and as seen in Fig. \ref{NTC}, increasing this parameter supports the stability of the system. 

Before moving on to discuss the right panel in Fig. \ref{growingRate}, let us consider the restrictions on the wavenumber $q$. As we have already mentioned, the PN correction terms must be smaller than the Newtonian terms. So Eq. \eqref{PN growing rate} is limited and does not include all wavelengths  of perturbations.  We find the relevant conditions on $q$ and discard the forbidden wavenumbers. Considering the dashed curves in the left panel of Fig. \ref{growingRate}, allowed ranges of $q$ from up to down are $0.168<q<5.046$, $0.207<q<4.348$, $0.242<q<3.472$, and $0.307<q<2.918$ respectively.

In the right panel of Fig. \ref{growingRate} $\mu$ is fixed, e.g., $\mu =10^{-4}$. By increasing $\eta$, not only the growth rate increases but also the instability interval gets larger. This result is expected since, as already discussed in Fig. \ref{Dens_plots},  larger $\eta$ means more instability and naturally higher growth rate. This result is also in agreement with Fig. \ref{NTC}. Regarding the dashed curves, the allowed ranges of dimensionless wavenumber in this panel, from up to down, are $0.108<q<8.670$, $0.132<q<7.221$, $0.165<q<5.887$, and $0.207<q<4.745$ respectively. Therefore for both panels, except a small interval, the PN approximation holds for most of the instability region.

So far, we have shown that for both Newtonian and PN limits, $\mu$ and $\eta$, in general, have stabilizing and destabilizing effects respectively.  However, we have not yet compared growth rates in Newtonian and PN limits with each other. It is clear from both panels of Fig. \ref{growingRate} that, at the same radius and with the same $\eta$ and $\mu$, the instability interval is wider  and  includes smaller wavelengths in the Newtonian case.  In the Newtonian picture, the growth rates also are larger. Therefore, one may  conclude that the system is more stable in the PN limit.
Moreover, both panels of Fig. \ref{growingRate} shows that the difference between the growth rate of Newtonian and PN systems is significant at inner radii and  for shorter wavelengths. Also, for a fixed $\eta$ and $\mu$,  the growth rate decreases with radius.

\subsection{Boundary of stability in ($Q_s$,$\Lambda$) plane}
Finally, it is  interesting to determine the boundary of the local gravitational stability of the two-dimensional self-gravitating gaseous disk in the PN context. We recall  that  the boundary of stable and unstable perturbations can be found by  setting $\omega=0$ in Eq. \eqref{abb PN dis}. So we have
\begin{align}\label{PN dimless dis}
\frac{\pi^2 G_{\text{p}}^2 \Sigma^{*2}}{\kappa_{\text{p}}^4}Q_{\text{p}}^2k^2-\frac{2\pi G_{\text{p}} \Sigma^{*}}{\kappa_{\text{p}}^2}\left|k\right|+1=0
\end{align}
 By substituting the effective quantities $\kappa_{\text{p}}$, $G_{\text{p}}$, $\Sigma^*$, and $Q_{\text{p}}$ into Eq. \eqref{PN dimless dis}, we can solve this equation in terms of $Q$.  After expanding the solution up to $c^{-2}$ terms, we obtain
\begin{align}\label{Q(Lambda)}
\nonumber Q & =\sqrt{4 \Lambda-4 \Lambda^2}+\frac{\Lambda}{ c^2 \kappa ^2 R \Sigma   \Omega \sqrt{4 \Lambda-4 \Lambda^2}}\Big\lbrace 4 \kappa ^2 p R \Omega-6 \gamma  \kappa ^2 R \Sigma  \Omega  U_{\text{N}}+4 \kappa ^2 R \Sigma  \Omega  U_\text{N} -\gamma  \kappa ^2 R^3 \Sigma  \Omega^3+4 \kappa ^2 \Pi  R \Sigma  \Omega\\\nonumber
&+\Lambda\Big( (\gamma +2) \kappa ^2 R^3 \Sigma  \Omega ^3 -2 H \kappa ^2 R \Sigma -2 \kappa ^2 p R \Omega-2 \kappa ^2 \Pi  R \Sigma  \Omega
  -4 R \Sigma  \Omega ^2 \left(R H'+H\right)-4 R^3 \Sigma  \Omega ^5+6 \gamma  \kappa ^2 R \Sigma  \Omega  U_\text{N}\\
&+ 2 \Sigma  \Omega\Big( R \big(-R \left(\kappa ^2+4 \Omega ^2\right) U_\text{N}'+U_\text{u}''+U_\text{v}''-4 \Omega  \left(R U_{\varphi}''+U_{\varphi}'\right)+\psi''\big)+4 \Omega  U_{\varphi}+3  (U_\text{u}'+U_\text{v}'+\psi')     \Big)   \Big) \Big\rbrace
\end{align}
where $\Lambda$ is a dimensionless wavelength defined as $\Lambda=\lambda/\lambda_{\text{crit}}$ in which $\lambda_{\text{crit}}$ is the largest unstable wavelength that propagates on the surface of the rotating disk with zero sound speed in Newtonian gravity, i.e., $\lambda_{\text{crit}}=4\pi^2 G \Sigma/\kappa^2$ \citep{binney2008galactic}. As expected, this equation coincides with the boundary of stable and unstable modes in Newtonian gravity by ignoring the PN correction terms. It is easy to show that $\Lambda<1$.
In fact, the line of the stability in the Newtonian approach can be defined as $k^2c_\text{s}^2-2\pi G\Sigma|k|+\kappa^2=0$. One can rewrite this equation in terms of wavelength and solve it for $\lambda$. If we expand the result as a series of $c_{\text{s}}$, we have
\begin{align}
\frac{4 \pi ^2 G \Sigma }{\kappa ^2}-\frac{c_\text{s}^2}{G \Sigma }+O\left(c_\text{s}^4\right)
\end{align}
It is clear that when $c_\text{s}$ vanishes, we will have $\lambda=\lambda_\text{crit}$. Naturally, we find smaller wavelength for the case of non zero sound speed. So we can see $\lambda/\lambda_\text{crit}<1$. For the PN approach, we deal with some small corrections, but the main point will be true. Therefore, we do not worry about radicals in Eq. \eqref{Q(Lambda)}.
\begin{figure*}[!]
\begin{center}
\includegraphics[scale=0.64]{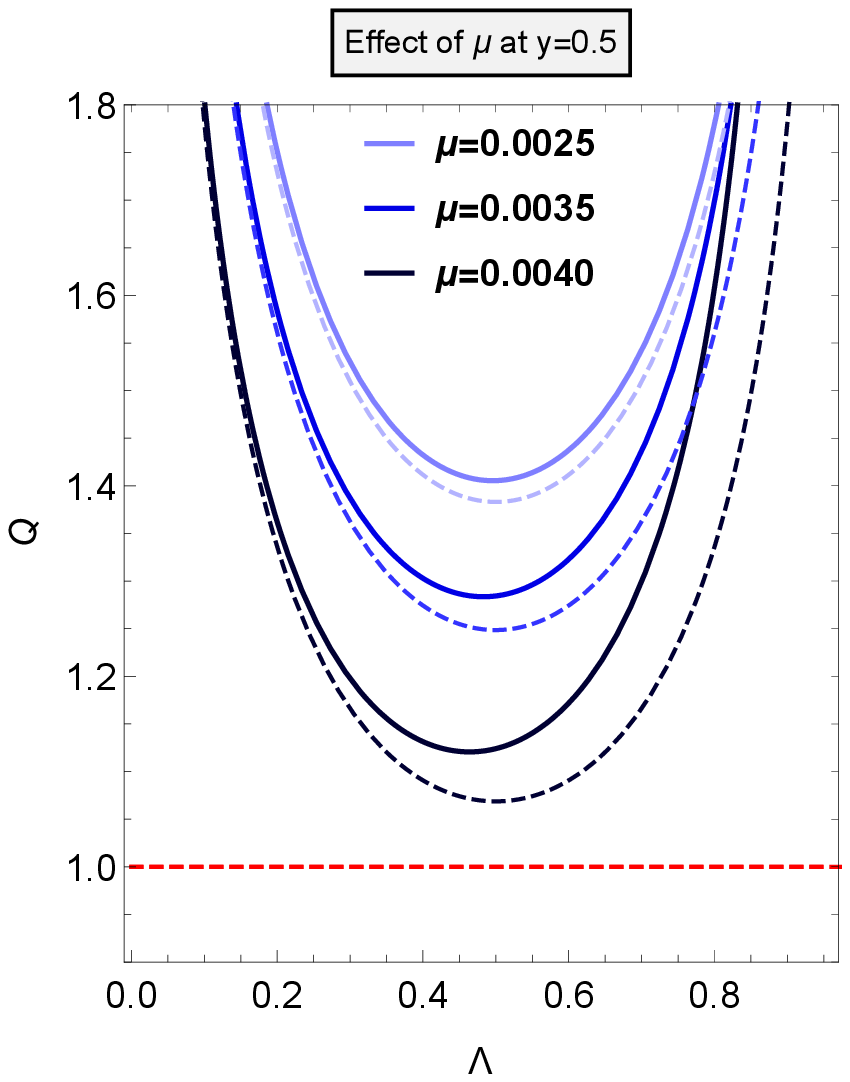}
\hspace{3mm}
\includegraphics[scale=0.64]{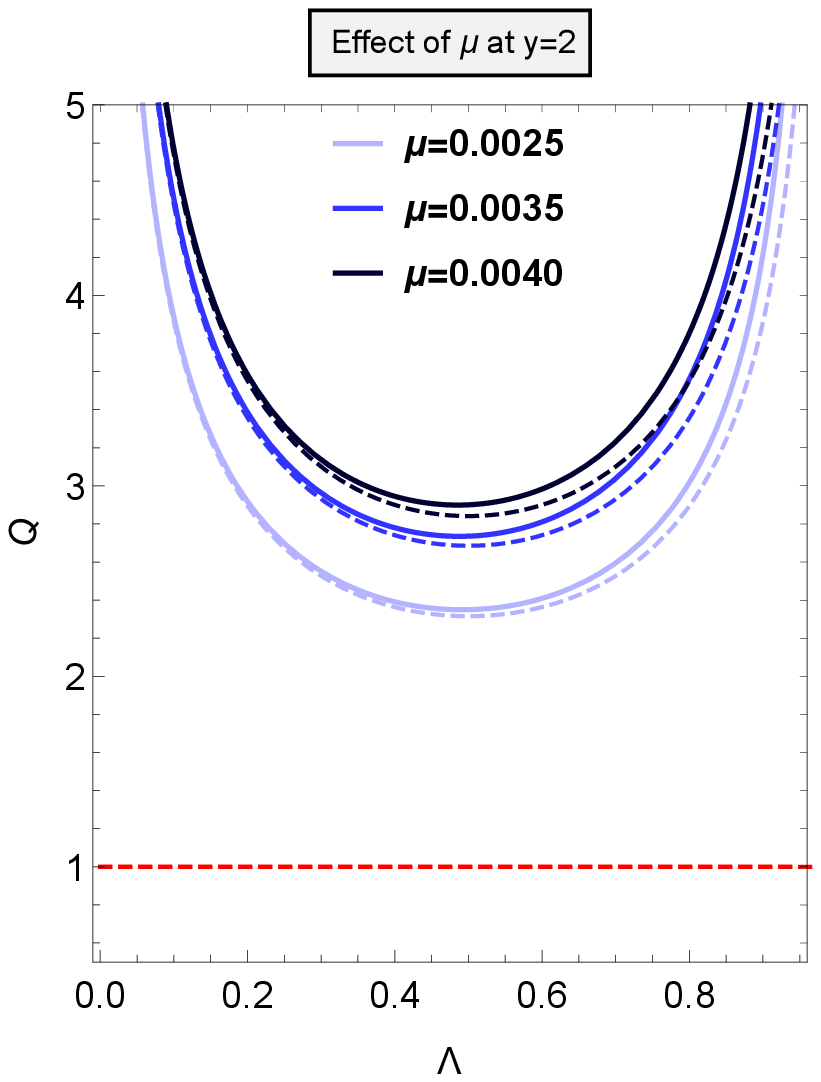}
\hspace{2mm}
\includegraphics[scale=0.64]{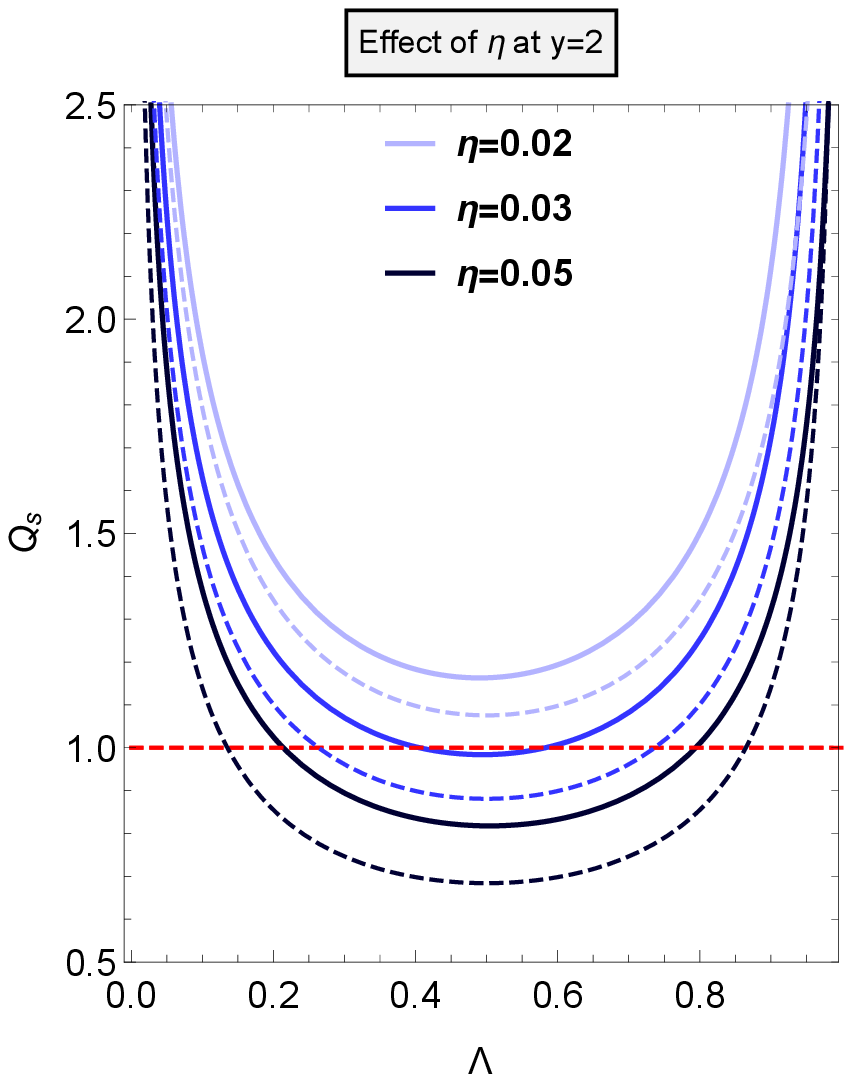}
\end{center}
\caption{The scaled boundary of stable and unstable axisymmetric disturbances in a fluid disk with $\gamma = 5/3$ in the Newtonian(dashed) and PN (solid) limits. 
The horizontal axes show the dimensionless wavelength $\Lambda=\lambda/\lambda_{\text{crit}}$.
The left and the middle panels show the behavior of the system in a fixed $\eta$ ( $=0.002$). The left (middle) panel shows $y=0.5$ ($y=2$). The right panel belongs to $\mu=0.005$ and $y=2$.
The red dashed line shows $Q_s=1$. 
\label{Boundary}}
\end{figure*}
By rewriting Eq. \eqref{Q(Lambda)} as $Q=\mathcal{F}\left(\Lambda,y,\eta,\mu \right) $, one can define $Q_s$ as
\begin{align}\label{Q_s}
Q_s=\frac{Q}{\mathcal{F}\left(\Lambda,y,\eta,\mu \right)}
\end{align}
Note that the Newtonian $Q_s$ can be found by turning off the $c^{-2}$ terms in Eq. \eqref{Q_s}. It is worth mentioning that, although it is convenient to plot the boundary in the ($Q$, $\Lambda$) plane, tracking the footprints of the PN parameters would be much easier in the ($Q_s$, $\Lambda$) plane for our case. Now, the question is how can these parameters influence the boundary of stable and unstable axisymmetric disturbances in a fluid disk?  Is it possible to infer the unexpected behavior of $\mu$ parameter from the changes in the boundary of stability in ($Q_s$, $\Lambda$) plane? To answer these questions, one can use Fig. \ref{Boundary}. The left panel in this figure shows that, at a small radius, increasing $\mu$ will make the system  more unstable in both Newtonian and PN limits. So the unexpected behavior of this parameter, which was mentioned before,  is obvious here too. On the other hand, the middle panel shows that the behavior of $\mu$ coincides with what is expected at a usual radius and increasing it leads to more stability in both viewpoints. 
In the right panel, one can see the destabilizing effect of $\eta$. This panel shows that the system moves to be more unstable by increasing this parameter.
Here, it is easy to check that in agreement with the results of Fig. \ref{Dens_plots}, the PN system  is more stable than the Newtonian case. One can see in all panels of Fig. \ref{Boundary} that,  for each pair of parameters, $Q_s$ is bigger in the context of PN gravity compared to the Newtonian one.

As a final remark in this section, it should be noted that considering the conditions \eqref{cond_eta} to \eqref{cond_em3}, we have used the allowed values of the parameters here. However, imposing a constraint on the PN correction term in Eq. \eqref{Q(Lambda)}, which should be smaller than the main Newtonian term, one may ignore some long wavelengths in the figure. In the left panel, for the curves from lighter to darker (this hierarchy is held to the end of this part),  we have $\Lambda<0.996$, $\Lambda<0.982$, and $\Lambda<0.961$, respectively. Similarly, for the middle panel we have $\Lambda<0.991$, $\Lambda<0.987$, and $\Lambda<0.985$. We also have $\Lambda<0.990$, $\Lambda<0.994$, and $\Lambda<0.997$ for the right panel.

\section{Application to the hyper massive neutron stars}\label{hmns}
 \begin{table}[t]
 \caption{Characteristics of various models of HMNS}
 \renewcommand{\arraystretch}{1.2}
\begin{center}
 \begin{tabular}{lccccc}
 \hline
 Model            &   $R_{\text{d}} \left[km \right] $  &  $\Sigma_0 \left[10^{22} \right]$  & $\overline{K}\left[10^{-8} \right]$  & $\eta$ &  $\mu$ \\
 \hline\hline
 GNH3-M125        &   $4.105$                     &  $4.8$         & $1.40$           & $0.146$     & $0.007$ \\              
 GNH3-M135        &   $3.432$                     &  $7.3$         & $1.98$           & $0.187$     & $0.016$ \\               

 H4-M125          &   $4.897$                     &  $3.4$         & $0.97$           & $0.125$     & $0.004$ \\                
 H4-M135          &   $3.639$                     &  $6.5$         & $1.80$           & $0.176$     & $0.012$ \\

 ALF2-M125        &   $4.726$                     &  $3.7$         & $1.00$           & $0.129$     & $0.004$ \\                    
 ALF2-M135        &   $2.680$                     &  $10.2$        & $3.24$           & $0.238$     & $0.043$ \\               

 SLy-M125         &   $2.847$                     &  $9.8$         & $2.88$           & $0.207$     & $0.031$ \\                
 SLy-M135         &   $2.911$                     &  $10.2$        & $2.75$           & $0.219$     & $0.031$ \\                

 APR4-M125        &   $3.078$                     &  $8.4$         & $2.46$           & $0.192$     & $0.023$ \\
 APR4-M135        &   $3.417$                     &  $7.4$         & $2.00$           & $0.187$     & $0.016$ \\

 \hline
 \end{tabular}
\end{center}
 \tablecomments{The stability parameters $\eta$ and $\mu$ for five different HMNS numeric models in \cite{hanauske2017rotational}. Each model has simulated for two different cases: high mass ($M=1.35 M_{\odot}$) and low mass ($M=1.25 M_{\odot}$) binaries. Note that the units of $\Sigma_0$ and $\overline{K}$ are $\text{kg}\text{m}^{-2}$ and $\text{m}^3 \text{kg}^{-1}\text{s}^{-2}$ respectively.}
 \label{table1}
 \end{table}

So far our consideration in the previous section has been devoted to an exponential toy disk. More specifically we self-consistently studied the stability of the disk in both Newtonian and PN descriptions. In this section, we apply the results to a more realistic system, i.e., HMNS, in which relativistic effects are important.
Although we  restrict our analysis to HMNS system, we should mention that there are some other environments where the PN corrections may be important. For example, sometimes a black hole-neutron star merger will tidally disrupt the neutron star outside the black hole's innermost stable circular orbit, forming a post-merger torus, and probably a short gamma-ray  burst. It would be interesting to study the stability of the  post-merger black hole-neutron star disk (\citealp{foucart2012black}). 
Moreover, some recent  works (for example see \citealp{inayoshi2016there})  have argued that the maximum observed mass for super-massive black holes (~$10^{10}\,M_{\odot}$) is due to the onset of Toomre instability near the innermost stable circular orbit. Considering the PN effects in this system may shed light on the dynamics of the system.  Also, a very speculative origin for the LIGO black hole-black hole binaries is gravitational fragmentation of the core of a rapidly spinning massive star, during a  core-collapse supernova. GR corrections to the gravitational instability criterion, in principle, can be relevant here (\citealp{d2018single}). In the following, we focus on HMNSs and leave the above-mentioned systems for future studies.

The resulting object in the merger of a binary system of neutron stars is an HMNS. The formation of the HMNSs is a complicated process and consequently, approximative and numerical methods are used to study them. 
The main properties of the HMNSs have been widely investigated in the literature, using numerical simulations in GR. It is worth mentioning that, these simulations cover a wide range of initial conditions and various equations of state and masses (for example see \citealp{hanauske2017rotational}; \citealp{hotokezaka2013remnant}). 

We should also remark that some dynamical instabilities may occur in an HMNS. For example, the magnetorotational instability (MRI) can be established in the HMNS, and can substantially change the strength of the magnetic field (\citealp{duez2006evolution}, \citealp{siegel2013magnetorotational}). Furthermore, as we already mentioned, other dynamical instabilities like Kelvin-Helmholtz and bar like $m=2$ instabilities can also occur in this system. It is also important mentioning that the Papaloizou-Pringle instability has been seen in numerical relativity simulations of post-merger black hole-neutron star pairs (\citealp{mewes2016numerical}).

Although we have restricted ourselves to the linear stability analysis, it is important to mention that the thermodynamical properties of the gas can play an important role in developing the nonlinear gravitational instability (\citealp{gammie2001nonlinear}). 
Numerical experiments of \cite{gammie2001nonlinear} show that the local fragmentation can happen if the cooling is efficient in the system. In other words, there will be a possibility for gravitational instability when the gas can be cooled on a timescale less than $3\Omega^{-1}$. This is known as Gammie criterion. On the other hand, for the cases that the cooling timescale is longer than $3\Omega^{-1}$, the disk reaches a steady \textit{gravitoturbulent} state in which $Q\approx 1$ and the disk is stable against fragmentation, see also (\citealp{rafikov2009properties}). We will discuss this fact with more details  at the end of this section.

Here using a fully general-relativistic simulation done by \citealp{hanauske2017rotational}, we apply our exponential toy model to a set of simulated HMNSs. To do so, we present a rough estimation and search for the footprint of the local gravitational instability. In order to apply our exponential disk toy model to the above-mentioned simulations, we  utilize some simplifications and assumptions as follows:

$i)$ We know that there is no exact EOS for a high-density matter, especially in neutron stars. One type of EOS that is used in the simulations is the piecewise polytropic EOS (\citealp{read2009constraints}). In this EOS, the neutron stars are assumed to be cold, and the rest-mass density determines the other thermodynamical quantities. However, a modified version of this kind of EOS approximately considers also the thermal effects. In this EOS, the pressure and specific internal energy divided into two cold and thermal parts, where the thermal part is considered to be an ideal fluid component. Albeit we follow the analysis presented in \cite{hanauske2017rotational}, that shows that the effect of the thermal component can be ignored throughout the disk. Using the values of physical quantities of HMNSs introduced for five different models in \cite{hanauske2017rotational}, and assuming the general form of the EOS as $p=K\Sigma^{\Gamma}$, one can approximately study the local gravitational instability in the HMNS. It is important mentioning that, as a crude estimation we choose $\Gamma=2$, which is widely used in neutron star studies (e.g.  \citealp{siegel2013magnetorotational}).

$ii)$ The typical value for the polytropic constant $K'$ in the polytropic EOS, i.e., $p=K' \rho ^{\Gamma}$, is $0.014 \, \text{m}^5 \text{kg}^{-1}\text{s}^{-2}$ (or $100$ in the units that $c=G=M_{\odot}=1$, see \citealt{siegel2013magnetorotational}). Here we use this typical value along with the other physical properties of the torus to find an approximate magnitude  of the polytropic constant $K$ for our two-dimensional model. To do so, we assume a scale length $h$ for the thickness of the torus as $h=R_{\text{d}}/i$, where to find a reliable estimation we vary $i$ as $i=3,4,5$. In this case, one may straightforwardly write the polytropic constant as
\begin{align}
K_i \simeq 0.014\left(\frac{R_{\text{d}}}{i} \right) ^{-2}
\end{align}
 Note that we will use the averaged value of $K_i$ denoted by $\overline{K}$.

$iii)$ The mass and radius of the disk and the total mass of the HMNS for each model have been measured in \cite{hanauske2017rotational}. However, the mentioned models also have a core in the center. Nevertheless,  in Fig. 16  of \cite{hanauske2017rotational}, it is seen that at least in the disk, the density profile is more or less exponential. 
Therefore we fit an exponential surface density, i.e., $\Sigma=\Sigma_0 e^{-2y}$ where $y=\frac{R}{2R_{\text{d}}}$, to the models. We find unknown parameters $\Sigma_0$ and $R_{\text{d}}$ by imposing the following restrictions on the surface density
\begin{align}
\int\limits_{0}^{R_{disk}}\Sigma dA=M_{core} \qquad,\qquad \int\limits_{R_{disk}}^{R_t}\Sigma dA=M_{disk}
\end{align}
where $R_t$ is the total radius of the HMNS and we take it as $R_t\simeq 25 \, \text{km}$. We remind that $R_{\text{d}}$ is the characteristic length of the disk. Also, $R_{disk}$ is the radial position where the disk starts and $M_{disk}$ ($M_{core}$) is the total rest mass outside (inside) $R_{disk}$. By solving these two integral equations we can find $\Sigma_0$ and $R_{\text{d}}$. 
Consequently, after finding $\Sigma_0$, $R_{\text{d}}$ and $K$, for each model, one can find stability parameters $\eta$ and $\mu$. Therefore, following our toy model in the previous section, the local stability of each model can be readily checked at radii where the disk is in the PN regime. The results have been summarized in  Tables \ref{table1} and \ref{table2}.
 \begin{table}[t]
 \caption{Toomre's criterion for the allowed models}
 \renewcommand{\arraystretch}{1.2}
 \begin{center}
 \begin{tabular}{lccccc}
  \hline
  Model                            &   $y$          &  $Q$         & $1+C(y,\eta,\mu)$      & NTC           & PNTC             \\
  \hline\hline
 \multirow{2}{*}{GNH3-M125}        &   $4.9$        &  $1.03$      & $0.22$         & \checkmark    & \checkmark      \\    
 \vspace{1.6mm}                    &   $5$          &  $1.11$      & $0.17$         & \checkmark    & \checkmark      \\
 \hline
 \multirow{3}{*}{H4-M125}          &   $4.3$        &  $0.56$      & $0.57$         & --            & --              \\  
                                   &   $4.6$        &  $0.68$      & $0.50$         & --            & \checkmark      \\
 \vspace{1.5mm}                    &   $4.9$        &  $0.84$      & $0.44$         & --            & \checkmark      \\ 
 \hline           
 \multirow{3}{*}{ALF2-M125}        &   $4.4$        &  $0.59$      & $0.54$         & --            & \checkmark      \\ 
                                   &   $4.7$        &  $0.72$      & $0.48$         & --            & \checkmark      \\
                                   &   $4.9$        &  $0.83$      & $0.42$         & --            & \checkmark      \\ 
                     
  \hline
  \end{tabular}
 \end{center}
 \tablecomments{ Investigating Toomre's criterion in the Newtonian and PN viewpoints. The first column is the model name. The second column shows different radii $y$. The third (fourth) column shows the value of Toomre's parameter (R.H.S. of PN Toomre's criterion). The  check marks in the fifth and the sixth columns show the establishing Newtonian and PN Toomre's criterion respectively. Note that, for each model, radii less than the smallest specified radius do not satisfy the conditions \eqref{cond_eta} to \eqref{cond_em3}. \label{table2}}
 \end{table}
 
Table \ref{table1} shows $\eta$ and $\mu$ for five models with different  EOSs.
One can see from this table that the value of $\eta$ varies in the interval  $0.125<\eta<0.238$. On the other hand, regarding the constraint that PN theory imposes on $\eta$ (see Sec. \ref{etamu}), the maximum allowed value of this parameter in the interval $0<y<5$ is $\eta=0.150$. So, the models with a higher value of $\eta$ are not inside the realm of validity of PN theory. Therefore in  Table \ref{table2}, we deal with the allowed models. 

On the other hand, the minimum value of $\eta$ in Table \ref{table1} ($\eta=0.125$) corresponds to $y=4.04$. Therefore, using the ascending nature of Eq. \eqref{cond_eta}, we can show that only the radii $y>4.04$ lie in the PN regime. In other words, at smaller radii, the weak field limit is not satisfied. 

One can check Newtonian as well as PN Toomre's criterion in Table \ref{table2}. In this table,  we have written these criteria as ``NTC" and ``PNTC" for Newtonian and PN Toomre's criterion respectively. We consider the allowed models at different radii. This table shows that in the Newtonian viewpoint, the local fragmentation is possible because, at most radii, the Newtonian Toomre's criterion is not satisfied. On the other hand, at some radii, it is possible for both Newtonian and PN disks to be stable. For example, for the first model, at $y=4.9$ and $y=5$, it is clear from  Table \ref{table2} that the stability condition is satisfied in both cases.

It should be noted that, in the PN description at most allowed radii, the PN Toomre's criterion is established. More specifically, the PN effects stabilize the disk against local fragmentations. On the other hand, considering the physical properties of the HMNS, the Newtonian viewpoint is not trustworthy in this case. Consequently the usual Toomre's criterion cannot be used in this system. In fact, we have explicitly shown that the standard Newtonian stability criterion leads to wrong predictions. Moreover, to the best of our knowledge, no gravitational local fragmentation has been reported in the relevant relativistic simulations. Therefore one may conclude that this PN viewpoint is in agreement with the relevant relativistic numerical simulations.

The stabilizing effects of the relativistic corrections may seem surprising. In fact, considering the complex combination of the physical quantities in the PN calculations, it is not simple to explain the mentioned behavior of PN corrections. However, the effective form of the PN criterion is helpful for a better understanding of this issue. For our models, we have considered the effective PN quantities, i.e., $\kappa_{\text{p}}$, $G_{\text{p}}$, and $c_{{\text{sp}}}$, and compared them with those of Newtonian theory. It turns out that $G_{\text{p}}$ plays a more significant role. In fact, it is clear from Eq. \eqref{eff1} that $G_{\text{p}}$ can be smaller or larger than the Newtonian constant of gravitation. In our toy models, the PN corrections effectively reduce the gravitational ``constant" $G_{\text{p}}$ and consequently yield a higher value for $Q_{\text{p}}$ compared to the Newtonian Toomre's parameter $Q$. Of course, for a more precise interpretation, one should take into account all the effective quantities, and see what is the overall outcome.

As our final remark in this section, and for the sake of completeness, let us consider the Gammie criterion. Using Eq. \eqref{OmegaPN}, one can find  $\Omega_{\text{p}}$ for the allowed models in Table \ref{table2}.  Therefore, the mentioned dynamical timescale $3\Omega^{-1}$ can be calculated. It is straightforward to show that, for the cases studied in Table \ref{table2}, this timescale is shorter than $\sim 1~\text{ms}$. 
On the other hand, for an HMNS the typical timescale of neutrino cooling is $\sim 2 \text{s}$ (\citealp{paschalidis2012importance}). Therefore the cooling  timescale is much larger than the dynamical timescale $3\Omega^{-1}$. This rough estimation, in agreement with the before-mentioned results, means that the local fragmentation cannot occur in an HMNS system.
\section{Discussion and Conclusion}\label{conclusion}
We have studied the local gravitational stability of a relativistic two-dimensional self-gravitating and differentially rotating fluid disk in the context of PN theory. In this approximation, one can investigate the  relativistic effects on the dynamics of the system. 

In fact, we have introduced a relativistic version of Toomre's criterion including relativistic effects up to 1\tiny PN \normalsize  approximation.
For this purpose, we have studied the dynamics of a gaseous disk in Sec. \ref{dispersion_relation_section}, and found the hydrodynamics equation in the PN approximation.
Then we have linearized these equations, and applied the tight winding approximation (WKB) to investigate the local stability of the system. It provides a considerable simplicity in the calculations. Then after finding the PN potentials of the WKB density wave, we solved the linearized PN equations in order to find the dispersion relation for the propagation of small perturbations on the surface of the disk.

Using the PN dispersion relation and introducing some effective quantities as $G_{\text{p}}$, $c_{\text{sp}}$, and $\kappa_{\text{p}}$, we have found the local stability criterion for the fluid disk in the context of PN theory in Sec. \ref{Toomre}. In order to compare this criterion with the Newtonian case, we  have rewritten the PN Toomre's parameter in terms of the Newtonian quantities. The PN Toomre's criterion can be expressed as $Q>1+C(y,\eta,\mu)$, where $C(y,\eta,\mu)$ is the PN correction term.
To complete our result and investigate the effect of each PN correction on the local stability, we  have chosen an exponential disk toy model. 
Also, we have introduced two dimensionless parameters, i.e., $\eta$ and $\mu$, which are made up of a combination of $c$, $G$, $R_{\text{d}}$, $\Sigma_0$, and $K$, to study the behavior of the system more clearly. In fact, these parameters represent the strength of gravity and pressure in the system. 
Also, some restrictions on these parameters have been imposed to satisfy slow-motion conditions in the 1\tiny PN \normalsize approximation.

At first, after finding an allowed range for $\eta$ and $\mu$ which satisfies the PN conditions at all radii, we have studied the general effects of these parameters on the stability of the system. It is clear from the Fig. \ref{Dens_plots} that the role of $\eta$ ($\mu$) is destabilizing (stabilizing) in both Newtonian and PN limits. However, at small radii, increasing $\mu$ can unexpectedly, destabilize both Newtonian and the PN system.  Furthermore, we have shown that the system is more stable in the PN limit compared with the Newtonian case. However, the difference between the two theories gets negligible for small values of both parameters.

Also, we have studied the growth rate of the perturbation in both regimes. For this aim, the dimensionless growth rate, see Eq. \eqref{PN growing rate},  has been plotted versus the dimensionless wavenumber $q$. Fig. \ref{growingRate} shows that the growth rate decreases (increases) and the instability interval gets shorter (larger), by increasing $\mu$ ($\eta$), in a fixed radius. So, one can see that in both Newtonian and PN limits, $\mu$ ($\eta$) has stabilizing (destabilizing) effects. It is worth mentioning that regarding the fact that the PN correction terms  in Eq. \eqref{PN growing rate} must be smaller than the main Newtonian contribution, we could impose some restrictions on $q$. 
Furthermore, one can see from Fig. \ref{growingRate} that at the same radius and with the same values for $\eta$ and $\mu$ parameters, the instability interval is wider in Newtonian limit. In this case the instability interval includes smaller wavelengths. Accordingly, the growth rates are larger in the Newtonian viewpoint. Therefore we conclude tha the system is more stable in the PN limit.

Finally, we have studied the boundary between stable and unstable modes, by using Eqs. \eqref{Q_s} and \eqref{Q(Lambda)}. This boundary is investigated for different radii and various values of the stability parameters. Fig. \ref{Boundary}, in agreement with Fig. \ref{Dens_plots}, shows that the system is more stable in the context of PN theory. 

Finally, as an application, we have applied our calculations to the relativistic disk around an HMNS. We have used a rough estimation, and studied the possibility of occurrence of the local gravitational instability in this system. In fact, we have shown that the standard Newtonian Toomre's criterion allows the system to be locally fragmented. Of course, the Newtonian description is not a reliable way to investigate such a relativistic system. On the other hand, we have shown that, in agreement with the relativistic simulations, the local fragmentation could not happen in the HMNS when we use the PN version of Toomre's criterion.

\section*{acknowledgment}
We would like to appreciate the anonymous referee for constructive and helpful comments. MR would like to thank Shahram Abbassi for continuous encouragement and support during this work. This work is supported by Ferdowsi University of Mashhad under Grant NO. 43710 (26/02/1396).

\bibliographystyle{apj}
\bibliography{short,Toomre}

\begin{thebibliography}{}

\bibitem[\protect\citeauthoryear{Abbott et~al.}{Abbott
  et~al.}{2017}]{abbott2017gw170817}
Abbott, B.~P., et~al. 2017, PhRvL, 119, 161101

\bibitem[\protect\citeauthoryear{Anderson et~al.}{Anderson
  et~al.}{2008}]{anderson2008magnetized}
Anderson, M., Hirschmann, E.~W., Lehner, L., Liebling, S.~L., Motl, P.~M.,
  Neilsen, D., Palenzuela, C.,  \& Tohline, J.~E. 2008, PhRvL, 100, 191101

\bibitem[\protect\citeauthoryear{Bertin \& Romeo}{Bertin \&
  Romeo}{1988}]{bertin1988global}
Bertin, G.,  \& Romeo, A.~B. 1988, A\&A, 195, 105

\bibitem[\protect\citeauthoryear{Binney \& Tremaine}{Binney \&
  Tremaine}{2008}]{binney2008galactic}
Binney, J.,  \& Tremaine, S. 2008, Galactic dynamics (Princeton university
  press)

\bibitem[\protect\citeauthoryear{Blanchet}{Blanchet}{2006}]{blanchet2006}
Blanchet, L. 2006, LRR, 9, 4

\bibitem[\protect\citeauthoryear{Blanchet}{Blanchet}{2014}]{blanchet2014gravitational}
Blanchet, L. 2014, LRR, 17, 2

\bibitem[\protect\citeauthoryear{Blanchet \& Damour}{Blanchet \&
  Damour}{1989}]{blanchet1989post}
Blanchet, L.,  \& Damour, T. 1989, Ann. Inst. Henri Poincar{\'e}, A, 50, 377

\bibitem[\protect\citeauthoryear{Blanchet, Damour, \& Iyer}{Blanchet
  et~al.}{1995}]{blanchet1995gravitational}
Blanchet, L., Damour, T.,  \& Iyer, B.~R. 1995, PhRvD, 51, 5360

\bibitem[\protect\citeauthoryear{Blanchet \& Sch{\"a}fer}{Blanchet \&
  Sch{\"a}fer}{1989}]{blanchet1989higher}
Blanchet, L.,  \& Sch{\"a}fer, G. 1989, MNRAS, 239, 845

\bibitem[\protect\citeauthoryear{Blandford \& Teukolsky}{Blandford \&
  Teukolsky}{1976}]{blandford1976arrival}
Blandford, R.,  \& Teukolsky, S.~A. 1976, ApJ, 205, 580

\bibitem[\protect\citeauthoryear{Burke}{Burke}{1971}]{burke1971gravitational}
Burke, W.~L. 1971, JMP, 12, 401

\bibitem[\protect\citeauthoryear{Chandrasekhar}{Chandrasekhar}{1965}]{chandrasekhar1965post}
Chandrasekhar, S. 1965, ApJ, 142, 1488

\bibitem[\protect\citeauthoryear{Chandrasekhar}{Chandrasekhar}{1967}]{chandrasekhar1967post}
Chandrasekhar, S. 1967, ApJ, 148, 621

\bibitem[\protect\citeauthoryear{Chandrasekhar}{Chandrasekhar}{1969}]{chandrasekhar1969conservation}
Chandrasekhar, S. 1969, ApJ, 158, 45

\bibitem[\protect\citeauthoryear{Chandrasekhar \& Esposito}{Chandrasekhar \&
  Esposito}{1970}]{chandrasekhar1970212}
Chandrasekhar, S.,  \& Esposito, F.~P. 1970, ApJ, 160, 153

\bibitem[\protect\citeauthoryear{Chandrasekhar \& Nutku}{Chandrasekhar \&
  Nutku}{1969}]{chandrasekhar1969Second}
Chandrasekhar, S.,  \& Nutku, Y. 1969, ApJ, 158, 55

\bibitem[\protect\citeauthoryear{Damour \& Taylor}{Damour \&
  Taylor}{1991}]{damour1991orbital}
Damour, T.,  \& Taylor, J.~H. 1991, ApJ, 366, 501

\bibitem[\protect\citeauthoryear{Demianski \& Ivanov}{Demianski \&
  Ivanov}{1997}]{demianski1997dynamics}
Demianski, M.,  \& Ivanov, P. 1997, A\&A, 324, 829

\bibitem[\protect\citeauthoryear{Duez et~al.}{Duez
  et~al.}{2006}]{duez2006evolution}
Duez, M.~D., Liu, Y.~T., Shapiro, S.~L., Shibata, M.,  \& Stephens, B.~C. 2006,
  PhRvD, 73, 104015

\bibitem[\protect\citeauthoryear{D’Orazio \& Loeb}{D’Orazio \&
  Loeb}{2018}]{d2018single}
D’Orazio, D.~J.,  \& Loeb, A. 2018, PhRvD, 97, 083008

\bibitem[\protect\citeauthoryear{Ellis et~al.}{Ellis
  et~al.}{2018}]{ellis2017search}
Ellis, J., Hektor, A., H{\"u}tsi, G., Kannike, K., Marzola, L., Raidal, M.,  \&
  Vaskonen, V. 2018, PhLB

\bibitem[\protect\citeauthoryear{Elmegreen}{Elmegreen}{1994}]{elmegreen1994supercloud}
Elmegreen, B. 1994, ApJ, 433, 39

\bibitem[\protect\citeauthoryear{Elmegreen}{Elmegreen}{1987}]{elmegreen1987supercloud}
Elmegreen, B.~G. 1987, ApJ, 312, 626

\bibitem[\protect\citeauthoryear{Epstein}{Epstein}{1977}]{epstein1977binary}
Epstein, R. 1977, ApJ, 216, 92

\bibitem[\protect\citeauthoryear{Fan \& Lou}{Fan \& Lou}{1997}]{fan1997swing}
Fan, Z.,  \& Lou, Y.-Q. 1997, MNRAS, 291, 91

\bibitem[\protect\citeauthoryear{Fock}{Fock}{1959}]{fock1959theory}
Fock, V. 1959, The theory of space time and gravitation pergamon

\bibitem[\protect\citeauthoryear{Foucart}{Foucart}{2012}]{foucart2012black}
Foucart, F. 2012, PhRvD, 86, 124007

\bibitem[\protect\citeauthoryear{Gammie}{Gammie}{1996}]{gammie1996linear}
Gammie, C.~F. 1996, ApJ, 462, 725

\bibitem[\protect\citeauthoryear{Gammie}{Gammie}{2001}]{gammie2001nonlinear}
Gammie, C.~F. 2001, ApJ, 553, 174

\bibitem[\protect\citeauthoryear{Goodman \& Xu}{Goodman \&
  Xu}{1994}]{goodman1994parasitic}
Goodman, J.,  \& Xu, G. 1994, ApJ, 432, 213

\bibitem[\protect\citeauthoryear{Hanauske et~al.}{Hanauske
  et~al.}{2017}]{hanauske2017rotational}
Hanauske, M., Takami, K., Bovard, L., Rezzolla, L., Font, J.~A., Galeazzi, F.,
  \& St{\"o}cker, H. 2017, PhRvD, 96, 043004

\bibitem[\protect\citeauthoryear{Hotokezaka et~al.}{Hotokezaka
  et~al.}{2013}]{hotokezaka2013remnant}
Hotokezaka, K., Kiuchi, K., Kyutoku, K., Muranushi, T., Sekiguchi, Y.-i.,
  Shibata, M.,  \& Taniguchi, K. 2013, PhRvD, 88, 044026

\bibitem[\protect\citeauthoryear{Hulse \& Taylor}{Hulse \&
  Taylor}{1975}]{hulse1975deep}
Hulse, R.,  \& Taylor, J. 1975, ApJ, 201, L55

\bibitem[\protect\citeauthoryear{Inayoshi \& Haiman}{Inayoshi \&
  Haiman}{2016}]{inayoshi2016there}
Inayoshi, K.,  \& Haiman, Z. 2016, ApJ, 828, 110

\bibitem[\protect\citeauthoryear{Jalali}{Jalali}{2007}]{jalali2007unstable}
Jalali, M.~A. 2007, ApJ, 669, 218

\bibitem[\protect\citeauthoryear{Jog}{Jog}{1996}]{jog1996local}
Jog, C.~J. 1996, MNRAS, 278, 209

\bibitem[\protect\citeauthoryear{Jog \& Solomon}{Jog \&
  Solomon}{1984}]{jog1984two}
Jog, C.~J.,  \& Solomon, P. 1984, ApJ, 276, 114

\bibitem[\protect\citeauthoryear{Julian \& Toomre}{Julian \&
  Toomre}{1966}]{julian1966non}
Julian, W.~H.,  \& Toomre, A. 1966, ApJ, 146, 810

\bibitem[\protect\citeauthoryear{Kalnajs}{Kalnajs}{1977}]{kalnajs1977dynamics}
Kalnajs, A. 1977, ApJ, 212, 637

\bibitem[\protect\citeauthoryear{Kalnajs}{Kalnajs}{1972}]{kalnajs1972equilibria}
Kalnajs, A.~J. 1972, ApJ, 175, 63

\bibitem[\protect\citeauthoryear{Kim \& Ostriker}{Kim \&
  Ostriker}{2001}]{kim2001amplification}
Kim, W.-T.,  \& Ostriker, E.~C. 2001, ApJ, 559, 70

\bibitem[\protect\citeauthoryear{Kormendy \& Kennicutt}{Kormendy \&
  Kennicutt}{2004}]{kormendy2004secular}
Kormendy, J.,  \& Kennicutt, R.~C. 2004, ARA\&A, 42, 603

\bibitem[\protect\citeauthoryear{Leroy et~al.}{Leroy
  et~al.}{2008}]{leroy2008star}
Leroy, A.~K., Walter, F., Brinks, E., Bigiel, F., De~Blok, W., Madore, B.,  \&
  Thornley, M. 2008, AJ, 136, 2782

\bibitem[\protect\citeauthoryear{Lin \& Pringle}{Lin \&
  Pringle}{1987}]{lin1987formation}
Lin, D.,  \& Pringle, J. 1987, ApJ, 320, L87

\bibitem[\protect\citeauthoryear{McKee \& Ostriker}{McKee \&
  Ostriker}{2007}]{mckee2007theory}
McKee, C.~F.,  \& Ostriker, E.~C. 2007, ARA\&A, 45, 565

\bibitem[\protect\citeauthoryear{Mewes et~al.}{Mewes
  et~al.}{2016}]{mewes2016numerical}
Mewes, V., Font, J.~A., Galeazzi, F., Montero, P.~J.,  \& Stergioulas, N. 2016,
  PhRvD, 93, 064055

\bibitem[\protect\citeauthoryear{Narayan \& Yi}{Narayan \&
  Yi}{1994}]{narayan1994advection}
Narayan, R.,  \& Yi, I. 1994, ApJ, 428, L13

\bibitem[\protect\citeauthoryear{Nazari et~al.}{Nazari
  et~al.}{2017}]{nazari2017post}
Nazari, E., Kazemi, A., Roshan, M.,  \& Abbassi, S. 2017, ApJ, 839, 75

\bibitem[\protect\citeauthoryear{Obergaulinger et~al.}{Obergaulinger
  et~al.}{2009}]{obergaulinger2009semi}
Obergaulinger, M., Cerd{\'a}-Dur{\'a}n, P., M{\"u}ller, E.,  \& Aloy, M. 2009,
  A\&A, 498, 241

\bibitem[\protect\citeauthoryear{Paschalidis, Etienne, \& Shapiro}{Paschalidis
  et~al.}{2012}]{paschalidis2012importance}
Paschalidis, V., Etienne, Z.~B.,  \& Shapiro, S.~L. 2012, PhRvD, 86, 064032

\bibitem[\protect\citeauthoryear{Poisson \& Will}{Poisson \&
  Will}{2014}]{poisson2014gravity}
Poisson, E.,  \& Will, C.~M. 2014, Gravity: Newtonian, Post-Newtonian,
  Relativistic (Cambridge University Press)

\bibitem[\protect\citeauthoryear{Rafikov}{Rafikov}{2001}]{rafikov2001local}
Rafikov, R.~R. 2001, MNRAS, 323, 445

\bibitem[\protect\citeauthoryear{Rafikov}{Rafikov}{2009}]{rafikov2009properties}
Rafikov, R.~R., 704, 281

\bibitem[\protect\citeauthoryear{Read et~al.}{Read
  et~al.}{2009}]{read2009constraints}
Read, J.~S., Lackey, B.~D., Owen, B.~J.,  \& Friedman, J.~L. 2009, PhRvD, 79,
  124032

\bibitem[\protect\citeauthoryear{Rezania}{Rezania}{2000}]{rezania2000normal}
Rezania, V. 2000, arXiv preprint gr-qc/0002070

\bibitem[\protect\citeauthoryear{Romeo}{Romeo}{1990}]{romeo1990thesis}
Romeo, A.~B. 1990, Ph.D. thesis, Scuola Internazionale Superiore di Studi
  Avanzati

\bibitem[\protect\citeauthoryear{Romeo}{Romeo}{1992}]{romeo1992stability}
Romeo, A.~B. 1992, MNRAS, 256, 307

\bibitem[\protect\citeauthoryear{Romeo}{Romeo}{1994}]{romeo1994faithful}
Romeo, A.~B. 1994, A\&A, 286

\bibitem[\protect\citeauthoryear{Safronov}{Safronov}{1960}]{safronov1960gravitational}
Safronov, V. 1960, in Annales d'Astrophysique, Vol.~23, 979

\bibitem[\protect\citeauthoryear{Sellwood}{Sellwood}{2014}]{sellwood2014secular}
Sellwood, J. 2014, RvMP, 86, 1

\bibitem[\protect\citeauthoryear{Shadmehri \& Khajenabi}{Shadmehri \&
  Khajenabi}{2012}]{shadmehri2012gravitational}
Shadmehri, M.,  \& Khajenabi, F. 2012, MNRAS, 421, 841

\bibitem[\protect\citeauthoryear{Shibata, Baumgarte, \& Shapiro}{Shibata
  et~al.}{2000}]{shibata2000bar}
Shibata, M., Baumgarte, T.~W.,  \& Shapiro, S.~L. 2000, ApJ, 542, 453

\bibitem[\protect\citeauthoryear{Siegel et~al.}{Siegel
  et~al.}{2013}]{siegel2013magnetorotational}
Siegel, D.~M., Ciolfi, R., Harte, A.~I.,  \& Rezzolla, L. 2013, PhRvD, 87,
  121302

\bibitem[\protect\citeauthoryear{Silk \& Norman}{Silk \&
  Norman}{1981}]{silk1981dissipational}
Silk, J.,  \& Norman, C. 1981, ApJ, 247, 59

\bibitem[\protect\citeauthoryear{Thorne \& Will}{Thorne \&
  Will}{1971}]{thorne1971theoreticali}
Thorne, K.~S.,  \& Will, C.~M. 1971, ApJ, 163, 595

\bibitem[\protect\citeauthoryear{Toomre}{Toomre}{1964}]{toomre1964gravitational}
Toomre, A. 1964, ApJ, 139, 1217

\bibitem[\protect\citeauthoryear{Vandervoort}{Vandervoort}{1970}]{vandervoort1970equilibria}
Vandervoort, P.~O. 1970, ApJ, 161, 67

\bibitem[\protect\citeauthoryear{Wang \& Silk}{Wang \&
  Silk}{1994}]{wang1994gravitational}
Wang, B.,  \& Silk, J. 1994, ApJ, 427, 759

\bibitem[\protect\citeauthoryear{Will}{Will}{1994}]{will1994proceedings}
Will, C. 1994, in Proceedings of the Eighth Nishinomiya-Yukawa Memorial
  Symposium on Relativistic Cosmology, ed. M.~Sasaki (Tokyo: Universal Academy
  Press)

\bibitem[\protect\citeauthoryear{Will}{Will}{1971a}]{will1971theoreticalii}
Will, C.~M. 1971a, ApJ, 163, 611

\bibitem[\protect\citeauthoryear{Will}{Will}{1971b}]{will1971theoreticaliii}
Will, C.~M. 1971b, ApJ, 169, 125

\bibitem[\protect\citeauthoryear{Will}{Will}{1987}]{thorne1987300}
Will, C.~M. 1987, in 300 years of gravitation, ed. K.~Thorne, S.~Hawking, \&
  W.~Israel (Cambridge, England: Cambridge University Press), 80

\bibitem[\protect\citeauthoryear{Will}{Will}{2014}]{will2014confrontation}
Will, C.~M. 2014, LRR, 17, 4

\bibitem[\protect\citeauthoryear{Will \& Wiseman}{Will \&
  Wiseman}{1996}]{will1996gravitational}
Will, C.~M.,  \& Wiseman, A.~G. 1996, PhRvD, 54, 4813

\end{thebibliography}
\appendix
\section{the forth-degree dispersion relation}\label{exact quadratic function}
In this Appendix, we show the complete form of Eq. \eqref{abb form omega}. This equation is the fourth-degree dispersion relation in PN theory. As we mentioned in Sec. \ref{dis PN}, the set of three equations \eqref{Eul R}, \eqref{Eul phi} and \eqref{vRa} for variables $v_{Ra}$, $v_{\varphi a}$, and $\Sigma_a^*$ has nontrivial solutions when
\begin{eqnarray}\label{omega4app}
\omega ^4 \left(A_4+\frac{B_4}{c^2}\right)+\omega ^3 \left(A_3+\frac{B_3}{c^2}\right)+\omega ^2 \left(A_2+\frac{B_2}{c^2}\right)+\omega\left(A_1+\frac{B_1}{c^2}\right)+A_0+\frac{B_0}{c^2}=0
\end{eqnarray}
We show each coefficient of $\omega^i$ (where $i=0$ to $4$) by $A_i$ (the Newtonian part) and $B_i$ (the coefficients of $c^{-2}$). After applying the WKB approximation, the Newtonian and PN coefficients of the  fourth-order equation read
\begin{align}
A_0=2\,m^2\,\pi\,G\,\Omega_{\text{p}}^2\,\text{sign}(k)-\frac{m^2\,\Omega_{\text{p}}^2}{k\,\Sigma^*} \Big(k^2\,c_{\text{s}}^{*2}+\left(2-m^2\right)\Omega_{\text{p}}^2+2\Omega_{\text{p}}\left(R\,\Omega_{\text{p}}\right)'\Big)
\end{align}
\begin{align}
B_0= & -m^2\pi\,G\,\Omega_{\text{p}}^2\,\text{sign}(k)\Big( 3\,R^2\,\Omega_{\text{p}}^2-\frac{2\,p}{\Sigma^*}-2\,\Pi+10\,U\Big)+\frac{m^2\,k\,\Omega_{\text{p}}^2\,c_{\text{s}}^{*2}}{\Sigma^*}\Big(\frac{R^2\,\Omega_{\text{p}}^2}{2}+\frac{p}{\Sigma^*}+\Pi+U\Big)+\frac{m^2}{k\,R\,\Sigma^*}\\\nonumber
&\times\Big[4\big(4\,\Omega_{\text{p}}^3+R\,\Omega_{\text{p}}^2\Omega_{\text{p}}'\big)\big(\frac{U_{\varphi}}{R}+U_{\varphi}'\big)-2\,R\Omega_{\text{p}}\left(6\,\Omega_{\text{p}}^3+R\,\Omega_{\text{p}}^2\Omega_{\text{p}}'\right)U'\Big]
\end{align}

\begin{align}
A_1=-4\,m\,\pi\,G\,\Omega_{\text{p}}\,\text{sign}(k)+\frac{2\,m\,\Omega_{\text{p}}}{k\,\Sigma^*}\Big(k^2\,c_{\text{s}}^{*2}+\left(2-2\,m^2\right)\Omega_{\text{p}}^2+2\,\Omega_{\text{p}}\left(R\,\Omega_{\text{p}}\right)'\Big)
\end{align}
\begin{align} 
B_1= & 2\,m\,\pi\,G\,\Omega_{\text{p}}\,\text{sign}(k)\Big(3\,R^2\,\Omega_{\text{p}}^2-\frac{2\,p}{\Sigma^*}-2\,\Pi+10\,U\Big)-\frac{m\,k\,\Omega_{\text{p}}\,c_{\text{s}}^{*2}}{\Sigma^*}\Big(R^2\,\Omega_{\text{p}}^2+\frac{2\,p}{\Sigma^*}+2\,\Pi+2\,U\Big)-\frac{m\,\Omega_{\text{p}}}{k\,\Sigma^*}\\\nonumber
&\times\Big[8\big(4\,\Omega_{\text{p}}+R\,\Omega_{\text{p}}'\big)\big(\frac{U_{\varphi}}{R}+U_{\varphi}'\big)-4\,R\Omega_{\text{p}}\left(6\,R\Omega_{\text{p}}+R^2\Omega_{\text{p}}'\right)U'\Big]
\end{align}

\begin{align}
A_2=2\,\pi\,G\,\text{sign}(k)-\frac{1}{k\,\Sigma^*}\Big(k^2\,c_{\text{s}}^{*2}+\left(2-6\,m^2\right)\,\Omega_{\text{p}} ^2+2\,\Omega_{\text{p}}\left(R\,\Omega_{\text{p}}\right)'\Big)
\end{align}
\begin{align} 
B_2= & -\pi\,G\,\text{sign}(k)\Big(3\,R^2\,\Omega_{\text{p}}^2-\frac{2p}{\Sigma^*}-2\,\Pi+10\,U\Big)+\frac{k\,c_{\text{s}}^{*2}}{\Sigma^*} \Big(\frac{R^2\,\Omega_{\text{p}} ^2}{2}+\frac{p}{\Sigma^*}+\Pi+U\Big)+\frac{1}{k\,\Sigma^*}\\\nonumber
&\times \Big[4\big(4\,\Omega_{\text{p}}+R\,\Omega_{\text{p}}'\big)\big(\frac{U_{\varphi}}{R}+U_{\varphi}'\big)-2\,\Omega_{\text{p}}\left(6\,R\Omega_{\text{p}}+R^2\Omega_{\text{p}}'\right)U'\Big]
\end{align}
\begin{align}
 A_3=-\frac{4\,m\,\Omega_{\text{p}} }{k\,\Sigma^*},~~~ B_3=\frac{28\, m\pi G \Omega_{\text{p}}}{k\,\left|k\right|},~~~ A_4=\frac{1}{k\,\Sigma^* },~~~ B_4=-\frac{6\,\pi\,G}{k\,\left|k\right| }
\end{align} 
As mentioned before, the solutions of this fourth-degree equation are long and complicated. Therefore we do not write them here. Nevertheless, neglecting the $O(c^{-2})$ terms, these solutions reproduce the standard case written as
\begin{eqnarray}
\nonumber
\omega_{1,2}&=& m\Omega \mp\sqrt{k^2\,c_{\text{s}}^{*2}-2\,\pi\,G\,|k|\,\Sigma_0^*+\kappa^2}\\\nonumber
\omega_{3,4}&=&  m\Omega
\end{eqnarray}
One can easily verify that $\omega_{1}$ and $\omega_{2}$ can be expressed as the compact form $(m\Omega-\omega)^2=k^2\,c_{\text{s}}^{*2}-2\,\pi\,G\,|k|\,\Sigma_0^*+2\,R\,\Omega\,\Omega'+4\,\Omega^2$, which is the well-known Newtonian dispersion relation for non-axisymmetric WKB density waves.

\end{document}